\documentclass[11pt,prd,showpacs,
tightenlines, 
a4paper,preprintnumbers,
superscriptaddress,
nofootinbib]{revtex4}

\usepackage[T1]{fontenc}
\usepackage{charter}

\usepackage{graphicx}
\usepackage{epsfig}
\usepackage{amsmath}
\usepackage{mathrsfs}
\usepackage{xcolor}
\usepackage{slashed}
\usepackage{multirow}
\usepackage{diagbox}

\usepackage[colorlinks,citecolor=blue,linktoc=all,linkcolor=cyan,urlcolor=blue]{hyperref}
\usepackage[a4paper,portrait,top=1.4cm, bottom=1.8cm,left=1.3cm,right=1.4cm]{geometry}
\usepackage{tabularx}
\newcolumntype{Y}{>{\centering\arraybackslash}X}
\def\beq{\begin{equation}}
\def\eeq{\end{equation}}
\def\bea{\begin{eqnarray}}
\def\eea{\end{eqnarray}}
\def\beqa{\begin{equation}\begin{array}{l}}
\def\eeqa{\end{array}\end{equation}}
\def\eqlab#1{\label{eq:#1}}

\def\eref#1{(\ref{eq:#1})}
\def\eqref#1{Eq.~(\ref{eq:#1})}

\def\Figref#1{Fig.~\ref{fig:#1}}

\def\secref#1{\ref{sec:#1}}

\def\half{\mbox{\small{$\frac{1}{2}$}}}

\def\barr{\left(\begin{array}{c}}
\def\earr{\end{array}\right)}
\def\bmat{\left(\begin{array}{cc}}
\def\emat{\end{array}\right)}
\def\al{\alpha}
\def\be{\beta}
\def\ga{\gamma} 
\def\de{\delta} \def\De{\Delta}\def\vDe{\varDelta}
\def\veps{\varepsilon}  \def\eps{\epsilon}

 \def\Si{{\it\Sigma}}
\def\th{\theta}  
\def\w{\omega}

\def\nn{\nonumber}

\def\rN{{\rm N}}

\def\cO{\mathscr{O}}

\def\MA{{\mathcal A}}

\def\N{{\scriptscriptstyle \rm N}}

\def\piN{\ensuremath{\pi \rN}}

\def\3d{3-D}

\def\hpm#1{\hphantom{#1}}
\def\textfrac#1#2{{\textstyle\frac{#1}{#2}}}
\def\etal{{\it et al}}
\def\MN{M_\N}
\def\EN{E_\N}
\def\mpi{m_\pi}
\def\alphae{\alpha_{E1}}
\def\betam{\beta_{M1}}

\DeclareMathOperator{\re}{Re}
\DeclareMathOperator{\im}{Im}
\DeclareMathOperator{\Max}{Max}

\def\AnswerYes{y}
\def\ShowLabelsVersion{n}         
\def\ShowChangesVersion{n}        
\def\ShowAnnotationsVersion{n}    
\ifx\ShowLabelsVersion\AnswerYes
 \usepackage{showkeys}
\fi
\ifx\ShowAnnotationsVersion\AnswerYes
  \newcommand{\comment}[1]{{\color{blue}\textit{#1}}}
  \newcommand{\margin}[1]{\marginpar{\scriptsize\sffamily\bfseries{#1}}}
\else
  \newcommand{\comment}[1]{}
  \newcommand{\margin}[1]{}
  
\fi
\ifx\ShowChangesVersion\AnswerYes
  \usepackage{ulem}
   
  \newcommand{\delete}[1]{\sout{#1}}            
  \renewcommand{\emph}[1]{\textit{#1}}           
\else

  \newcommand{\sout}[1]{}
  \newcommand{\xout}[1]{}

  \newcommand{\delete}[1]{}
\fi

\begin{document}
\preprint{MITP/15-017}


\title{Predictions of covariant chiral perturbation theory for nucleon polarisabilities and polarised Compton scattering}

\author{Vadim Lensky}\email{lensky@itep.ru}
\affiliation{Institut f\"ur Kernphysik \& PRISMA  Cluster of Excellence, Johannes Gutenberg Universit\"at Mainz,  D-55128 Mainz, Germany}
\affiliation{Institute for Theoretical and Experimental Physics, 117218 Moscow, Russia}
\affiliation{National Research Nuclear University MEPhI (Moscow Engineering Physics Institute), 115409 Moscow, Russia}
\affiliation{Theoretical Physics Group, School of Physics and Astronomy, University of Manchester, Manchester, M13 9PL, United Kingdom}

\author{Judith A. McGovern}
\affiliation{Theoretical Physics Group, School of Physics and Astronomy, University of Manchester, Manchester, M13 9PL, United Kingdom}

\author{Vladimir Pascalutsa}
\affiliation{Institut f\"ur Kernphysik \& PRISMA  Cluster of Excellence, Johannes Gutenberg Universit\"at Mainz,  D-55128 Mainz, Germany}

\date{\today}

\begin{abstract}
We update the predictions of the SU(2) baryon chiral perturbation theory
 for the dipole polarisabilities of the proton, $\{ \alpha_{E1} , \, \beta_{M1}\}_p 
 = \{ 11.2(0.7), \, 3.9(0.7)\} \times 10^{-4}$ fm$^3$, and obtain the corresponding predictions for 
 the quadrupole, dispersive, and spin polarisabilities: 
$ \{ \alpha_{E2} , \, \beta_{M2} \}_p
 = \{ 17.3(3.9), \, -15.5(3.5)\} \times 10^{-4}$ fm$^5$, $\{\alpha_{E1\nu} , \, \beta_{M1\nu}\}_p  = \{ -1.3(1.0), \, 7.1(2.5)\} \times 10^{-4}$ fm$^5$,
 and $ \{ \gamma_{E1E1} , \, \gamma_{M1M1},\, \gamma_{E1M2} , \, \gamma_{M1E2}  \}_p
 = \{ -3.3(0.8), \, 2.9(1.5),\, 0.2(0.2),\, 1.1(0.3) \} \times 10^{-4}$ fm$^4$. The results for the scalar polarisabilities are
in significant disagreement with semi-empirical analyses based on dispersion relations, however the results for the spin
polarisabilities agree remarkably well. 
Results for proton Compton-scattering multipoles and polarised observables up to the Delta(1232) resonance region are presented
too. The asymmetries $\Sigma_3$ and $\Sigma_{2x}$ reproduce 
the experimental data from LEGS and MAMI. Results for $\Si_{2z}$ agree with a recent sum rule
evaluation in the forward kinematics. The asymmetry $\Si_{1z}$ near the pion production threshold shows a large sensitivity to
chiral dynamics, but no data is available for this observable.
We also provide the predictions for the polarisabilities of the neutron, the numerical values
being
$\{ \alpha_{E1} , \, \beta_{M1}\}_n 
 = \{ 13.7(3.1), \, 4.6(2.7)\} \times 10^{-4}$ fm$^3$,
 $ \{ \alpha_{E2} , \, \beta_{M2} \}_n
 = \{ 16.2(3.7), \, -15.8(3.6)\} \times 10^{-4}$ fm$^5$, $\{\alpha_{E1\nu} , \, \beta_{M1\nu}\}_n  = \{ 0.1(1.0), \, 7.2(2.5)\} \times 10^{-4}$ fm$^5$,
 and $ \{ \gamma_{E1E1} , \, \gamma_{M1M1},\, \gamma_{E1M2} , \, \gamma_{M1E2}  \}_n
 = \{ -4.7(1.1), \, 2.9(1.5),\, 0.2(0.2),\, 1.6(0.4) \} \times 10^{-4}$ fm$^4$.
The neutron dynamical polarisabilities and multipoles are examined too.
We also discuss subtleties related to matching dynamical and static
polarisabilities.
\end{abstract}


                             
\date{\today}
\maketitle

\newpage
\tableofcontents

\newpage
\section{Introduction}
Low-energy  Compton scattering off the nucleon is traditionally used to access the {\it nucleon polarisabilities}, but the
relation between the observables and the polarisabilities is not straightforward; see refs.~\cite{Griesshammer:2012we,Holstein:2013kia,Hagelstein:2015} for recent reviews. A direct relation only exists at the level
of the low-energy expansion (LEX), which is a polynomial expansion in the photon energy $\w$. The LEX validity is limited to very
low energies, well below the pion production threshold ($\w \ll m_\pi$). Most of the experimental data are, however, obtained
at energies above 100 MeV. Chiral perturbation theory ($\chi$PT) 
\cite{Weinberg:1978kz,Gasser:1983yg} in the single-nucleon sector \cite{GSS89}, as a low-energy
effective-field theory of QCD, can provide the necessary model-independent link between the nucleon Compton-scattering data and polarisabilities
beyond the LEX applicability. 
In recent years  $\chi$PT calculations have been performed
 using both the heavy-baryon (HB$\chi$PT)
and manifestly-Lorentz-covariant (B$\chi$PT) formulations;
see respectively \cite{McGovern:2001dd,Beane:2002wn,Beane:2004ra, Pascalutsa:2002pi,McGovern:2012ew}
and \cite{Lensky:2008re,Lensky:2009uv,Lensky:2014efa}; some details of the differences between the  two approaches can be found in \cite{Lensky:2012ag}. These calculations are  not only useful for precision determination of the  
polarisabilities from experiment --- they  also
provide a testing ground for $\chi$PT in the single
nucleon sector.\footnote{Previous work on predictions for the  polarisabilities of the nucleons in the framework of $\chi$PT is contained in refs.~\cite{Bernard:1991rq,Bernard:1993bg,Bernard:1995dp,Hemmert:1997tj,Holstein:1999uu,Ji:1999sv,VijayaKumar:2000pv,Gellas:2000mx,Bernard:2002pw,Bernard:2012hb,Blin:2015era}, and in a chiral framework in~\cite{Gasparyan:2011yw}.
}

In this paper we follow up on the manifestly-covariant calculation of ref.~\cite{Lensky:2009uv}, where the next-to-next-to-leading order (NNLO) calculation of proton Compton scattering
was carried out in SU(2) B$\chi$PT with pion, nucleon and Delta(1232) degrees of freedom. This NNLO calculation provides
a prediction for all the nucleon polarisabilities and scattering observables; however only
the proton scalar dipole polarisabilities and the unpolarised differential cross sections were examined
in ref.~\cite{Lensky:2009uv}. In the present paper
we consider a number of the other predictions. We correct the numerical results for the dipole polarisabilities of the proton and present 
the corresponding results for the neutron. We present
the NNLO results, both analytically and numerically, for the quadrupole polarisabilities and the four dipole spin polarisabilities.
We also extend the calculation of~\cite{Lensky:2009uv} to the Delta(1232)-resonance region and
present results for the Compton multipoles and dynamical polarisabilities. In doing so, we shall remark
on a subtlety in the matching of the multipole expansion and static polarisabilities. Finally, we shall consider the  B$\chi$PT predictions
for the polarised proton observables, namely the asymmetries $\Si_3$, $\Si_{2x}$, $\Si_{2z}$, and $\Si_{1z}$.
These results should be of particular interest to
the ongoing Compton-scattering experiments at MAMI (Mainz) and HIGS (Duke) facilities, which aim to 
determine the scalar and spin polarisabilities of the proton and neutron in polarised measurements \cite{MainzA2-11,Weller:2009zza,Martel:2014pba}.

The paper is structured as follows. In Sect.~\ref{sec:amp} we recall the main ingredients of the NNLO B$\chi$PT calculation of Compton scattering.  
In Sect.~\ref{sec:amp:tensor} we introduce various parametrisations of the Compton amplitude and establish relations between them.
We provide the predictions of B$\chi$PT for the nucleon static polarisabilities in Sect.~\ref{sec:pols:static}, and the Compton multipoles and dynamical polarisabilities in Sect.~\ref{sec:pols:dyn}, and compare them with various empirical and theoretical results. Section~\ref{sec:obs} 
presents our predictions for some of the polarised observables, compared
with experimental data where possible. We conclude with  Sect.~\ref{sec:concl}.

\section{Compton scattering in \texorpdfstring{B$\boldsymbol{\chi}$PT}{BChPT}}
\label{sec:amp}

%
Following refs.~\cite{Lensky:2008re,Lensky:2009uv,Lensky:2014efa}, where one can find  details
such as the relevant $\chi$PT Lagrangians, we consider low-energy Compton 
scattering in B$\boldsymbol{\chi}$PT, i.e., a manifestly-covariant formulation of 
$\chi$PT with pion, nucleon and Delta(1232) isobar degrees of freedom; see ref.~\cite[Sect.\ 4]{Pascalutsa:2006up} for review. 
Our present calculation of the Compton amplitude follows that  of ref.~\cite{Lensky:2009uv} below photoproduction threshold (and hence for the static polarisabilities), but improves the treatment of the Delta-excitation near the resonance, as described below.

\subsection{Power counting}
Our EFT expansion uses the $\delta$-counting~\cite{Pascalutsa:2002pi}, where the mass difference between the nucleon and the Delta, $\varDelta=M_\Delta-\MN$, is considered
an intermediate scale, so that \beq
m_\pi/\varDelta\simeq\varDelta/\Lambda_\chi \equiv\delta,
\eeq
and hence the usual chiral expansion scale $m_\pi/\Lambda_\chi$ is counted as $\delta^2$.
An important feature of the $\delta$-counting is that the characteristic momentum $p$ distinguishes two regimes:
{\it low energy:} $p\simeq m_\pi$, and 
{\it Delta-resonance:} $p\simeq\varDelta$.

Since the Delta propagators go as $1/(p\pm \vDe)$, rather 
than simply $1/p$ as for the nucleon, the counting of the graphs
with Deltas is different in the two regimes. For example, replacing a
nucleon propagator by a Delta changes the power of the graph by $p/\Delta$, which
is $O(p^{1/2})$ in the low-energy regime and $O(1)$ in the resonance region. In addition, 
the counting demands that the one-Delta-reducible ($1\De$R) loop graphs, which contain the pole at $p=\vDe$, are resummed in the regime $p\simeq \varDelta$,
which gives rise to the Delta-resonance width and results in a natural description of the resonance peak.

In general, a graph with $L$ loops, $N_\pi$ pion propagators, $N_N$ nucleon propagators, $N_\De$ Delta propagators and $V_k$ vertices of order $k$ counts in the low-energy regime as $\cO(p^n)$, with
\beq 
n = 4L - 2N_\pi -N_N -\half N_\De + k V_k .
\eeq
In the resonance regime, one identifies the number of the $1\De$R
propagators, $N_{1\De\mathrm{R}}$,  then 
\beq 
n = 4L - 2N_\pi -N_N - N_\De - 2 N_{1\De\mathrm{R}} + k V_k .
\eeq

\begin{figure}[t]
\includegraphics[width=0.6\columnwidth]{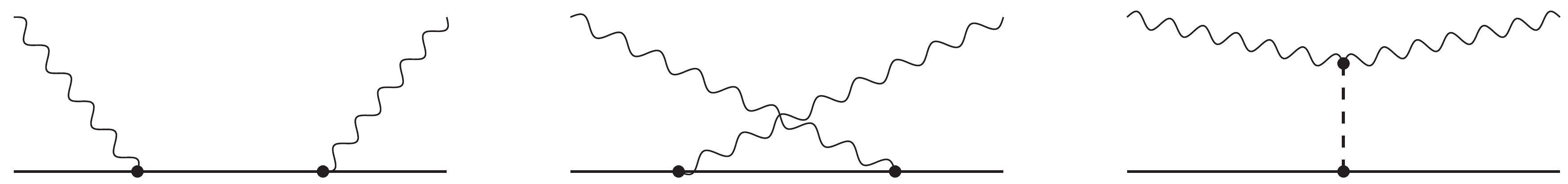}
\caption{Born graphs and the anomaly graph. Dots are vertices from the lowest-order Lagrangians.}
\label{fig:Born}
\end{figure}

In the low-energy regime, which is most relevant for the extraction of nucleon polarisabilities, 
the diagrams with $N_\De$ Delta propagators are suppressed by 
$\de^{N_\De}$, or equivalently by $p^{N_\De/2}$, with respect to the same diagram but with nucleon propagators.  The leading order (LO) contribution in this regime hence
comes from the nucleon Born term (first two diagrams in \Figref{Born}) which are responsible, for instance, for the correct Thomson limit. 
Inclusion of the nucleon anomalous magnetic moment in these graphs is warranted at $\cO(p^3)$, which here is the next-to-leading order (NLO).

\begin{figure}[ht]
 \includegraphics[width=0.65\columnwidth]{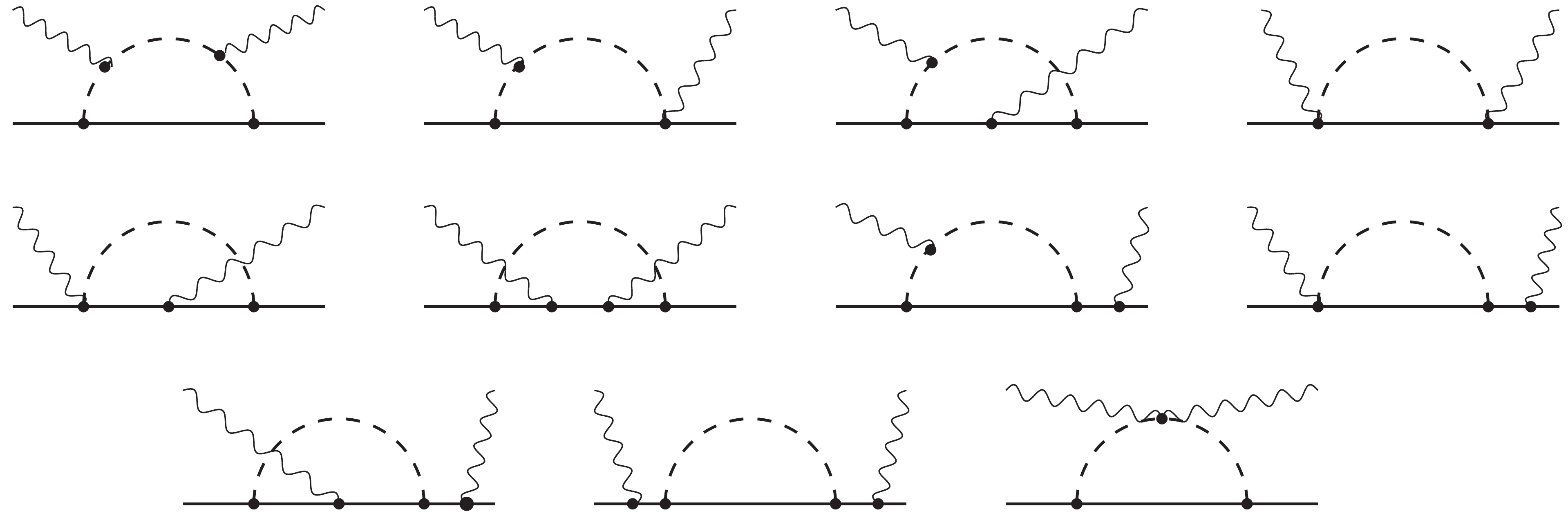}
\caption{Pion-nucleon loops that contribute to nucleon polarisabilities at leading order. Crossed and time-reversed
graphs are not shown but are included in the calculation.}
\label{fig:piNloops}
\end{figure}

\begin{figure}[hbt]
\includegraphics[width=0.7\columnwidth]{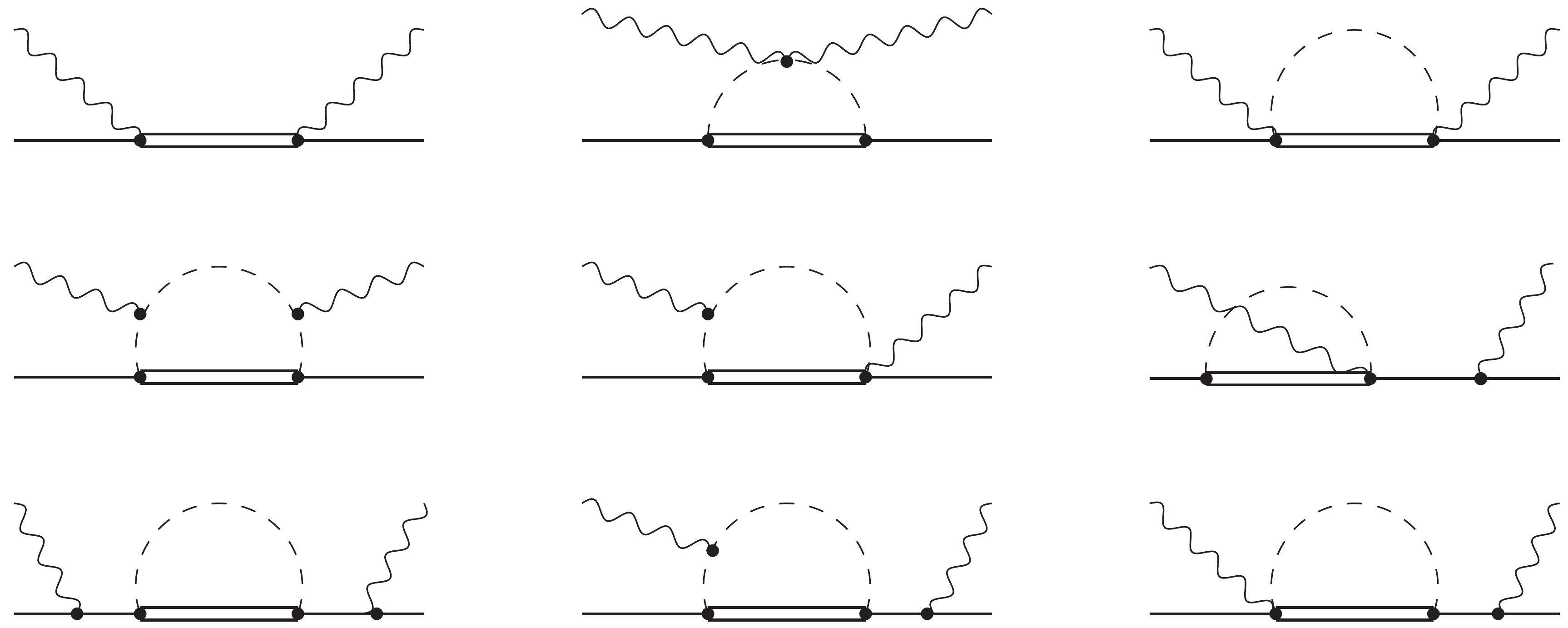}
\caption{Pion-Delta loops and the Delta tree graph that contribute to nucleon polarisabilities at next-to-leading order.
 Crossed and time-reversed graphs are not shown but are included in the calculation. 
Double lines denote the propagator of the Delta.}
\label{fig:piDloops}
\end{figure}

Also of $\cO(p^3)$ is the $\pi^0$-pole graph in \Figref{Born}, and 
all of the pion-nucleon loops in Fig.~\ref{fig:piNloops}.
These graphs give the LO contribution to the polarisabilities, as
the Born graphs do not contribute to polarisabilities by definition.
The Delta-pole and pion-Delta loop graphs, Fig.~\ref{fig:piDloops}, enter at $\cO(p^{7/2})$, here the next-to-next-to-leading order (NNLO), and give the NLO contribution to polarisabilities.
We note that the Delta-pole term, despite being formally suppressed compared to the pion-nucleon loops, has long been known to give a large contribution to the magnetic polarisabilities of the nucleon.

The NNLO amplitude thus receives contributions from the nucleon Born and $\pi^0$ pole graphs of Fig.~\ref{fig:Born},
the $\pi{\rm N}$ loop graphs of Fig.~\ref{fig:piNloops} and the Delta graphs in Fig.~\ref{fig:piDloops}; all of these were computed 
in \cite{Lensky:2009uv}. 

\subsection{Complete NLO calculation in the Delta-region}
\begin{figure}[ht]
 \includegraphics[width=0.45\columnwidth]{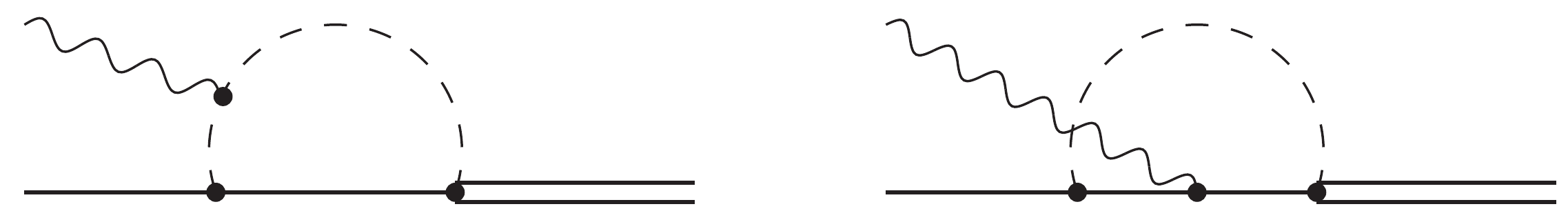}
\caption{$\gamma$N$\Delta$ vertex with NLO loop corrections. Only the imaginary parts of the loops shown here
lead to contributions at next-to-leading order around the Delta peak.}
\label{fig:vpiNloops}
\end{figure}

In order to extend this calculation to the Delta-resonance region, we
resum the $\piN$ loop contributions to the Delta propagator in the Delta-pole graphs, as explained in ref.~\cite{Pascalutsa:2002pi}. In addition we take into account the leading pion-loop corrections to the $\gamma$N$\Delta$ vertex, shown
in Fig.~\ref{fig:vpiNloops}. Those additions are important for  unitarity above the pion threshold.
At NLO in the regime $p\simeq \varDelta$, only the imaginary part of these vertex loops is relevant (specific remarks on the real part are given in Sect.~\secref{remarks}); this means that these corrections
vanish below the pion production threshold and hence do not affect the static polarisabilities. We will, however, investigate their importance for the dynamical polarisabilities and Compton multipoles; see Sect.~\ref{sec:pols:dyn}.
The technical details of the counting and the evaluation of these $\ga N\De$-vertex corrections are found in refs.~\cite{Pascalutsa:2005vq,McGovern:2012ew}.

\subsection{Remarks on higher-order Delta contributions}
\label{sec:remarks}
The effects of running of the
$\gamma$N$\Delta$ coupling constants $g_M$ and $g_E$, 
arising due to the real parts of the  $\gamma$N$\Delta$ vertex corrections
Fig.~\ref{fig:vpiNloops},
are beyond the order we are working at in the Delta region~\cite{Pascalutsa:2002pi}. Nevertheless,
we briefly consider the potential impact of these
effects, together with the effects of 
the running of the Delta mass and field renormalisation constants. None of these effects are included in subsequent section; they are also not the only higher-order effects which could impact the polarisabilities.  Of the latter, the inclusion of the anomalous magnetic moments of the proton and neutron within $\piN$ loop graphs may well be significant \cite {Ji:1999sv,VijayaKumar:2000pv,Gellas:2000mx,Bernard:2002pw}.

\begin{figure}[ht]
 \includegraphics[width=\columnwidth]{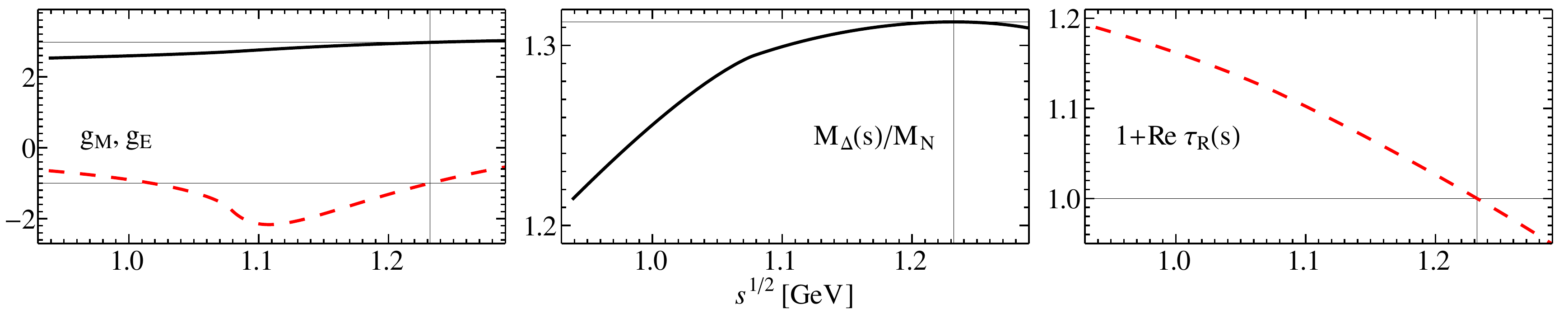}
\caption{  Running of the $\gamma$N$\Delta$ coupling constants $g_M$ and $g_E$ (left panel, solid/dashed line, respectively),
of the Delta mass (centre panel), and of the field renormalisation constant (right panel), as functions of $\sqrt{s}$ in GeV. Grid lines
represent the values renormalised at the Delta peak and the corresponding value of $\sqrt{s}$.}
\label{fig:running}
\end{figure}

The running of the real part of the $\gamma N \Delta$ couplings from the graphs of Fig.~\ref{fig:vpiNloops} was calculated in  \cite{Pascalutsa:2005vq,McGovern:2012ew}.
The corresponding numerical results
are shown in the left panel of Fig.~\ref{fig:running}.
Both coupling are seen to be reduced, in absolute
value, at low energy, and hence the Delta-pole
contribution to the polarisabilities will be reduced.
On the other hand, this effect is partially compensated by the one-loop running of the Delta propagator, which reads:   
\beq
S_{\mu\nu}(p)=-\frac{1+\tau_R(s)}{\slashed{p}-M_\Delta(s)+i
\Gamma_{\Delta}(s)/2}\mathcal{P}^{3/2}_{\mu\nu}(p)\,,
\eeq
where $\mathcal{P}^{3/2}_{\mu\nu}(p)$ is the spin-$3/2$ projection operator, $M_\Delta(s)$ is the running Delta mass, and $\Gamma_{\Delta}(s)$ is the width of the Delta.  The renormalised self-energy and field renormalisation are calculated as shown in Appendix~\ref{sec:remarks-details}.

The resulting $M_\Delta(s)$ and $1+\re\tau_R(s)$ are shown Fig.~\ref{fig:running} and are seen thus 
to strengthen the Delta-pole effects at lower energies.
A complete $\mathcal{O}(p^4)$ B$\chi$PT calculation will need
to be done to correctly quantify the effect of these contributions.
This calculation is, however, beyond the scope of the present work.

\subsection{Summary}
With these ingredients, our calculation of the amplitudes is NNLO [$\cO(p^{7/2})$] at low energies and  NLO [$\cO(p^2)$] in the Delta resonance
region, as summarised in Table~\ref{tab:counting}. 
\begin{table}[h]
\begin{tabularx}{0.7\textwidth}{|c|Y|Y|Y|Y|Y|Y|}
\hline
\diagbox{Regime}{Source} & $N\,  \&\,  \pi^0$ pole & $\piN$ loops & $\pi \Delta$ loops & $\Delta$ pole & $\Delta$-pole corr. \\
\hline
 $p \simeq m_\pi$          &    $p^2 + p^3 $      &  $p^3$        &          $ p^{7/2} $           &   $ p^{7/2} $             &   $ p^4 $                   \\
\hline
 $p \simeq \varDelta$         &    $p^2 + p^3 $     & $p^3$      &              $ p^{3} $    &      $ p $          &      $ p^{2} $                \\
\hline
\end{tabularx}
\caption{Counting the graphs in Figs.~\ref{fig:Born}--\ref{fig:vpiNloops} in the two different regimes.}
\label{tab:counting}
\end{table}

The numerical values of the physical constants used in the present calculation are given in  Table~\ref{tab:constants}. They are only marginally different from the ones used in~\cite{Lensky:2009uv}, and 
the same as in~\cite{Lensky:2012ag}.
\begin{table}[bh]
\begin{tabular}{l l l}
\hline
$\alpha_{\rm em}=1/137.04$ & $g_A=1.270$& $f_\pi = 92.21$ MeV  \\
$m_{\pi^{\pm}}= 139.57$ MeV & $m_{\pi^0}=134.98$ MeV\quad &\\
$\MN\equiv M_p= 938.27$ MeV \qquad& $\kappa_{p}=1.793$ & $M_\Delta= 1232$ MeV \\
$h_A\equiv 2g_{\piN \Delta}=2.85$ & $g_M=2.97$ & $g_E=-1.0$\\
\hline
\end{tabular}
\caption{Parameters used in our calculation (the $\pi^0$ mass is only used for the computation of the $t$-channel pion-pole graph). 
Most of the values are from  Particle Data Group~\cite{Agashe:2014kda}. 
The $\pi$N$\Delta$ coupling constant $h_A$ is fit to the the experimental Delta width and the magnetic and electric $\gamma$N$\Delta$ coupling constants $g_M$ and $g_E$ are taken from the pion photoproduction study of ref.~\cite{Pascalutsa:2005vq}. More details
can be found in ref.~\cite{McGovern:2012ew}.}
\label{tab:constants}
\end{table}

\section{Decompositions of the Compton amplitude}
\label{sec:amp:tensor}

Before presenting the results for the proton polarisabilities and Compton observables, we provide details of the tensor decomposition of the reaction amplitude.
To compute the Compton-scattering amplitude for a spin-$1/2$ target in a manifestly Lorentz- and gauge-invariant form, we use a set of eight covariant  tensors~\cite{Pascalutsa:2002pi}:
\beq
\eqlab{csampl}
T_{fi}= 
{\cal E'}_{\!\!\mu}^{\ast}(q')\, {\cal E}_\nu(q)\,\sum\limits_{i=1}^{8}\MA_i(s,t)\,\bar u_{s'} (p')\,  O_i^{\mu\nu}\,
 u_{s}(p)
\,,
\eeq
where $\MA_1\dots \MA_8$ are the invariant scalar amplitudes with the Mandelstam variables $s$, $t$, $u$ defined as usual.
The final and initial 4-momenta of the nucleon (and photon) are denoted as $p'$, $p$ (and $q'$, $q$), respectively, and obey the on-mass-shell conditions: $p^{\prime\,2}=p^2=\MN^2$, $q^{\prime\,2}=q^2=0$.
The free nucleon spinor $u_s(p)$ is  normalised to $2\MN$, whereas 
 ${\cal E}_\mu$ is a modified photon polarisation vector:
\beq
{\cal E}_\mu(q) = \veps_\mu  - \frac{P\cdot \veps}{P\cdot q}\, q_\mu\,,
\eeq
with $P=p+p'$. The tensors $O_i$ are given by:
\bea
O_1^{\mu\nu}&=& -g^{\mu\nu} \nn\\
O_2^{\mu\nu}&=& q^\mu {q'}^\nu  \nn\\
O_3^{\mu\nu}&=& - \ga^{\mu\nu}\nn\\
O_4^{\mu\nu}&=& g^{\mu\nu}\,(q'\cdot\ga \cdot q) \nn\\
O_5^{\mu\nu}&=& q^\mu {q'}_{\!\!\al}\ga^{\al\nu} +
\ga^{\mu\al} q_\al {q'}^\nu \\
O_6^{\mu\nu}&=& q^\mu q_{\al}\ga^{\al\nu} +
\ga^{\mu\al} q_\al' {q'}^\nu \nn\\
O_7^{\mu\nu}&=& q^\mu {q'}^\nu\,  (q'\cdot\ga \cdot q)\nn\\
O_8^{\mu\nu}&=& \ga^{\mu\nu\al\be} q_\al q_\be' =
-i\ga_5 \eps^{\mu\nu\al\be} q_\al' q_\be\nn\,,
\eea
where $\gamma^{\mu\nu}=\frac{1}{2}\left[\gamma^\mu,\gamma^\nu\right]$, $\gamma^{\mu\nu\alpha\beta}=\frac{1}{2}\left[\gamma^{\mu\nu\alpha},\gamma^\beta\right]$ with $\gamma^{\mu\nu\alpha}=\frac{1}{2}\left\{\gamma^{\mu\nu},\gamma^\alpha\right\}$,
and $\eps^{0123}=-1$.
The representation~\eref{csampl} is obtained by writing down the most general covariant structure (for the on-shell case) and imposing
the electromagnetic current-conservation condition. The amplitudes $A_i$ are most easily computed in the following
Lorenz-invariant gauge: 
\beq
P\cdot \veps =0 = P\cdot \veps'\,.
\eeq
This condition can also be achieved in the Coulomb gauge ($\veps_0=0=\veps_0'$) by going to the Breit frame: $\vec P=0$.

The set of tensors $O_i$ is overcomplete.  Alternative covariant bases containing only six basis tensors have been used in the literature, in particular the 
Hearn-Leader basis, a slight variant of which is used in ref.~\cite{Babusci:1998ww} with corresponding amplitudes denoted $T_i$.
From these though a more convenient set of amplitudes  can be constructed, which we will refer to as the L'vov amplitudes $A^{\mathrm L}_i$
(they are simply  $A_i$ in~\cite{Babusci:1998ww}).  These crossing-symmetric amplitudes are routinely
used in dispersion-relation calculations of Compton amplitudes and are a convenient bridge between other sets of amplitudes,
cf.~Appendix~\ref{app:matAtoA}.

In addition, by representing the nucleon Dirac spinors via their upper and lower Pauli spinors,
the tensors $O_i$ can be decomposed into a set of non-relativistic basis tensors $t_i$.  In the centre-of-mass (c.m.)
or Breit frames (in both of which the incoming and outgoing photon energies are equal) there are six such tensors,
and one obtains the following decomposition of the amplitude~\cite{Hemmert:1997tj}:
\beq
\begin{split}
 T_{fi}=&2\MN \sum\limits_{i=1}^6 A_i(\omega,\theta)\,\chi_{s'}^\dagger\, t_i\,\chi_s \\
=&2\MN\chi_{s'}^\dagger\Bigr(\hphantom{+}
A_1(\omega,\theta)\,\vec\epsilon^{\,\,'\!*}\cdot\vec\epsilon+A_2(\omega,\theta)\,\vec\epsilon^{\,\,'\!*}\cdot\hat q\,\,\vec\epsilon\cdot\hat q'\\ 
&\hspace*{1.5cm} +i A_3(\omega,\theta)\,\vec\sigma\cdot\bigl(\vec\epsilon^{\,\,'\!*}\times\vec\epsilon\,\bigr)
+i A_4(\omega,\theta)\,\vec\sigma\cdot\bigl(\hat q'\times \hat q\bigr)\,\,\vec\epsilon^{\,\,'\!*}\cdot\vec\epsilon \\
&\hspace*{1.5cm}+ i A_5(\omega,\theta)\,\vec\sigma\cdot\left[\bigl(\vec\epsilon^{\,\,'\!*}\times \hat q\bigr)\,\vec\epsilon\cdot \hat q'
                                             - \bigl(\vec\epsilon \times \hat q'\bigr)\,\vec\epsilon^{\,\,'\!*}\cdot \hat q\right]\\
&\hspace*{1.5cm}+ i A_6(\omega,\theta)\,\vec\sigma\cdot\left[\bigl(\vec\epsilon^{\,\,'\!*}\times \hat q'\bigr)\,\vec\epsilon\cdot \hat q'
                                             - \bigl(\vec\epsilon \times \hat q\bigr)\,\vec\epsilon^{\,\,'\!*}\cdot \hat q\right]\Bigl)\chi_s\,,
\end{split}
\label{eq:TviaAH}
\eeq
where now $\vec\epsilon^{\,\,'}$ ($\vec\epsilon\,$) are the final (initial) photon polarisation vectors, $\hat q'$ ($\hat q$) are the final (initial)
photon momentum unit vectors, $\omega$ and $\theta$ are the photon energy and scattering angle,  $\vec \sigma$ the Pauli matrices, and $\chi_s$ are Pauli spinors.  The $A_i$ are of course frame-dependent, and we will denote them $A_i^{\mathrm {cm}}$ and $A_i^{\mathrm {Br}}$ for the c.m.\ and Breit frames respectively.

In the present calculation, our starting point is the eight Compton amplitudes $\MA_i(s,t)$, as obtained in ref.~\cite{Lensky:2009uv}
with the modifications described in Sect.~\ref{sec:amp}.   These can be transformed into the any of the minimal sets $A_i$ as follows:
\beq
A_i(\omega,\theta)=\sum\limits_{j=1}^{8}C_{ij}^{\MA\to A}(s,t) \MA_j(s,t)\,;
\label{eq:AtoAH}
\eeq
the explicit expression for the $6\times 8$ matrix $C^{\MA\to A}(s,t)$ for various sets of amplitudes
are discussed in Appendix~\ref{app:matAtoA}.   For the study of Compton multipoles and dynamical nucleon polarisabilities
the c.m.\ frame amplitudes are the natural basis, while the connection to static polarisabilities is more easily made via the 
L'vov or Breit amplitudes.

\section{Static polarisabilities}
\label{sec:pols:static}
\subsection{Definitions}
\label{sec:pols:static:def}

The polarisabilities characterise the quadratic response of the external electric  and magnetic fields, $\vec E$ and $\vec H$.
In moderate fields the response of the nucleon  can be described 
by the following effective Hamiltonians~\cite{Babusci:1998ww,Holstein:1999uu}:\footnote{The notation differs slightly; for instance ref.~\cite{Holstein:1999uu} used  $\gamma_{E2}$ for $\gamma_{M1E2}$ and $\gamma_{ET}$ for $\gamma_{E2E2}$.}
\begin{align}
 \mathcal{H}_{\rm eff}^{(2)} &= -\frac12 \, 4\pi
    (\alpha_{E1} \vec E^2 + \beta_{M1} \vec H^2 ),\label{eq:H-eff1}\displaybreak[0]\\
  \mathcal{H}_{\rm eff}^{(3)} &=   -\frac12\,4\pi \Big(
     \gamma_{E1E1} \vec\sigma \cdot \vec E \times \dot{\vec E}
   + \gamma_{M1M1} \vec\sigma \cdot \vec H \times \dot{\vec H}
   -2 \gamma_{M1E2} E_{ij}\sigma_i H_j
   +2 \gamma_{E1M2} H_{ij}\sigma_i E_j \Big),\label{eq:H-eff2}\displaybreak[0]\\
 \mathcal{H}_{\rm eff}^{(4)} &= -\frac12\, 4\pi
    (\alpha_{E1\nu} \dot{\vec E}^2 + \beta_{M1\nu} \dot{\vec H}^2)
    -\frac1{12}\, 4\pi (\alpha_{E2} E_{ij}^2 + \beta_{M2} H_{ij}^2),\label{eq:H-eff3}\displaybreak[0]\\
\mathcal{H}_{\rm eff}^{(5)}&=-\frac12\, 4\pi\left(\gamma_{E1 E1\nu}\vec{\sigma}\cdot\dot{\vec{E}}
\times\ddot{\vec{E}}+\gamma_{M1 M1\nu}\vec{\sigma}\cdot\dot{\vec{H}}\times
\ddot{\vec{H}}-2\gamma_{M1E2\nu}\sigma_i\dot{E}_{ij}\dot{H}_j+2\gamma_{E1M2\nu}
\sigma_i\dot{H}_{ij}\dot{E_j}\right.\nonumber\\
&+\left.4\gamma_{E2E2}\epsilon_{ijk}\sigma_iE_{j\ell}\dot{E}_{k\ell}
+4\gamma_{M2M2}\epsilon_{ijk}\sigma_iH_{j\ell}\dot{H}_{k\ell}
-6\gamma_{M2E3}\sigma_iE_{ijk}H_{jk}+6\gamma_{E2M3}\sigma_iH_{ijk}E_{jk}
\right).\label{eq:H-eff4}
\end{align} 
\def\heffrefs{\ref{eq:H-eff1}--\ref{eq:H-eff4}}
where  
\begin{align}
  E_{ij} =& \frac12 (\nabla_i E_j + \nabla_j E_i), \quad\nonumber\\
{E}_{ijk}=&\frac13\, (\nabla_i\nabla_j{E}_k+\nabla_i\nabla_k{E}_j
+\nabla_j\nabla_k{E}_i)-\frac1{15}(\delta_{ij}\nabla^2E_k+\delta_{jk}
\nabla^2E_i+\delta_{ik}\nabla^2E_j),
\end{align}
and similarly for $H$. 

The non-relativistic Hamiltonian is defined only in a specific reference frame.  As far as the Compton-scattering process is concerned, the
Breit frame yields the most natural description because of the manifest crossing symmetry of the corresponding amplitudes.  
The contribution of $\cal H_{\rm eff}$ to the Breit-frame amplitudes reads: 
\begin{align}\label{eq:polamp}
A_1^\mathrm{Br}=&+\frac{4\pi\EN}{\MN}\left( \bigl(\alpha_{E1}+z\beta_{M1}\bigr)\w_B^2+
\left(\alpha_{E1\nu}+ z\beta_{M1\nu}
-\textfrac{1}{ 12}\beta_{M2}+z \textfrac{1}{ 12}\alpha_{E2} +\textfrac{1}{ 6}z^2\beta_{M2}\right)\w_B^4
\right) \nonumber\\
A_2^\mathrm{Br}=&+\frac{4\pi\EN}{\MN}\left(-\beta_{M1}\w_B^2
+\left(-\beta_{M1\nu}+\textfrac{1}{12}\alpha_{E2}-\textfrac{1}{ 6}z\beta_{M2}\right)\w_B^4
\right)\nonumber\\
A_3^\mathrm{Br}=&-\frac{4\pi\EN}{\MN}\left(\left(\gamma_{E1E1}+ \gamma_{E1M2}+z( \gamma_{M1M1}+ \gamma_{M1E2})\right)\w_B^3
+\left(\gamma_{E1E1\nu}+\gamma_{E1M2\nu} +z(\gamma_{M1M1\nu} +\gamma_{M1E2\nu})
\right.\right.\nonumber\\  &\qquad\qquad \left.\left.
-\textfrac{12}{ 5}\gamma_{M2E3}-
3\gamma_{M2M2}+z(\gamma_{E2E2}+\textfrac{8}{ 5}\gamma_{E2M3})+4z^2(\gamma_{M2M2}+\gamma_{M2E3})\right)\w_B^5
\right)\nonumber\\
A_4^\mathrm{Br}=&+\frac{4\pi\EN}{\MN}\left((\gamma_{M1E2} -\gamma_{M1M1})\w_B^3+(\gamma_{M1E2\nu}-\gamma_{M1M1\nu}-\gamma_{E2E2}-\textfrac{2}{ 5}\gamma_{E2M3}+z(2\gamma_{M2E3}-4\gamma_{M2M2}))\w_B^5
\right)\nonumber\\
A_5^\mathrm{Br}=&+\frac{4\pi\EN}{\MN}\left(\gamma_{M1M1}\w_B^3+
(\gamma_{M1M1\nu}-\gamma_{E2E2}-\gamma_{E2M3}+z(4\gamma_{M2M2}+\gamma_{M2E3}))\w_B^5
\right)\nonumber\\
A_6^\mathrm{Br}=&+ \frac{4\pi\EN}{\MN} \left(\gamma_{E1M2}\w_B^3+(\gamma_{E1M2\nu}-\textfrac{7}{ 5}\gamma_{M2E3}-
2\gamma_{M2M2}+3z\gamma_{E2M3})\w_B^5
\right)
\end{align}
where $\w_B$ and $z=\cos\theta_B$ refer here to the Breit-frame photon energy and scattering angle, and
$\EN$ is the nucleon energy. This is a complete low-energy expansion of the
non-Born part of the Compton amplitudes to order $\w_B^5$. 

In addition to the above polarisabilities, we shall examine the
forward and backward spin polarisabilities  $\gamma_0$ and $\gamma_\pi$ defined as:
\beq
\begin{split}
\gamma_0   &= -\gamma_{E1E1}-\gamma_{M1M1}-\gamma_{E1M2}-\gamma_{M1E2}\,,\\
\gamma_\pi &= \gamma_{M1M1}+\gamma_{M1E2}-\gamma_{E1E1}-\gamma_{E1M2}\,,
\end{split}
\eeq
and the so-called ``higher-order forward spin polarisability'' $\bar{\gamma}_0$, defined as a linear combination of the quadrupole spin
polarisabilities~\cite{Pasquini:2010zr}:
\beq\label{eq:gamma-bar}
\bar{\gamma}_0=-\gamma_{E1E1\nu}-\gamma_{M1M1\nu}-\gamma_{E1M2\nu}-\gamma_{M1E2\nu}-\gamma_{E2E2}-
\gamma_{M2M2}-\frac 8 5(\gamma_{E2M3}+\gamma_{M2E3})\,.
\eeq

The covariant $\chi$PT expressions for the polarisabilities are given in Appendix \ref{app:pols}, with the
corresponding numerical values given below, for the proton and neutron separately.
These expressions were obtained by computing the covariant amplitudes $\MA_i$, converting them to the Breit-frame amplitudes  as discussed above, expanding in $\w_B$ and $z$ and identifying the coefficients  with the polarisabilities.  Alternatively, one
 can go via the L'vov amplitudes, as shown in Appendix~\ref{app:hipols}  where 
 we give their complete and consistent relation to the static polarisabilities  of Eq.~(\heffrefs).  
For example,   $\bar{\gamma}_0=a_{4,\nu}/2\pi \MN$, where $a_{4,\nu}$ is a coefficient of the Taylor expansion
of $A_4^{\mathrm{L}}$, \eqref{lvov}.

\subsection{Proton}
\label{sec:pols:static:proton}
Our results for the scalar dipole and quadrupole and spin dipole
static polarisabilities are shown in Tables~\ref{tab:polsscalar} and~\ref{tab:polsspin}.
The analytical expressions for the $\piN$ loop, $\pi \Delta$ loop, and Delta-pole contributions to these polarisabilities are
given in Appendix~\ref{app:pols}. 
We also show results obtained
by other authors.  The latter are a mixture of predictions and fits to Compton-scattering data.
 In most cases the rather well-established constraints obtained via the optical theorem from photoproduction
 cross sections via the Baldin sum rule for $\alpha_{E1}+\beta_{M1}$ \cite{Baldin:1960} have been imposed,
 and the GDH-like sum rule for $\gamma_0$ is also sometimes used.

\begin{table}[ht]
\begin{ruledtabular}
\begin{tabular}{c|c|c|c|c|c|c}
  Source      &$\alpha_{E1}$    & $\beta_{M1}$     &$\alpha_{E2}$  &$\beta_{M2}$ &$\alpha_{E1\nu}$ &  $\beta_{M1\nu}$\\
\hline
$\cO(p^3)\ \piN$ loops & $\hpm{-}6.9 $   &      $-1.8$     & $\hpm{-}13.5$       & $-8.4$  & $\hpm{-}0.7$ & $1.8$   \\
$\cO(p^{7/2})\ \pi\Delta$
        loops & $\hpm{-}4.4 $   &    $-1.4$       & $\hpm{-}3.2$        & $-2.7$  & $-0.6$ & $0.6$     \\
$\De$ pole &   $-0.1 $       & $\hpm{-}7.1$    & $\hpm{-}0.6$        & $-4.5$      & $-1.5$ & $4.7$ \\
Total         & $ 11.2\pm0.7 $  &$\hpm{-}3.9\pm0.7$& $\hpm{-}17.3\pm3.9$       & $-15.5\pm3.5$    & $-1.3\pm 1.0$ & $7.1\pm 2.5$   \\
\hline
\hline
Fixed-$t$ DR~\cite{Babusci:1998ww,Olmos:2001}
              &  $12.1^*$         & $\hpm{-}1.6^*$            &  $\hpm{-}27.5$       & $-22.4$        & $-3.8$ & $9.1$       \\
\hline
Fixed-$t$ DR~\cite{Holstein:1999uu,Hildebrandt:2003fm}
              &     $\cdots$    &     $\cdots $     &  $\hpm{-}27.7$       & $-24.4$        & $-3.9$ & $9.3$       \\
\hline
HB$\chi$PT fit~\cite{McGovern:2012ew}
              & $10.65\pm 0.50$ & $3.15\pm 0.50$   & $\cdots $         & $\cdots $           & $\cdots $ & $\cdots $   \\
\hline
B$\chi$PT fit~\cite{Lensky:2014efa}
              & $10.6\pm 0.5 $  & $3.2\pm 0.5$     & $\cdots $         & $\cdots $         & $\cdots $  & $\cdots $   \\
\hline
PDG~\cite{Agashe:2014kda}
             & $11.2\pm 0.4$   & $2.5\pm 0.4$     &     $\cdots $    &  $\cdots $     & $\cdots $   & $\cdots $  \\

\end{tabular}
\caption{Values of {\it proton} static dipole, quadrupole, and dispersive polarisabilities, in units of $10^{-4}$~fm$^{3}$ (dipole) and $10^{-4}$~fm$^{5}$ (quadrupole and dispersive),
in comparison with the fixed-$t$ DR extraction of refs.~\cite{Babusci:1998ww,Holstein:1999uu,Hildebrandt:2003fm} and $\chi$PT fits. The results of the calculation
of ref.~\cite{Hildebrandt:2003fm} are kindly provided by B.~Pasquini~\cite{barbaraprivate}. \\
$^*$ Only $\alpha_{E1}+\beta_{M1}$ is predicted in DR; the difference is taken from fits to Compton-scattering data.
}
\label{tab:polsscalar}
\end{ruledtabular}
\end{table}

Table~\ref{tab:polsscalar} shows the values of the dipole, quadrupole, and dispersive scalar polarisabilities obtained in our calculation,
compared with the results of the DR calculations of refs.~\cite{Babusci:1998ww,Hildebrandt:2003fm}, as well as with the results
of the $\chi$PT fits~\cite{McGovern:2012ew,Lensky:2014efa} (in these calculations, only the dipole polarisabilities were fit to the data).  
The sum of the dipole scalar polarisabilities
$\alpha_{E1}+\beta_{M1}$ is well constrained by the Baldin sum rule to be $14.0\pm 0.2$~\cite{Gryniuk:2015eza}, and even though
our central value
of 15.1 is somewhat higher than that, they are not in contradiction given the uncertainty of our result.

The  fixed-$t$ DR calculations encounter difficulties in describing 
backward scattering (large $t$). As a result they seem to 
disagree with $\chi$PT in the value of $\alpha_{E1}-\beta_{M1}$. Fits to experimental data in a DR framework have given results around 10.0--10.5~\cite{Babusci:1998ww,Olmos:2001}, substantially higher than our value of 7.3. The chiral fits give results closer to 7.5 \cite{McGovern:2012ew,Lensky:2014efa}, in much better agreement with our prediction.

The values of $\alpha_{E2}$ and $\beta_{M2}$ resulting from our calculation are significantly below the corresponding DR
values. 
The same situation is observed in the values of $\alpha_{E1\nu}$ and $\beta_{M1\nu}$; the former is about a factor of two below the DR value. 

We note that these significant differences
in quadrupole polarisabilities do not show up, in any dramatic fashion,
in the Compton observables; both DR and $\chi$PT calculations
provide a good description of the available data.
It could be that the differences in polarisabilities
are negated in observables. 
This is certainly the case for forward scattering,
where both calculations agree on the Baldin sum rule, 
as well as on the (fourth-order) sum rule involving
the higher scalar polarisabilities~\cite{Gryniuk:2015eza}.

\begin{table}[ht]
\begin{ruledtabular}
\begin{tabular}{c|c|c|c|c|c|c|c}
  Source           & $\gamma_{E1E1}$ & $\gamma_{M1M1}$& $\gamma_{E1M2}$& $\gamma_{M1E2}$ &\hfill$\gamma_{0}\hfill$  &\hfill$\gamma_{\pi}\hfill$ &\hfill$\bar{\gamma}_0\hfill$ \\
\hline
$\cO(p^3)\ \piN$ loops      &   $-3.4 $      &     $-0.1$    &     $\hpm{-} 0.5  $  &   $\hpm{-} 0.9  $     & $\hpm{-} 2.0$                  & $\hpm{-} 3.6$                   &   $ \hpm{-}2.1$    \\
$\cO(p^{7/2})\ \pi\Delta$ loops  &    $\hpm{-}0.4 $      &     $-0.2$    &     $\hpm{-} 0.1  $  &   $-0.2  $     & $-0.1$                  & $-0.9$                   &   $-0.01$    \\
$\Delta$ pole      &   $-0.4 $      &     $\hpm{-}3.3 $    &     $-0.4  $  &   $\hpm{-} 0.4  $     & $-2.8$                  & $\hpm{-} 4.4$                   &   $-1.0$    \\
Total              &$-3.3\pm 0.8$   &$2.9\pm 1.5$   &$ 0.2\pm 0.2 $ & $1.1 \pm0.3 $  & $-0.9\pm1.4$            & $ 7.2\pm1.7$             &   $ 1.1\pm 0.5$    \\
\hline
\hline
Fixed-$t$ DR~\cite{Babusci:1998ww}                   &  $-3.4      $   &    $    2.7$  &   $\hpm{-}0.3$       &   $1.9       $  &      $-1.5$             &       $7.8$             &   $\cdots$    \\
\hline
Fixed-$t$ DR~\cite{Holstein:1999uu,Hildebrandt:2003fm,Pasquini:2007hf}
                   &  $-4.3      $   &    $     2.9$  &   $-0.02$       &   $2.2       $  &      $-0.8$             &       $9.4$             &   $0.6$    \\
\hline
HB$\chi$PT~\cite{McGovern:2012ew,Griesshammer:2015ahu}
                   &  $-1.1\pm 1.9$         &$2.2\pm0.8$           &  $-0.4\pm0.6$        &  $1.9\pm0.5$          &      $-2.6\pm1.9$              &       $5.6\pm1.9$               &   $\cdots$     \\
\hline
\hline
MAMI 2015~\cite{Martel:2014pba}
                    &  $-3.5\pm 1.2$  &$3.16\pm 0.85$  & $-0.7\pm 1.2$  & $1.99\pm 0.29$   &  $-1.01\pm 0.13$        &    $8.0\pm 1.8$           &     $\cdots$   \\
\end{tabular}
\caption{Values of {\it proton} static spin polarisabilities, in units of $10^{-4}$~fm$^{4}$ (except for $\bar{\gamma}_0$ which is in units of $10^{-4}$~fm$^6$),
in comparison with the fixed-$t$ DR extraction of refs.~\cite{Babusci:1998ww,Holstein:1999uu,Hildebrandt:2003fm,Pasquini:2007hf} and $\chi$PT fits.
Ref.~\cite{Martel:2014pba} is an extraction from asymmetry data in the DR framework of ref.~\cite{Drechsel:2002ar}, with
$\gamma_0$ and $\gamma_\pi$ as input. Note that the up-to-date fixed-$t$ DR values
of the proton spin polarisabilities corresponding to refs.~\cite{Holstein:1999uu,Hildebrandt:2003fm,Pasquini:2007hf} are also taken from ref.~\cite{Martel:2014pba}.
The fixed-$t$ DR value of $\bar{\gamma}_0$ is from ref.~\cite{Pasquini:2010zr}.
}
\label{tab:polsspin}
\end{ruledtabular}
\end{table}

Table~\ref{tab:polsspin} displays the values of the spin dipole polarisabilities predicted in our calculation,
compared with the DR results of refs.~\cite{Babusci:1998ww,Holstein:1999uu,Hildebrandt:2003fm,Pasquini:2007hf}, and the results of the $\chi$PT fit~\cite{McGovern:2012ew}.
Here, our results for all the leading spin dipole polarisabilities agree well with the DR ones. The only somewhat discrepant polarisability is $\gamma_{M1E2}$
which is smaller by a factor of two in our calculation. This, in turn, results in our value of $\gamma_\pi$ being somewhat smaller. At the same time,
  our value of $\gamma_0$ is very close to that given by the sum 
rule.\footnote{We have checked that our value of $\gamma_0$ differs from that of ref.~\cite{Blin:2015era} only because the input parameters, particularly $g_M$, are slightly different.} The HB$\chi$PT fit of ref.~\cite{McGovern:2012ew}, on the other hand, gives the values
of the spin polarisabilities that also are close to the DR values except for  $\gamma_{E1E1}$. Thus, the central values of some of the spin polarisabilities
are predicted to be different in B$\chi$PT and in HB$\chi$PT --- this can be seen, for instance, in $\gamma_{E1E1}$ or in the forward spin polarisability $\gamma_0$---though it should be noted that they do all agree within the (substantial) combined errors.
It is interesting to note that the $\chi$PT fits to the unpolarised data done in the two frameworks~\cite{McGovern:2012ew,Lensky:2014efa}, where only
 $\alpha_{E1}-\beta_{M1}$ was fit, resulted in almost identical values of the scalar dipole polarisabilities (cf.\ Table~\ref{tab:polsscalar}).
This demonstrates that the unpolarised low-energy data
are not sensitive to the values of the spin polarisabilities (at least to the extent these differ between the B and the HB calculations on which the fits were based).

\vspace*{-0.5em}
\subsection{Neutron}
\vspace*{-0.5em}
\label{sec:pols:static:neutron}
\begin{table}[ht]
\begin{ruledtabular}
\begin{tabular}{c|c|c|c|c|c|c}
  Source      &$\alpha_{E1}$    & $\beta_{M1}$     &$\alpha_{E2}$  &$\beta_{M2}$ &$\alpha_{E1\nu}$ &  $\beta_{M1\nu}$\\
\hline
$\cO(p^3)\ \piN$ loops & $\hpm{-}9.4$   &      $-1.1$     & $\hpm{-}12.4$       & $-8.7$  & $\hpm{-}2.1$ & $1.8$   \\
$\cO(p^{7/2})\ \pi\Delta$
        loops & $\hpm{-}4.4 $   &    $-1.4$       & $\hpm{-}3.2$        & $-2.7$  & $-0.6$ & $0.6$     \\
$\De$ pole &   $-0.1 $       & $\hpm{-}7.1$    & $\hpm{-}0.6$        & $-4.5$      & $-1.5$ & $4.7$ \\
Total         &  $ 13.7\pm3.1 $  &$\hpm{-}4.6\pm2.7$& $\hpm{-}16.2\pm3.7$       & $-15.8\pm3.6$    & $\hpm{-}0.1\pm1.0$ & $7.2\pm 2.5$   \\
\hline
\hline
Fixed-$t$ DR~\cite{Babusci:1998ww,Kossert:2002ws}
              &  $12.5^*$         & $\hpm{-}2.7^*$            &  $\hpm{-}27.2$       & $-23.5$        & $-2.4$ & $9.2$       \\
 \hline
 Fixed-$t$ DR~\cite{Drechsel:2002ar,Hildebrandt:2003fm,barbaraprivate}
              &     $\cdots$    &     $\cdots$     & $\hpm{-}27.9 $       & $-24.3$        & $-2.8$ & $9.3$       \\
\hline
HB$\chi$PT fit~\cite{Griesshammer:2012we,Myers:2014ace}
              & $11.55\pm 1.5$ & $3.65\pm 1.5$   & $\cdots $         & $\cdots $           & $\cdots $ & $\cdots $   \\
\hline
\hline
PDG~\cite{Agashe:2014kda}
             & $11.6\pm 1.5$   & $3.7\pm 2.0$     &     $\cdots $    &  $\cdots $     & $\cdots $   & $\cdots $  \\
\end{tabular}
\caption{Values of {\it neutron} static dipole, quadrupole, and dispersive polarisabilities, in units of $10^{-4}$~fm$^{3}$ (dipole) and $10^{-4}$~fm$^{5}$ (quadrupole and dispersive),
in comparison with the fixed-$t$ DR extraction of refs.~\cite{Babusci:1998ww,Hildebrandt:2003fm} and $\chi$PT fits.
The results of the calculation of ref.~\cite{Hildebrandt:2003fm} are kindly provided by B.~Pasquini~\cite{barbaraprivate}.\\$^*$ Only $\alpha_{E1}+\beta_{M1}$ is predicted in DR.
}
\label{tab:polsscalar_n}
\end{ruledtabular}
\end{table}

In Tables \ref{tab:polsscalar_n} and \ref{tab:polsspin_n} we, in turn, show our results for the neutron scalar and spin polarisabilities,
with the analytical expressions for the $\piN$ loop contributions to these polarisabilities given in Appendix~\ref{app:pols}.
The $\pi \Delta$ loop and Delta-pole contributions to the neutron polarisabilities are equal at this order to the corresponding
proton values. 

Table \ref{tab:polsscalar_n} shows the scalar polarisabilities of the neutron,
compared with the DR results of refs.~\cite{Babusci:1998ww,Drechsel:2002ar,Hildebrandt:2003fm,barbaraprivate} and $\chi$PT
fits of refs.~\cite{Griesshammer:2012we,Myers:2014ace}.
The Baldin sum rule result for the neutron is $\alpha_{E1}+\beta_{M1}=15.2\pm 0.4$~\cite{Levchuk:1999zy}, which is again somewhat lower than our central value.
Even though our error estimates of the neutron dipole scalar polarisabilities 
are rather large, one can see that, analogously to the proton, our prediction for $\beta_{M1}$
is higher than the DR values and closer to those of the chiral fits. The situation with the
neutron quadrupole and the dispersive polarisabilities is very similar to the proton case,
with B$\chi$PT predictions being significantly smaller than the DR results.

\begin{table}[ht]
\begin{ruledtabular}
\begin{tabular}{c|c|c|c|c|c|c|c}
  Source           & $\gamma_{E1E1}$ & $\gamma_{M1M1}$& $\gamma_{E1M2}$& $\gamma_{M1E2}$ &\hfill$\gamma_{0}\hfill$  &\hfill$\gamma_{\pi}\hfill$ &\hfill$\bar{\gamma}_0\hfill$ \\
\hline
$\cO(p^3)\ \piN$ loops      &    $-4.7 $      &     $-0.2$    &     $\hpm{-} 0.6  $  &   $\hpm{-} 1.3  $     & $\hpm{-} 3.0$                  & $\hpm{-} 5.3$                   &   $ \hpm{-}2.9$    \\
$\cO(p^{7/2})\ \pi\Delta$ loops  &    $\hpm{-}0.4 $      &     $-0.2$    &     $\hpm{-} 0.1  $  &   $-0.2  $     & $-0.1$                  & $-0.9$                   &   $-0.01$    \\
$\Delta$ pole      &   $-0.4 $      &     $\hpm{-}3.3 $    &     $-0.4  $  &   $\hpm{-} 0.4  $     & $-2.8$                  & $\hpm{-} 4.5$                   &   $-1.0$    \\
Total              &$-4.7\pm 1.1$   &$2.9\pm 1.5$   &$ 0.2\pm 0.2 $ & $1.6 \pm0.4 $  & $0.03\pm1.4$            & $ 9.0\pm2.0$             &   $ 1.9\pm 0.7$    \\
\hline
\hline
Fixed-$t$ DR~\cite{Babusci:1998ww}
                   &$-5.6          $ &$3.8         $ &   $-0.7        $  &   $2.9      $ &      $-0.4      $        &       $13.0      $        &   $\cdots$    \\
\hline
Fixed-$t$ DR~\cite{Drechsel:2002ar,Hildebrandt:2003fm,barbaraprivate}
                   &  $-5.9      $   &    $     3.8$  &   $-0.9$       &   $3.1       $  &      $-0.1$             &       $13.7$             &   $\cdots$    \\
 \hline
 HB$\chi$PT~\cite{McGovern:2012ew,Griesshammer:2015ahu}
                    &  $-4.0\pm 1.9$         &$1.3\pm0.8$           &  $-0.1\pm0.6$        &  $2.4\pm0.5$          &      $0.5\pm1.9$              &       $7.7\pm1.9$               &   $\cdots$     \\
\end{tabular}
\caption{Values of {\it neutron} static spin polarisabilities, in units of $10^{-4}$~fm$^{4}$ (except for $\bar{\gamma}_0$ which is in units of $10^{-4}$~fm$^6$),
in comparison with the fixed-$t$ DR extraction of refs.~\cite{Babusci:1998ww,Hildebrandt:2003fm}.
The results of the calculation of ref.~\cite{Hildebrandt:2003fm} are kindly provided by B.~Pasquini~\cite{barbaraprivate}.
}
\label{tab:polsspin_n}
\end{ruledtabular}
\end{table}
Table~\ref{tab:polsspin_n} shows the neutron spin polarisabilities, compared with the results of the HB$\chi$PT fit~\cite{McGovern:2012ew}, and of dispersive evaluations~\cite{Babusci:1998ww,Drechsel:2002ar,Hildebrandt:2003fm,barbaraprivate}  
(compatible dispersive evaluations of $\gamma_0$ and $\gamma_\pi$ for the proton and neutron may also be found in ref.~\cite{Schumacher:2011gs}).
Our results for the two largest neutron spin polarisabilities, $\gamma_{E1E1}$ and
$\gamma_{M1M1}$, agree well with the DR ones, whereas the magnitudes of $\gamma_{E1M2}$
and $\gamma_{M1E2}$ are predicted by our calculation to be smaller than is obtained
in the dispersive calculations, even with the uncertainties taken into account. This,
in turn, leads in our value of $\gamma_\pi$ being somewhat smaller than in the DR
framework, with $\gamma_0$ being consistent between all frameworks (although
not very well constrained). It is interesting that the HB$\chi$PT results follow
here a 
pattern different from those seen in the proton case: they tend to yield a smaller $\gamma_{M1M1}$, rather than
$\gamma_{E1E1}$, cf.\ Table~\ref{tab:polsspin} for the proton.

\vspace*{-0.5em}
\subsection{Error estimates}
\vspace*{-0.5em}
The uncertainties for the total values of static polarisabilities calculated in our work (see 4th row in Tables~\ref{tab:polsscalar}, \ref{tab:polsspin}, \ref{tab:polsscalar_n} and~\ref{tab:polsspin_n}) are
estimates of the contributions to the polarisabilities that come at NNLO.  With the exception of the proton $\alphae$ and $\betam$, they are calculated in the standard fashion (see e.g.\ \cite{Griesshammer:2012we,Griesshammer:2015ahu,Epelbaum:2014efa,Epelbaum:2014sza})
by considering the convergence
order-by-order in the chiral expansion.   As discussed above, our expansion parameter is $\delta= \mpi/\varDelta=0.48$, with $\mpi/\Lambda_\chi\sim\delta^2$.
The LO contribution to the polarisabilities is from the $\piN$ loops, and the NLO contributions are from the $\Delta$ diagrams.  A conservative estimate for the
uncertainty on a given polarisability is therefore given by 
\beq\label{eq:converge}
\sigma=\Max[\delta^2\times(\text{$\piN$ loops}),\delta\times(\text{$\Delta$ graphs}),\delta^2\times(\text{total})]
\eeq
where of course it is the absolute values which are compared.   
The question of ``theory errors''  in effective theories has recently been considered from a  Bayesian perspective by Furnstahl \etal~\cite{Furnstahl:2015rha}; briefly,
the actual calculation of  an observable to one or more orders tells us something about the size of the contributions, updating our prior expectations, e.g.\  of ``naturalness''
of the terms in the series.  With minimal prior assumptions, the quantity obtained from the equivalent of \eqref{converge} in a calculation to $k$ orders corresponds to the $k/(k+1)$ confidence limit. Hence in this case, with two orders calculated, $\sigma$ is a 67\% confidence limit, very close to the conventional ``one sigma'' band.  Furnstahl \etal\ also caution, however, that the corresponding probability distribution is in general quite non-Gaussian, and  the ``two sigma'' interval will generally correspond to much less than a 95\% probability interval.

The proton $\alphae$ and $\betam$ are a special case.  Were we to treat them in the same way, we would obtain substantial errors of 3.1 and 2.7 respectively.
However, alone of all the proton  polarisabilities considered, these have counterterms at NNLO.  In the absence of other information, one would expect their scale to be set by $\alpha_\mathrm{em} \Lambda_\chi^{-3}\sim 2.2$, which is entirely compatible with these errors.  But we do have other information, since a partial NNLO fit to proton Compton-scattering data was performed by some of the current authors in ref.~\cite{Lensky:2014efa}, obtaining values for $\alphae$ and $\betam$ much closer to the NLO predictions than these uncertainties would suggest (see the penultimate line of Table~\ref{tab:polsscalar}).  Although other, presumably similar-sized,  NNLO mechanisms will enter in a full calculation,  the counterterms in the fit will be readjusted and the net result is not expected to change significantly.  This can be regarded as a vindication of the error estimate of $\pm0.7$ on $\alphae$ and $\betam$ given in ref.~\cite{Lensky:2009uv}, and so we retain these errors in this work.

\section{Dynamical polarisabilities and multipoles}
\label{sec:pols:dyn}

\subsection{Definitions}
 \label{sec:pols:dyn:defs}

A widely-used parametrisation of the Compton amplitude is the multipole expansion in the centre-of-mass frame.
The initial and final photon-nucleon states in Compton scattering can be described using the angular momentum of the initial photon $l$,
 the index $\pm$ for $j=l\pm1$, and $E$ or $M$ to indicate initial and final photon parity: hence for each $l\ge1$ there are
 four multipole amplitudes $f^{l\pm}_{EE}, f^{l\pm}_{MM}, f^{l{+}}_{EM}$ and $f^{l{+}}_{ME}$.
As we are concerned only with the non-Born part of the Compton amplitude, the Born contribution is likewise implied to have been subtracted from the multipole amplitudes.
 The imaginary parts of the latter can be related to pion photoproduction multipoles via
\bea
&\im f_{EE}^{l\pm}=k\sum\limits_c\left|E^{(c)}_{(l\pm 1)\mp}\right|^2\,,\quad  & \im f_{MM}^{l\pm}=k\sum\limits_c\left|M^{(c)}_{l\pm}\right|^2\,,\label{eq:mult-f1}\\
&\im f_{EM}^{l\pm}=\pm k\sum\limits_c\re\left[E^{(c)}_{(l\pm 1)\mp}M^{(c)*}_{(l\pm 1)\mp}\right]\,,\quad  & \im f_{ME}^{l\pm}=\mp k\sum\limits_c\re\left[E^{(c)}_{l\pm}M^{(c)*}_{l\pm}\right]\,.
\label{eq:mult-f2}
\eea
\def\multrefs{\ref{eq:mult-f1}--\ref{eq:mult-f2}}
Here, the sum is over the production charge channels (i.e., $\pi^+n$ and $\pi^0 p$), $k$ is the pion momentum in the centre-of-mass
frame, and it is assumed that energy is sufficiently small so that only single pion production channel is significant.
 The dependence of each of the quantities in the r.h.s.\ and l.h.s.\ of these relations
on the photon energy $\omega$ is implied. In particular, the leading low-energy behaviour of the multipoles is~\cite{Guiasu:1978dz,Guiasu:1979sz}
\beq
f^{l\pm}_{EE}\sim f^{l\pm}_{MM}\sim \omega^{2l}, \qquad f^{l+}_{EM}\sim f^{l+}_{ME}\sim  \omega^{2l+1}\,.
\eeq 
For the details of the multipole decomposition of the Compton amplitude, in particular, expressions for multipoles in terms of the amplitudes $A^{\mathrm{cm}}$,
the reader is referred to  
ref.~\cite{Hildebrandt:2003fm} and references cited therein. 

The multipole decomposition of the non-Born Compton amplitude allows one to introduce the dynamical nucleon polarisabilities~\cite{Griesshammer:2001uw,Hildebrandt:2003fm}. 
The conventional definitions of these are 
\beq\label{eq:pol-f1}
\alpha_{El}(\omega)=(l (2 l - 1)!!)^2\left.\frac{(l+1)f^{l+}_{EE}+l f^{l-}_{EE}}{\omega^{2l}}\right.\,,\qquad
\beta_{Ml}(\omega)=(l (2 l - 1)!!)^2\left.\frac{(l+1)f^{l+}_{MM}+l f^{l-}_{MM}}{\omega^{2l}}\right.
\eeq
for the spin-independent dynamical polarisabilities, and
\beq\label{eq:pol-f2}
\gamma_{ElM(l+1)}(\omega)=2^{2 - l} (2 l + 1)!!\left.\frac{f^{l+}_{EM}}{\omega^{2l+1}}\right.\,,\qquad
\gamma_{MlE(l+1)}(\omega)=2^{2 - l} (2 l + 1)!!\left.\frac{f^{l+}_{ME}}{\omega^{2l+1}}\right.
\eeq
for the mixed spin dynamical polarisabilities. The unmixed spin dynamical polarisabilities are defined as
\beq \label{eq:pol-f3}
\gamma_{ElEl}(\omega)=(2 l - 1)\left.\frac{f^{l+}_{EE}- f^{l-}_{EE}}{\omega^{2l+1}}\right.\,,\qquad
\gamma_{MlMl}(\omega)=(2 l - 1)\left.\frac{f^{l+}_{MM}- f^{l-}_{MM}}{\omega^{2l+1}}\right.\,.
\eeq
\def\polrefs{\ref{eq:pol-f1}--\ref{eq:pol-f3}}

The Compton multipoles (or, equivalently, the dynamical polarisabilities) capture the underlying physics in the different energy regimes and multipolarities
(such as, e.g., the pion photoproduction cusp and the Delta peak). This provides an effective parameterisation of the Compton amplitude, working well in a wide range of energies~\cite{Hildebrandt:2003fm,Hildebrandt:2003md}.

For $l=1$ the dynamical polarisabilities, defined in Eqs.~(\polrefs), can be regarded as an extension of the static scalar and spin polarisabilities
to be functions of the photon energy, with the latter matching the limiting values of the former as the energy goes to zero. However, the fact that
the static polarisabilities are defined in the Breit frame causes this relation to be broken by the recoil terms for higher values of $l$.
In fact, for $l\ge 2$ even the definition of the unmixed spin polarisabilities, Eq.~(\ref{eq:pol-f3}), is problematic: while for $l=1$
the leading $\omega$ dependences cancel between the two unmixed multipoles, for $l\ge 2$  the recoil corrections fail to cancel, and cause the difference as well as the weighted sum to go as $\omega^{2l}$. 
This means that the unmixed $l\ge 2$ spin dynamical polarisabilities diverge as
$1/\omega$ in the zero-energy limit.  
In spite of this limitation, the (non-divergent) dynamic polarisabilities provide a compact way to summarise
the amplitudes and to compare the predictions of different frameworks.

As argued above, the static polarisabilities are best defined via the effective non-relativistic Hamiltonian in the Breit frame, 
and not as the zero-energy limit of the dynamical polarisabilities.  However
it is straightforward to calculate the relevant recoil corrections in the centre-of-mass frame
and hence match the low-energy expansion of the dynamical polarisabilities to the static ones. 
Details of this calculation are given in Appendix~\ref{app:hipols}; here we only note one result for the
zero-energy limit of the scalar quadrupole polarisabilities \cite{Babusci:1998ww}:
\beq
\alpha_{E2}(0)= \alpha_{E2}+\frac{3 \beta_{M1}}{2 \MN^2}  \qquad 
\beta_{M2}(0)= \beta_{M2}+\frac{3 \alpha_{E1}}{2 \MN^2}\,.
\label{eq:qpols}
\eeq

\subsection{Results}
\label{sec:pols:dyn:proton}

\begin{figure}[ht]
 \includegraphics[width=\columnwidth]{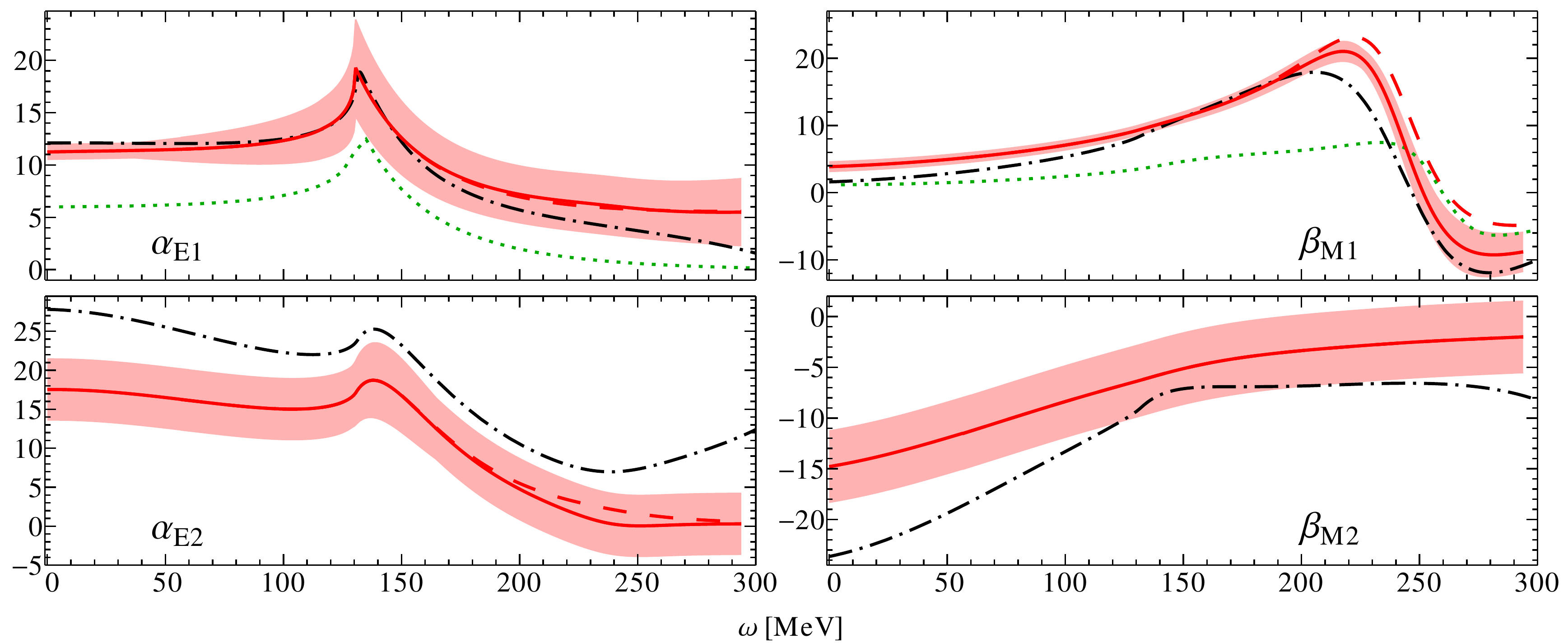}
\caption{  Real parts of {\it proton} scalar dipole and quadrupole dynamical polarisabilities, in units of $10^{-4}\mathrm{fm}^3$ and
$10^{-4}\mathrm{fm}^5$, respectively, as functions of photon cms energy in MeV. The curves are the results of this B$\chi$PT calculation
with or without the $\gamma$N$\Delta$ vertex running (red solid and red dashed, respectively),
compared with the results of the DR calculation of ref.~\cite{Hildebrandt:2003fm} (black dot-dashed)
and with the results of the Computational Hadronic Model~\cite{Aleksejevs:2013cda,Aleksejevs:2015} (green dotted, not shown for the quadrupole polarisabilities).}
\label{fig:pols_scalar}
\end{figure}

\begin{figure}[ht]
 \includegraphics[width=\columnwidth]{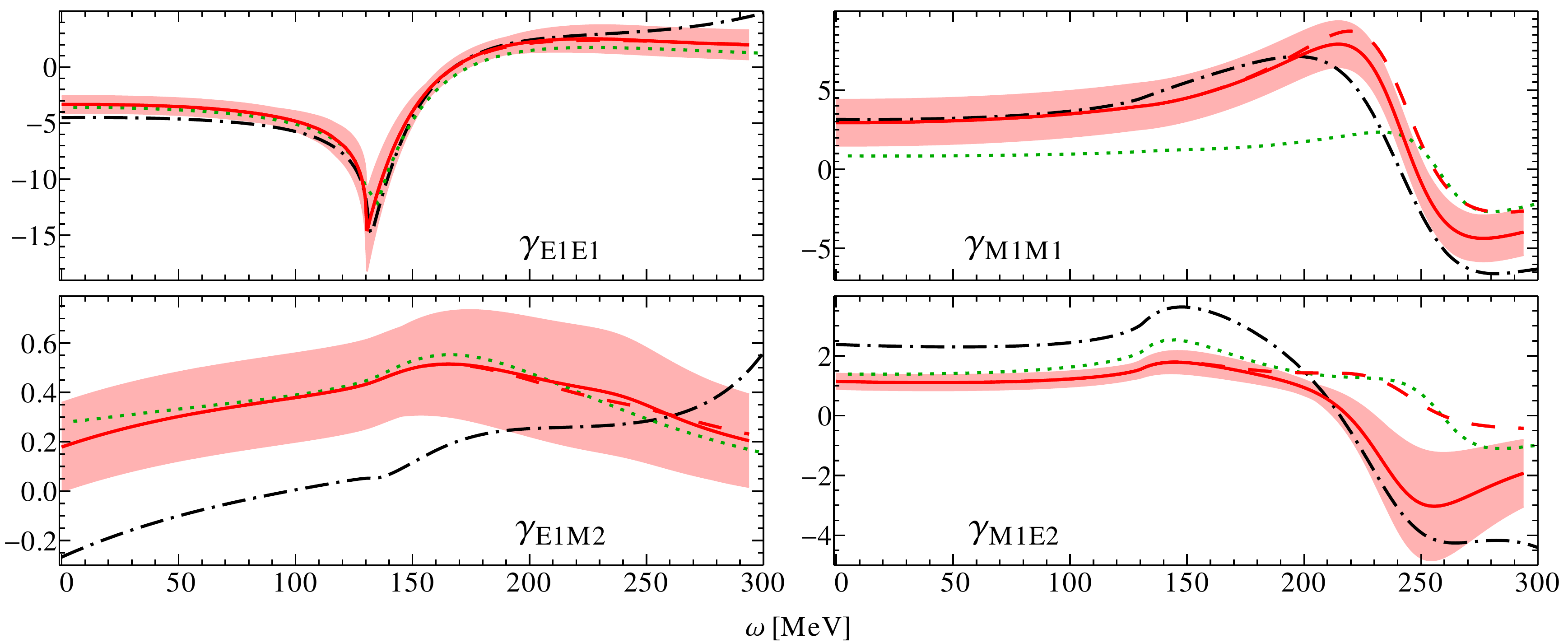}
\caption{  Real parts of {\it proton} spin dynamical polarisabilities in units of $10^{-4}\mathrm{fm}^4$
as functions of photon cms energy in MeV. The curves are the results of this B$\chi$PT calculation
with or without the $\gamma$N$\Delta$ vertex running (red solid and red dashed, respectively),
compared with the results of the DR calculation of ref.~\cite{Hildebrandt:2003fm} (black dot-dashed, curves from ref.~\cite{barbaraprivate}) and with the results
of the Computational Hadronic Model~\cite{Aleksejevs:2013cda,Aleksejevs:2015} (green dotted).}
\label{fig:pols_spin}
\end{figure}

\begin{figure}[ht]
 \includegraphics[width=\columnwidth]{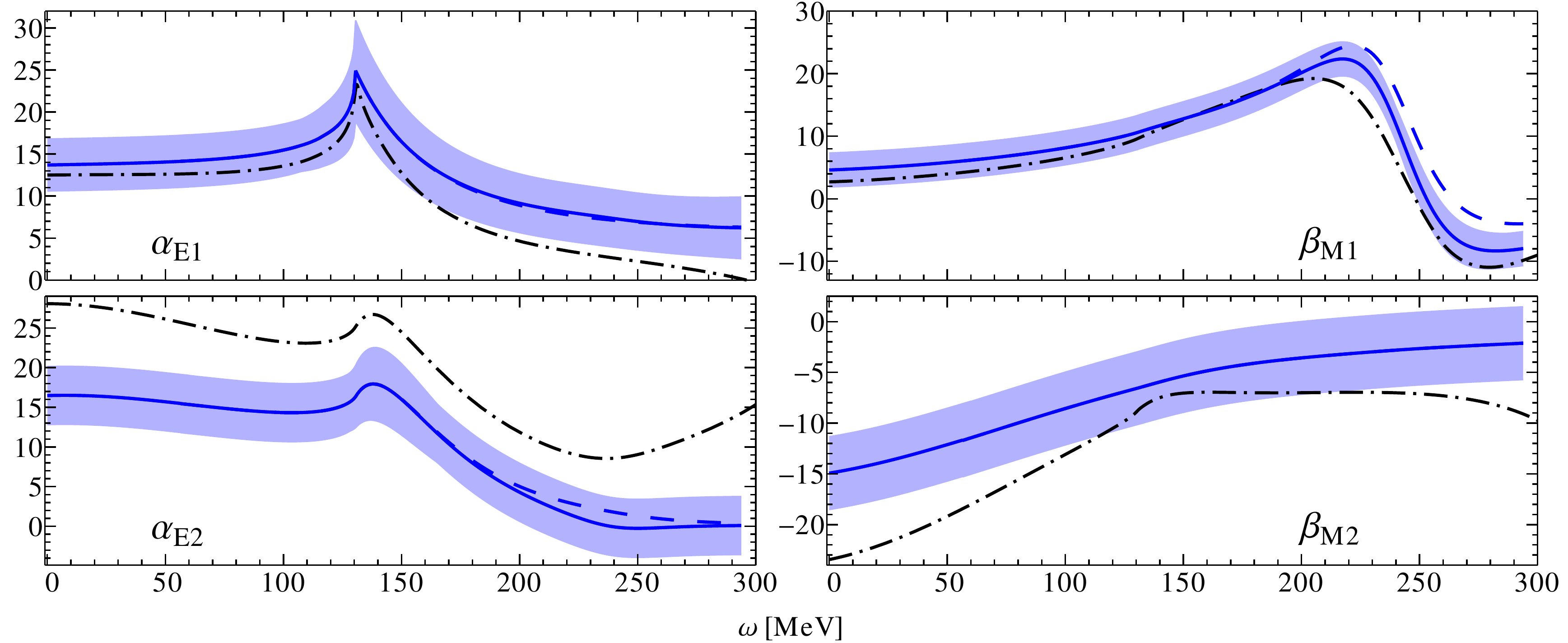}
\caption{  Real parts of {\it neutron} scalar dipole and quadrupole dynamical polarisabilities, in units of $10^{-4}\mathrm{fm}^3$ and
$10^{-4}\mathrm{fm}^5$, respectively, as functions of photon c.m.\ energy in MeV. The curves are the results of this B$\chi$PT calculation
with or without the $\gamma$N$\Delta$ vertex running (blue solid and blue dashed, respectively),
compared with the results of the DR calculation of ref.~\cite{Hildebrandt:2003fm} (black dot-dashed, curves from ref.~\cite{barbaraprivate}).}
\label{fig:pols_scalar_n}
\end{figure}

\begin{figure}[ht]
 \includegraphics[width=\columnwidth]{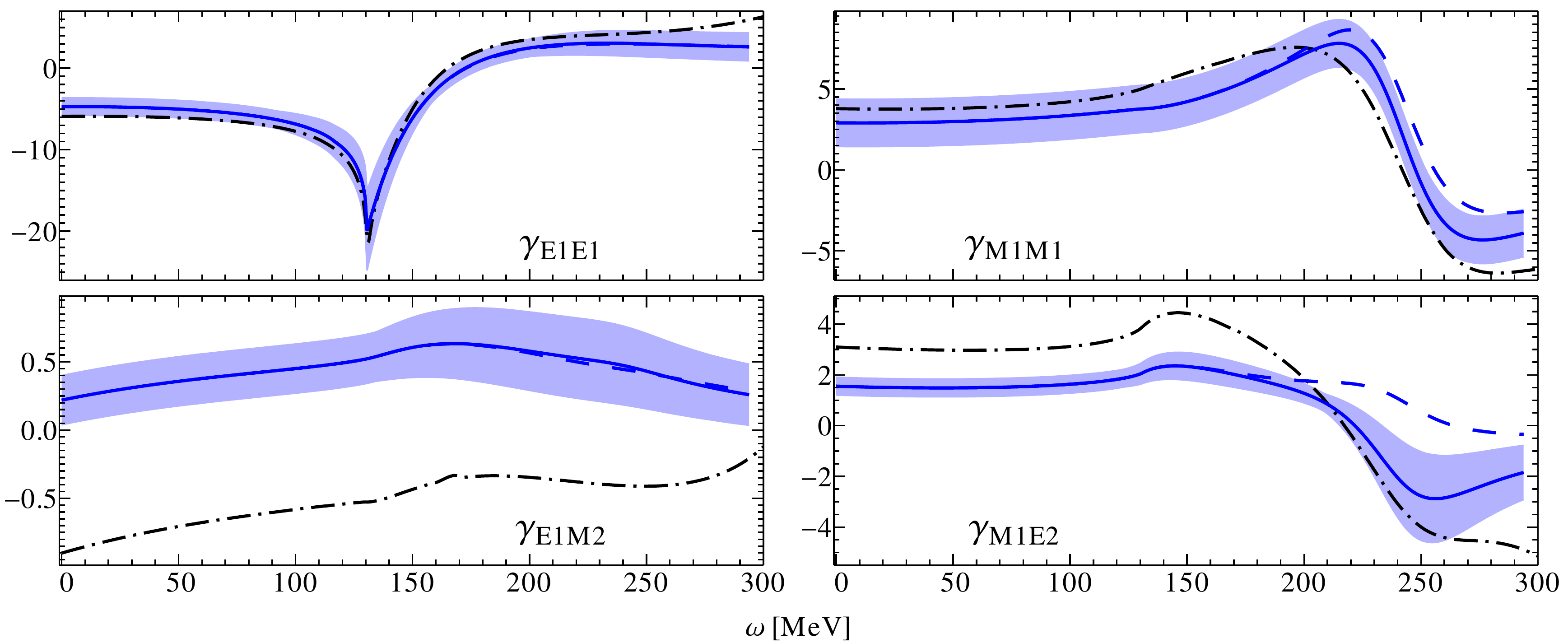}
\caption{  Real parts of {\it neutron} spin dynamical polarisabilities in units of $10^{-4}\mathrm{fm}^4$
as functions of photon c.m.\ energy in MeV. The curves are the results of this B$\chi$PT calculation
with or without the $\gamma$N$\Delta$ vertex running (blue solid and blue dashed, respectively),
compared with the results of the DR calculation of ref.~\cite{Hildebrandt:2003fm} (black dot-dashed, curves from ref.~\cite{barbaraprivate}).}
\label{fig:pols_spin_n}
\end{figure}

\begin{figure}[!htb]
 \includegraphics[width=\columnwidth]{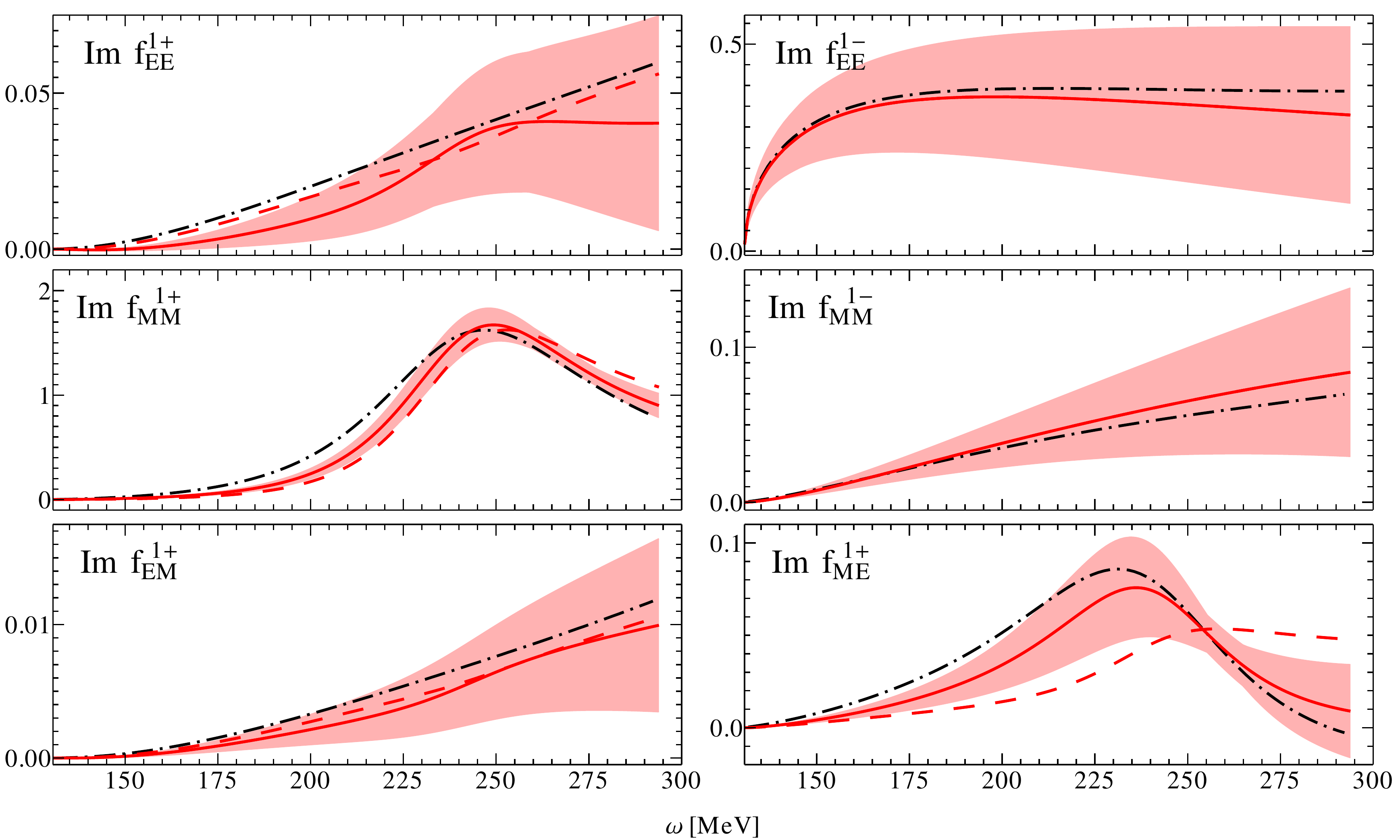}
\caption{  Imaginary parts of $l=1$  {\it proton} Compton multipoles in units of $10^{-3}m_{\pi}^{-1}$
as functions of photon c.m.\ energy in MeV. The curves are the results of this B$\chi$PT calculation,
with or without the $\gamma$N$\Delta$ vertex running (red solid/red dashed, respectively),
compared with the results of the MAID pion photoproduction analysis~\cite{Drechsel:2007if} (black dot-dashed).}
\label{fig:multl1}
\end{figure}

\begin{figure}[!htb]
 \includegraphics[width=\columnwidth]{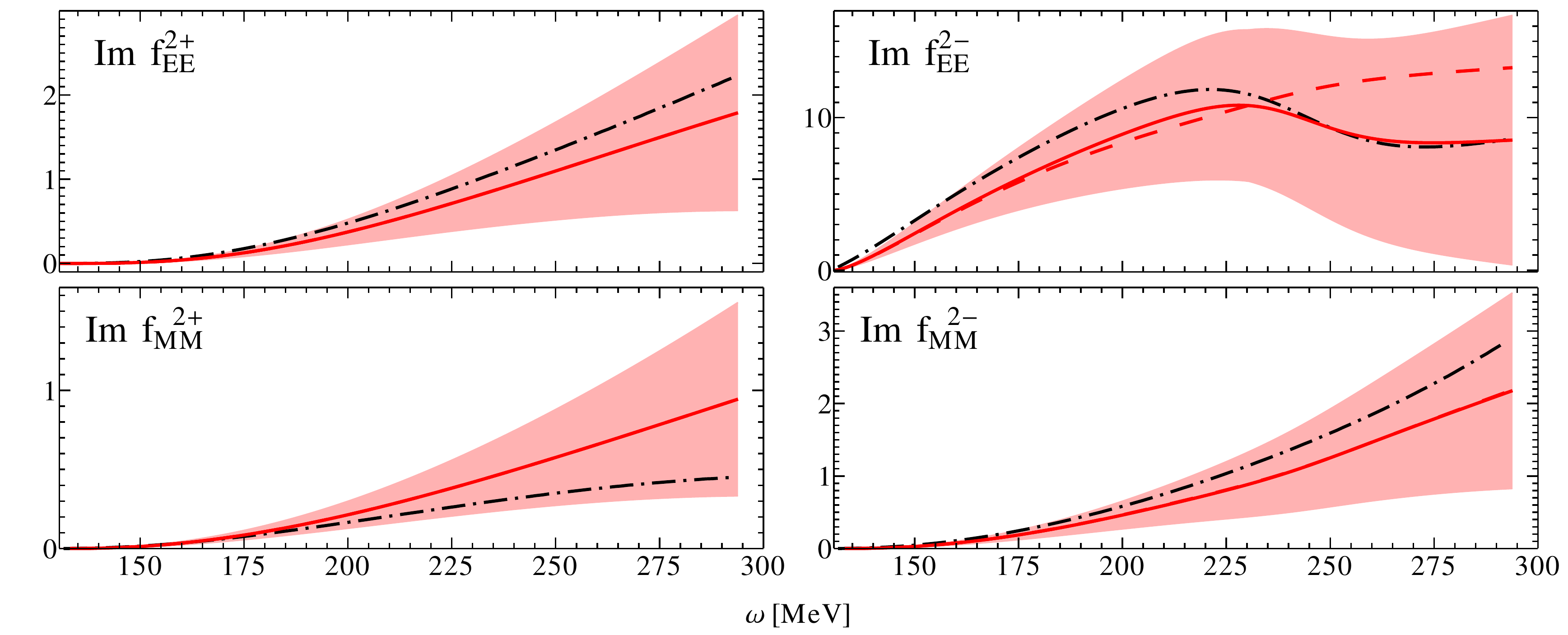}
\caption{  Imaginary parts of  $l=2$ {\it proton} Compton multipoles in units of $10^{-6}m_{\pi}^{-1}$
 as functions of photon c.m.\ energy in MeV. The curves are the results of this B$\chi$PT calculation
with or without the $\gamma$N$\Delta$ vertex running (red solid/red dashed, respectively),
 compared with the results of the MAID pion photoproduction analysis~\cite{Drechsel:2007if} (black dot-dashed).}
\label{fig:multl2}
\end{figure}

\begin{figure}[!thb]
 \includegraphics[width=\columnwidth]{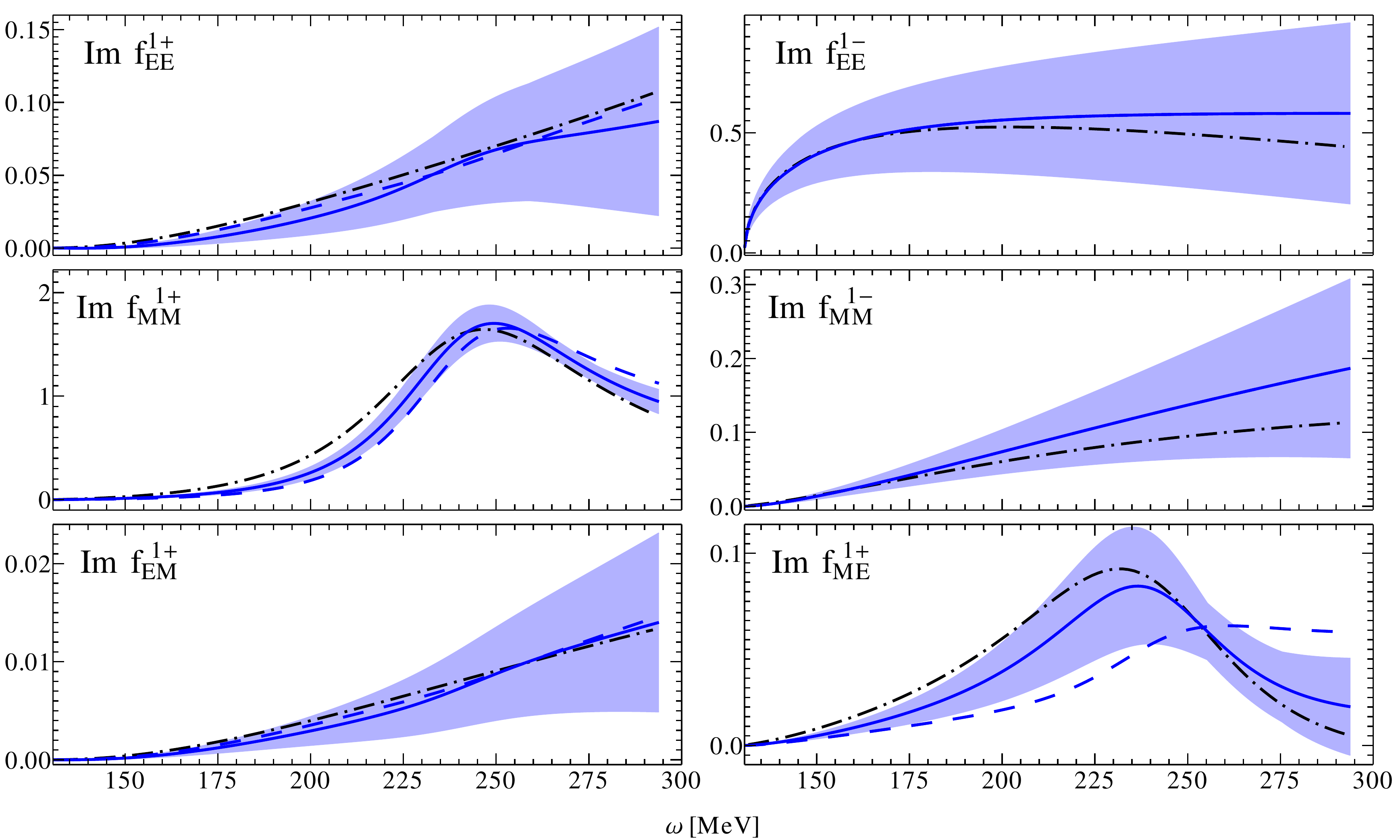}
\caption{  Imaginary parts of $l=1$ {\it neutron} Compton multipoles in units of $10^{-3}m_{\pi}^{-1}$
as functions of photon c.m.\ energy in MeV. The curves are the results of this B$\chi$PT calculation,
with or without the $\gamma$N$\Delta$ vertex running (blue solid/blue dashed, respectively),
compared with the results of the MAID pion photoproduction analysis~\cite{Drechsel:2007if} (black dot-dashed).}
\label{fig:multl1_n}
\end{figure}

\begin{figure}[!bht]
 \includegraphics[width=\columnwidth]{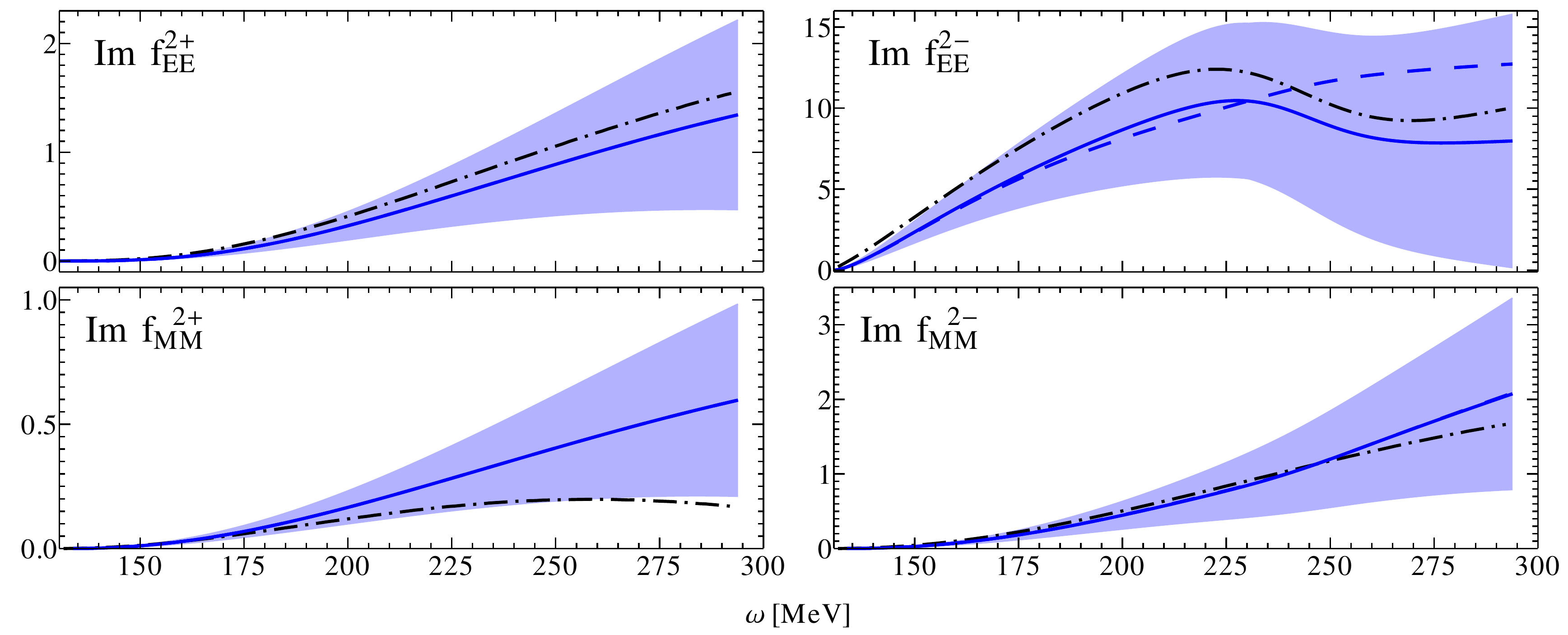}
\caption{  Imaginary parts of $l=2$ {\it neutron} Compton multipoles in units of $10^{-6}m_{\pi}^{-1}$
 as functions of photon cms energy in MeV. The curves are the results of this B$\chi$PT calculation
with or without the $\gamma$N$\Delta$ vertex running (blue solid/blue dashed, respectively),
 compared with the results of the MAID pion photoproduction analysis~\cite{Drechsel:2007if} (black dot-dashed).}
\label{fig:multl2_n}
\end{figure}

Here we provide predictions for the proton and neutron dynamical polarisabilities and  Compton
multipoles; our results are shown in Figs.~\ref{fig:pols_scalar}--\ref{fig:multl2_n}. Figs.~\ref{fig:pols_scalar} and~\ref{fig:pols_spin}
show the proton scalar (dipole and quadrupole) and spin dipole dynamical polarisabilities, in order. We compare our results with the dispersive calculation of ref.~\cite{Hildebrandt:2003fm}
and the results of the Computational Hadronic Model (CHM)~\cite{Aleksejevs:2013cda} (this framework is based on a covariant $\chi$PT,
and the main difference between our calculation and that of the CHM is the treatment of the Delta isobar; cf.\ the discussion below).
Note that the DR calculation of ref.~\cite{Hildebrandt:2003fm}
has been constrained to reproduce the values of $\alpha_{E1}$ and $\beta_{M1}$
of refs.~\cite{Olmos:2001} and~\cite{Kossert:2002ws} for the proton and the neutron,
respectively; see Tables~\ref{tab:polsscalar} and~\ref{tab:polsscalar_n}.
In addition, we show the effect of the loop corrections to the $\gamma$N$\Delta$ vertex.

The uncertainty bands in these figures are generated with a similar method to that used for the static polarisabilities,
in particular, at low energies these bands are defined by the corresponding  uncertainties on the static polarisabilities $\sigma$ (see \eqref{converge}).
At higher energies we estimate the errors due to higher-order contributions as
\beq\label{eq:converge_Delta_new}
\sigma(\omega)=\Max[\delta\times(\Delta\ \mathrm{pole\ corr.}),
\delta\times(\piN\ \mathrm{loops}),\delta\times(\pi\Delta\ \mathrm{loops}),(\mathrm{sum})]\,,
\eeq
with $\delta=\omega/\Lambda_\chi$ and the fourth term being the sum of the first three.
Besides that, since the shape of the Delta peak is well constrained and is well
reproduced in $\chi$PT, higher-order corrections would be expected to change mostly
its magnitude, in which case they can be absorbed
by means of fitting the $\gamma$N$\Delta$ coupling constants to experimental data.
We therefore treat the leading Delta-pole contribution differently in this regime, estimating the uncertainty $\sigma_{g_M}(\omega)$  of the dynamic polarisability from this contribution by varying $g_M$ by $\pm 0.1$.

The final uncertainty band is calculated at any energy by taking the largest of these quantities: $\Max[\sigma,\sigma(\omega),\sigma_{g_M}(\omega)]$.
Note that the nucleon Born graphs, which
do not enter the dynamical polarisabilities but contribute to
the Compton multipoles (and the observables), are also well constrained and so are not taken into account in the uncertainty estimates.

Only in those polarisabilities and multipoles that receive a significant contribution from the Delta-pole and only at energies around the Delta peak is
the uncertainty driven by $\sigma(\omega)$ or $\sigma_{g_M}(\omega)$, otherwise it coincides with the static uncertainty. The only notable exception from this rule are the proton $\alpha_{E1}(\omega)$ and $\beta_{M1}(\omega)$ --- these two polarisabilities have been assigned small static uncertainties, and the energy-dependent uncertainty estimate
exceeds them at relatively low energies, cf.\ Fig.~\ref{fig:pols_scalar}.

One can see from Figs.~\ref{fig:pols_scalar} and~\ref{fig:pols_spin} that our results for the dynamical polarisabilities agree qualitatively with the dispersive calculation of ref.~\cite{Hildebrandt:2003fm}.
There are, however, some differences, for instance, our scalar polarisability $\beta_{M1}(\omega)$ as well as the spin polarisabilities
$\gamma_{M1M1}(\omega)$ and $\gamma_{M1E2}(\omega)$ show sizeable deviation from the DR curves, both at low energies and in the region of the Delta peak.
The inclusion of the loop corrections to the $\gamma N \Delta$ vertex moves our curves closer to the DR ones, especially at energies around the Delta pole; this was first noted in ref.~\cite{McGovern:2013a}. The remaining difference,
especially the B$\chi$PT polarisabilities being rather high around the Delta peak, might be explained, as speculated in ref.~\cite{Lensky:2014efa}, if the leading $\gamma$N$\Delta$
coupling constant is a bit too large. One could fit this constant to see if a closer agreement with the Compton data results at the same time in a better agreement
between the $\chi$PT and DR calculations. Overall, the B$\chi$PT calculation reproduces all the characteristic features that show up
in the DR calculations, such as the pion production cusp and the Delta peak, which is not surprising given the physics contents of the chiral Lagrangians used in the $\chi$PT calculations.

The Computational Hadronic Model, on the other hand, is close to B$\chi$PT in those polarisabilities where the Delta pole does not contribute much to the energy dependence:
$\alpha_{E1}(\omega)$, $\gamma_{E1E1}(\omega)$ and $\gamma_{E1M2}(\omega)$. 
In  $\alpha_{E1}(\omega)$ the difference can be explained by the fact that the curves of ref.~\cite{Aleksejevs:2013cda} do not contain the $\pi\Delta$ loops whose contribution is nearly a constant function of $\omega$~\cite{Hildebrandt:2003fm,Lensky:2012ag}).
The CHM tends, however, to substantially underpredict the Delta-pole contribution to the Delta-dominated dynamical polarisabilities, especially $\beta_{M1}(\omega)$ and $\gamma_{M1M1}(\omega)$.

The treatment of the Delta isobar in the CHM differs from ours as follows: the $\gamma$N$\Delta$ Lagrangian used in the former calculation
contains only the LO term (corresponding to our $g_M$ coupling), and this LO Lagrangian does not possess the spin-$3/2$ gauge symmetry that ensures
the spurious spin-$1/2$ degrees of freedom do not contribute in the amplitudes containing the Delta (see the discussion in ref.~\cite{Pascalutsa:2002pi}
and references therein). In addition, the calculation of the CHM does not include pion loops with the Delta isobar.
Of these differences, only the admixture of a spin-$1/2$ component in the Delta isobar has the potential to account for the difference
in the results for $\beta_{M1}(\omega)$ and $\gamma_{M1M1}(\omega)$;
it remains to be seen, however, whether this is in fact the reason behind the underestimated Delta-driven polarisabilities in the calculation
of ref.~\cite{Aleksejevs:2013cda,Aleksejevs:2015}.

The theoretical uncertainty bands, together with the curves that do not contain the $\gamma$N$\Delta$ vertex corrections (which are NLO at $\omega\sim\varDelta$), illustrate the estimated contribution
of the terms that can enter at the next order.
$\gamma_{E1M2}(\omega)$ has the largest fractional uncertainty, because its small central value arises from cancellations between the leading and subleading contributions.

In Figs.~\ref{fig:pols_scalar_n} and~\ref{fig:pols_spin_n} we show the same polarisabilities as above in Figs.~\ref{fig:pols_scalar} and~\ref{fig:pols_spin},
respectively, but now for the neutron. They are compared with the neutron
results of the DR calculation of ref.~\cite{Hildebrandt:2003fm}, and we also
show in these figures the effect of the loop correction to the $\gamma$N$\Delta$ vertex,
analogously to the proton case. Our results for the neutron
demonstrate a qualitative agreement with the DR calculation comparable to that
observed above for the proton. This is not surprising: as pointed out above,
all the essential physics
included in the DR is also captured by the chiral calculation at this order.
The main features of the neutron curves are the same as have been described
above for the case of the proton. In particular, the most significant deviations
between the B$\chi$PT and the DR curves around the Delta peak occur in the scalar magnetic
dipole polarisability 
$\beta_{M1}(\omega)$ and in the two spin polarisabilities, $\gamma_{M1M1}(\omega)$ and $\gamma_{M1E2}(\omega)$,
where the Delta contribution is dominant. It is also noticeable that
the DR values of the neutron $\gamma_{E1M2}(\omega)$ are significantly different from the proton
ones, whereas the chiral calculation at this order gives essentially the same
values for the proton and the neutron. This difference makes the disagreement between the
DR and B$\chi$PT curves for the neutron $\gamma_{E1M2}(\omega)$ more prominent than in the proton case.

Moving from dynamical polarisabilities to Compton multipoles,
we show in Figs.~\ref{fig:multl1} and~\ref{fig:multl2}
our results for the imaginary parts of the proton multipoles with $l=1$ and $l=2$ respectively.
The analogous neutron curves are shown in Fig.~\ref{fig:multl1_n} for the
imaginary parts of the $l=1$ multipoles and in Fig.~\ref{fig:multl2_n} for those
with $l=2$.
They are compared with the results of the MAID multipole analysis~\cite{Drechsel:2007if}
of pion photoproduction, converted into the imaginary parts
of the Compton multipoles via Eqs.~(\multrefs). The error bands are calculated as explained above, and we again show the effect of the loop corrections to the $\gamma$N$\Delta$
vertex. Our results agree reasonably well with the MAID curves --- the latter lie within the calculated uncertainty bands in most cases for both the proton and the neutron. It can also be seen that the inclusion of the $\gamma$N$\Delta$ vertex
corrections improves the agreement with the MAID analysis in most cases. In some multipoles, however, there is a slight disagreement between the B$\chi$PT results and the MAID analysis,
most notably, in the leading slope of the resonance peak in $\im f^{1+}_{MM}(\omega)$ and in $\im f^{1+}_{ME}(\omega)$  as well as in $\im f^{1+}_{EE}(\omega)$ where the addition of the loop corrections
actually worsens the agreement between the B$\chi$PT and the MAID curves,
particularly for the proton.

\clearpage

\section{Polarised observables for the proton}
\label{sec:obs}
The wealth of world data on unpolarised Compton scattering has allowed for high-precision extractions of $\alpha_{E1}$ and $\beta_{M1}$ in chiral EFT
fits, for instance, in a $\cO(p^4)$ HB$\chi$PT fit~\cite{McGovern:2012ew}, and a partial $\cO(p^4)$ B$\chi$PT fit~\cite{Lensky:2014efa}
based on the current calculation. In contrast, the unpolarised cross section appears to be very weakly sensitive to the {\it spin} polarisabilities:
even though their values 
differ significantly in HB$\chi$PT  and in the current B$\chi$PT work, the resulting fits
to the unpolarised Compton cross section yield practically identical values of $\alpha_{E1}-\beta_{M1}$~\cite{Lensky:2014efa}.
On the other hand, (double) polarised observables, in particular, the beam asymmetry $\Sigma_3$, and the beam-target asymmetries $\Sigma_{2x}$ and
$\Sigma_{2z}$ are known to be sensitive to the values of the spin polarisabilities~\cite{Babusci:1998ww}. 
These asymmetries are defined as follows.\\
$\Sigma_3$ is the beam asymmetry for photons polarised linearly either parallel or perpendicularly to the scattering plane, with nucleons unpolarised:
\beq
\Sigma_3=\frac{d\sigma^{||}-d\sigma^{\perp}}{d\sigma^{||}+d\sigma^{\perp}}\,;
\eeq
$\Sigma_{2x}$ is the beam-target asymmetry for photons polarised circularly and nucleons polarised along the $x$ axis, i.e., in the reaction plane perpendicularly to the photon momentum (which is along the $z$ axis):
\beq
\Sigma_{2x}=\frac{d\sigma^{R}_{x}-d\sigma^{L}_{x}}{d\sigma^{R}_x+d\sigma^{L}_{x}}\,;
\eeq
and $\Sigma_{2z}$ is beam-target asymmetry for photons polarised circularly and nucleons polarised along the $z$ axis:
\beq
\Sigma_{2z}=\frac{d\sigma^{R}_{z}-d\sigma^{L}_{z}}{d\sigma^{R}_z+d\sigma^{L}_{z}}\,.
\eeq
In addition, we consider $\Sigma_{1z}$, the beam-target asymmetry for photons polarised linearly
with the polarisation directed under $\pm \pi/4$ with respect to the reaction plane
and nucleons polarised along the $z$ axis:
\beq
\Sigma_{1z}=\frac{d\sigma^{\pi/4}_{z}-d\sigma^{-\pi/4}_{z}}{d\sigma^{\pi/4}_z+d\sigma^{-\pi/4}_{z}}\,.
\eeq
The latter asymmetry vanishes below the pion production threshold.
The expressions for these asymmetries are given, e.g., in ref.~\cite{Babusci:1998ww}; though the authors of that reference
use a different tensor basis in the decomposition of the Compton amplitude, their expressions can be easily transformed to the basis used
in our work; see Appendix~\ref{app:matAtoA} (see also ref.~\cite{Hildebrandt:2003md,Hildebrandt:2005ix} for explicit expressions
 in terms of amplitudes $A_i^{\mathrm{cm}}$).

\begin{figure}[ht]
 \includegraphics[width=\columnwidth]{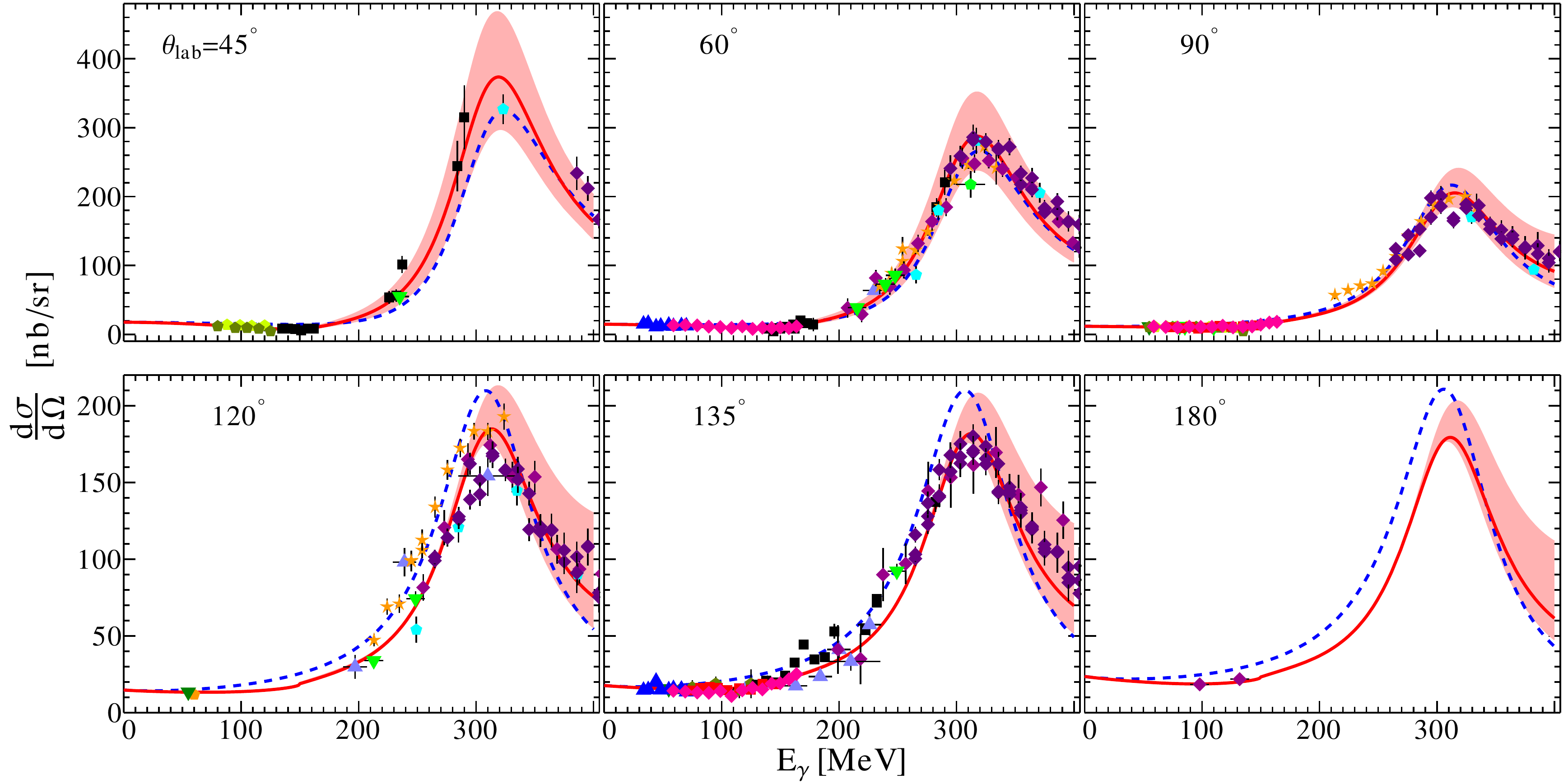}
\caption{  Unpolarised differential Compton-scattering cross section
in nb/sr as function of the photon energy $E_\gamma$ at fixed values
of the laboratory frame scattering angle $\theta_\mathrm{lab}$. The curves are the
complete results of this B$\chi$PT calculation (red solid) and the contribution
of only the nucleon Born and Delta-pole graphs (blue dashed).
For the data symbols see Table 3.1 of ref.~\cite{Griesshammer:2012we}.
}
\label{fig:XS}
\end{figure}

\begin{figure}[ht]
 \includegraphics[width=\columnwidth]{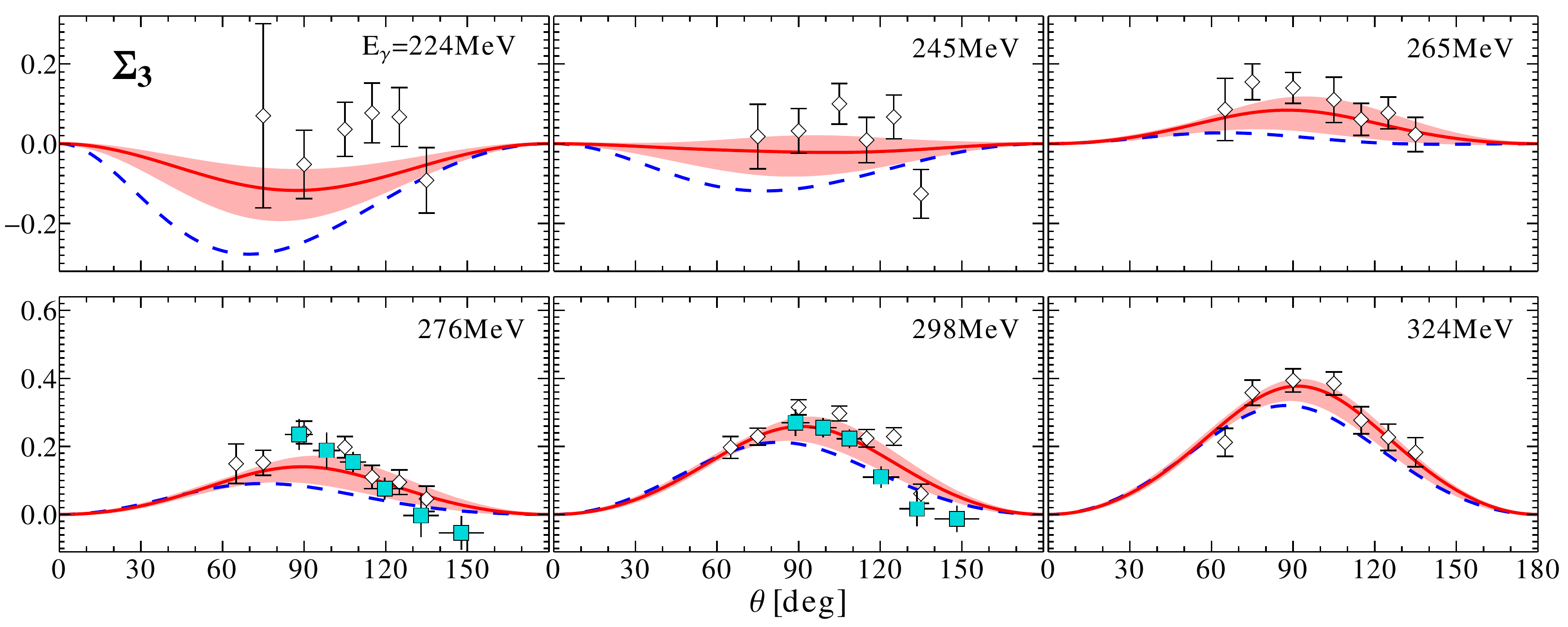}
\caption{  Reaction asymmetry $\Sigma_3$ as function of the c.m.\ angle at different values of $E_\gamma$ (annotated values are in MeV), compared with data
from LEGS~\cite{Blanpied:2001ae} (open diamonds) and MAMI~\cite{CC_thesis} (cyan squares).
The theoretical bands correspond to the full calculation and their width is determined as explained in the text. The blue dashed lines correspond to only the Born $+$ Delta
pole graphs included in the calculation.
}
\label{fig:S3}
\end{figure}

\begin{figure}[ht]
 \includegraphics[width=\columnwidth]{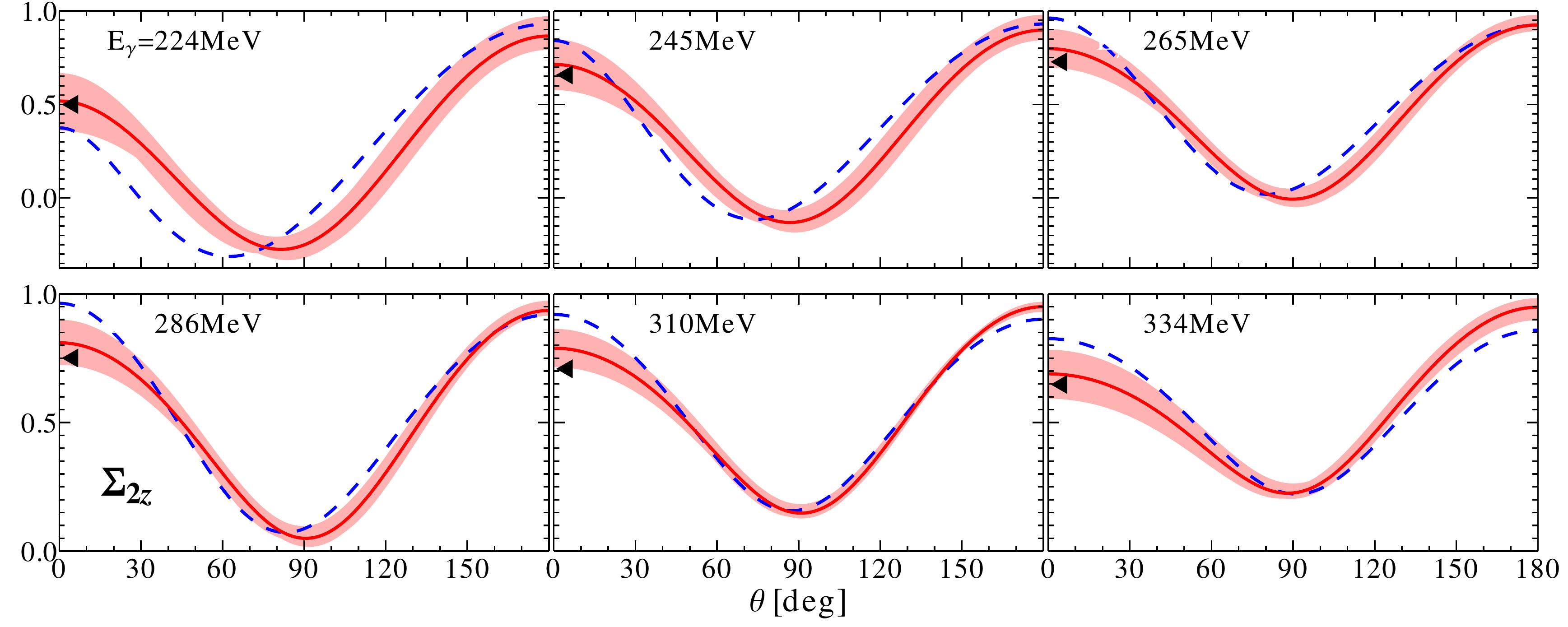}
\caption{  Reaction asymmetry $\Sigma_{2z}$ as function of the cms angle at the same values of $E_\gamma$ as in Fig.~\ref{fig:S3} (annotated values are in MeV).
 The black triangles show the results of the forward sum rule
evaluation of ref.~\cite{s2zfwd}.
The theoretical bands correspond to the full calculation and their width is determined as explained in the text.
 The blue dashed lines correspond to only the Born $+$ Delta
pole graphs included in the calculation.
}
\label{fig:S2z}
\end{figure}

\begin{figure}[ht]
 \includegraphics[width=\columnwidth]{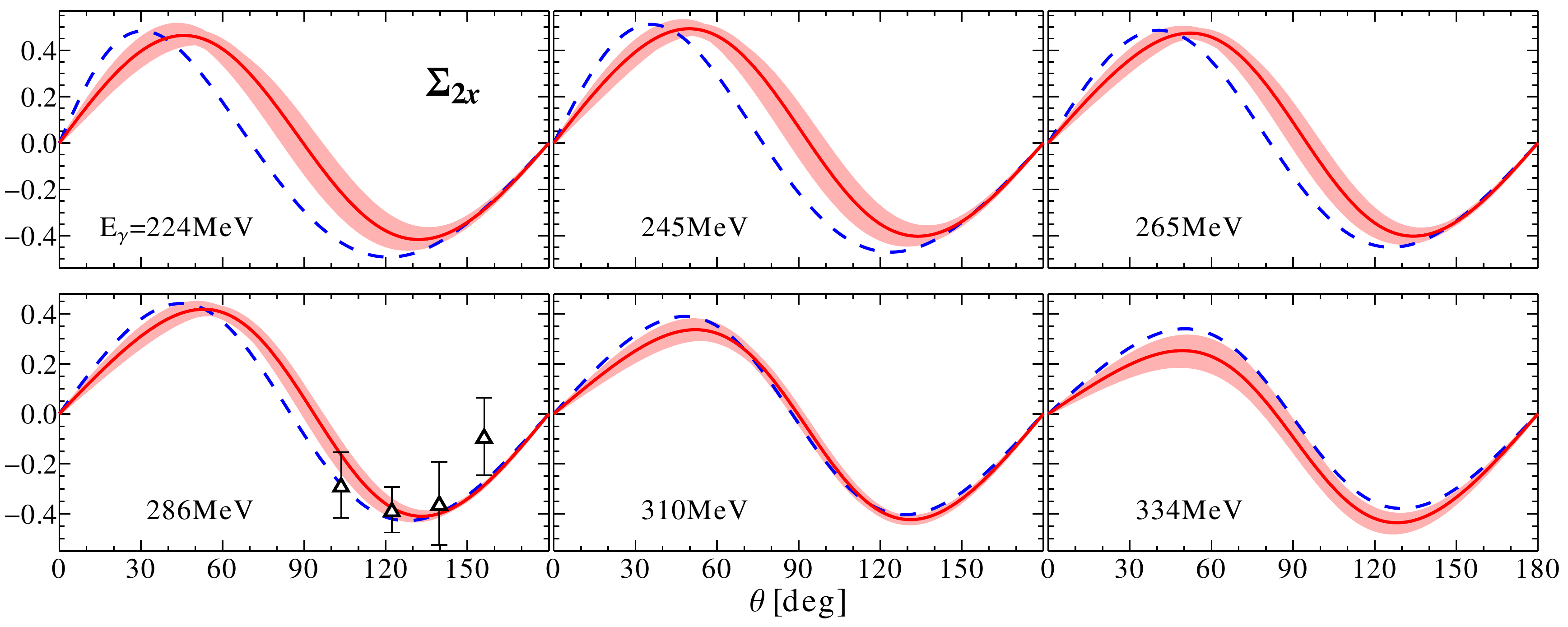}
\caption{  Reaction asymmetry $\Sigma_{2x}$ as function of the c.m.\ angle at the same values of $\omega_{\mathrm{lab}}$ as in Fig.~\ref{fig:S3} (annotated values are in MeV).
The data (open triangles) at $E_\gamma=286$~MeV are from the recent A2 experiment~\cite{Martel:2014pba}.
The theoretical bands correspond to the full calculation and their width is determined as explained in the text.
 The blue dashed lines correspond to only the Born $+$ Delta
pole graphs included in the calculation.
}
\label{fig:S2x}
\end{figure}

\begin{figure}[ht]
 \includegraphics[width=\columnwidth]{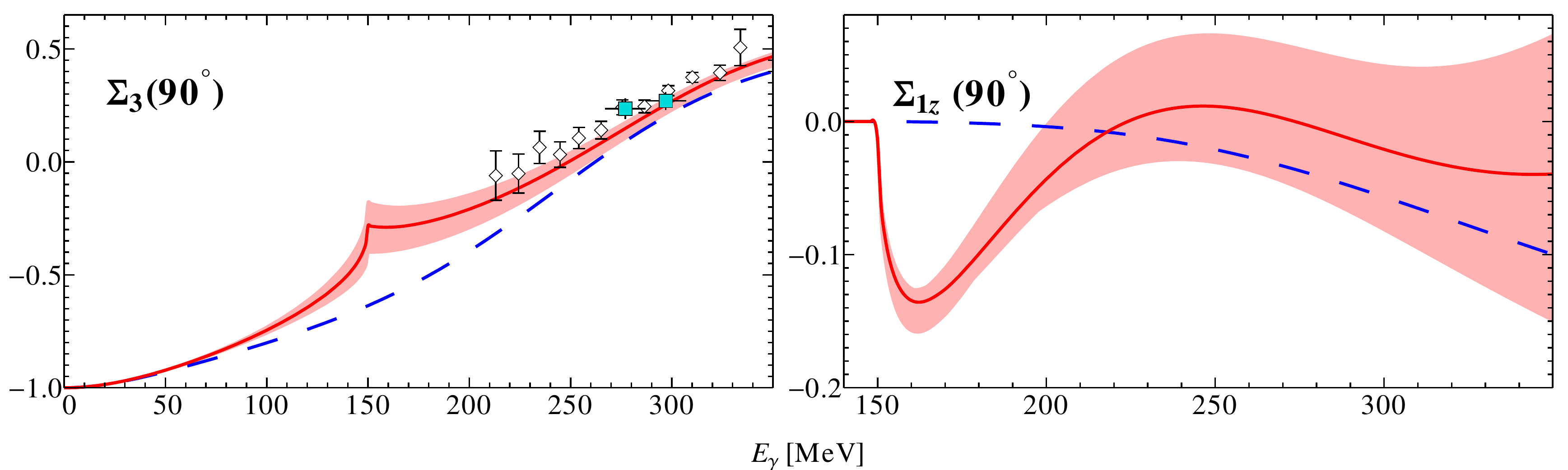}
\caption{  Left panel: reaction asymmetry $\Sigma_3$ as function of the laboratory frame photon energy at $\theta_\mathrm{cm}=90^\circ$, compared
with the data from LEGS~\cite{Blanpied:2001ae} (open diamonds) and MAMI~\cite{CC_thesis} (cyan squares). Right panel: reaction asymmetry $\Sigma_{1z}$
as function of the laboratory frame photon energy at $\theta_\mathrm{cm}=90^\circ$.
The theoretical bands correspond to the full calculation and their width is determined as explained in the text. The blue dashed lines correspond to only the Born $+$ Delta
pole graphs included in the calculation.
}
\label{fig:s3_s1z}
\end{figure}

As shown in ref.~\cite{Babusci:1998ww}, measurement of the double polarised observables $\Sigma_{2x}$ and $\Sigma_{2z}$
can provide information on all four spin polarisabilities, by analysing the angular dependence of the asymmetries.
In this section, we show our predictions for the asymmetries $\Sigma_3$, $\Sigma_{2x}$, and $\Sigma_{2z}$, together with theoretical uncertainty bands,
and compare them with the data from LEGS~\cite{Blanpied:2001ae} and with the recent data from Mainz~\cite{Martel:2014pba,CC_thesis}.
We also provide our predictions for $\Sigma_{1z}$ and discuss its features.
For the sake of completeness, we also show our predictions for the unpolarised cross section. Dispersion-relation
predictions for these asymmetries have been provided in ref.~\cite{Pasquini:2007hf}.

The uncertainty bands shown in figures are obtained similarly to those calculated for the (real parts of) Compton multipoles, namely, adding to the Compton amplitude any of the quantities in Eq.~\ref{eq:converge_Delta_new},
\beq
\delta A=\left\{\delta\times(\Delta\ \mathrm{pole\ corr.}),
\delta\times(\piN\ \mathrm{loops}),\delta\times(\pi\Delta\ \mathrm{loops}),(\mathrm{sum})
\right\}\,,
\label{eq:ampl}
\eeq
then calculating the observables using the shifted amplitudes $A\pm\delta A$ and taking
the one of the $\delta A$'s that provides the maximal width of the
band.
 Our estimate results in moderately wide uncertainty bands for all of the asymmetries,
especially in $\Sigma_{2z}$ and $\Sigma_{2x}$.
The band for the cross sections is also
rather wide. Since the amplitude is dominated
at these energies by the Delta pole, the main effect of adding the uncertainties as prescribed by \eqref{ampl} is to change the normalisation of
the predicted cross sections --- this change will mostly cancel in the asymmetries. 
By analogy, it also follows that some of the latter might be more sensitive
to the mechanisms other than the Delta-pole or Born graphs.

In Fig.~\ref{fig:XS}, we show the theoretical curves and bands for the unpolarised
cross section, together with experimental data from many different experiments,
see Table~3.1 of ref.~\cite{Griesshammer:2012we} for the list.
This figure shows that the B$\chi$PT prediction for the Compton unpolarised cross
section is consistent with the available data at energies up to the Delta peak.
We also include curves with only the nucleon Born and the Delta-pole graphs included, demonstrating the effect of the $\piN$ and $\pi\Delta$ loops on the unpolarised
cross section.

Our predictions for $\Sigma_3$ are shown in Fig.~\ref{fig:S3}, where we provide a comparison with the LEGS experimental data~\cite{Blanpied:2001ae}, at lab energies
$E_\gamma$ in the range 220--340~MeV. We also show the recent data
on this asymmetry from MAMI~\cite{CC_thesis}.
Our results are consistent with the LEGS data, especially at higher energies in this range. At lower energies the description of the data
is a bit worse but still broadly satisfactory.
The comparison between the dashed ans solid curves shows that the $\piN$ and $\pi\Delta$
 loops also become more important at lower energies, especially since $\Sigma_{3}$
 passes through zero at $E_\gamma\simeq 250$~MeV.
This view can also be supported by Fig.~\ref{fig:s3_s1z} (left panel) where we show $\Sigma_3$ as a function of $E_\gamma$
at $\theta=90^\circ$: the data, in particular, their energy dependence, are well reproduced by the theoretical curve.

Figures~\ref{fig:S2z} and~\ref{fig:S2x} show the other two asymmetries, $\Sigma_{2z}$ and $\Sigma_{2x}$ respectively,  at the same values of $\omega_\mathrm{lab}$
as in Fig.~\ref{fig:S3}. As pointed out above,  uncertainty bands for these observables are narrower than for $\Sigma_3$.

Figure~\ref{fig:S2z} includes the empirical extraction of $\Sigma_{2z}$ at $\theta=0^\circ$ performed via evaluation of the forward sum rule in ref.~\cite{s2zfwd}.
The B$\chi$PT theoretical curves show a remarkable agreement with the forward values extracted
via the sum rule. The dashed curve does not contain the
chiral loops, and the difference  shows the great importance of the latter
in this observable.

Figure~\ref{fig:S2x} shows the comparison of our prediction for $\Sigma_{2x}$ with the recent
experimental data from A2@MAMI~\cite{Martel:2014pba}.
These data correspond to the photon energy in the range between $272.7$ and $303.3$~MeV, 
and they are compared with our theoretical curve at $E_\gamma=286$~MeV,
which is almost equal to the central point of the experimental interval, 
$E_\gamma=288$~MeV.
One can see that our theoretical prediction also describes these new data on $\Sigma_{2x}$
well, the estimated theoretical uncertainty band being considerably smaller than the experimental errors. The typical width of the band is comparable to
the variation of the theoretical curve with energy over the A2 experimental range given above. 
Again, the curves that include only the nucleon Born and Delta-pole graphs
illustrate the relative importance of these and of the loop graphs in these
asymmetries. One can see that the loop graphs generally become more important at lower energies,
in accordance to what one would expect given that the Delta-pole amplitude,
dominating around the Delta peak, quickly falls off as the photon energy decreases.

Last but not least, the right panel of  Fig.~\ref{fig:s3_s1z}  shows the predictions for
the asymmetry $\Sigma_{1z}$ at $\theta=90^\circ$ as a function of the lab frame photon energy.
As pointed out above, $\Sigma_{1z}=0$ below the
pion production threshold. It is noticeable that this asymmetry is generated mostly
by the imaginary part of pion loops --- the Delta-pole contribution becomes important
only at energies around 220~MeV and higher. Note also that
this asymmetry can be measured in the same experiment as $\Sigma_{3}$.
Measuring this asymmetry
in the region just above the pion production threshold could thus be done simultaneously
with the planned measurements of $\Sigma_{3}$, and
it has a potential to provide new important information on the nucleon pion cloud.

\section{Conclusion}
\label{sec:concl}
We have considered  low-energy Compton scattering off the nucleon in the framework of manifestly-Lorentz-invariant $\chi$PT,
extending our previous calculation~\cite{Lensky:2009uv}
to the Delta-resonance region. 
We have examined the B$\chi$PT predictions for the static and 
dynamical polarisabilities of the nucleon and resolved their
matching in the low-energy expansion. 

These predictions are compared with models based
on fixed-$t$ dispersion relations (DRs) and the 
state-of-art heavy-baryon $\chi$PT results. 
Our results show both similarities and differences to each.
Without any exceptions we agree with HB$\chi$PT within the
combined errors. The agreement for the scalar dipole polarisabilities (which are fit in HB$\chi$PT but predicted here) is extremely good; however the central values of the spin polarisabilities can be quite different.
With DR it is harder to judge as the values do not have systematic uncertainties, but it seems that there is a genuine disagreement in the scalar quadrupole polarisabilities, and probably also in the mixed spin polarisabilities of the neutron.  There is also an apparent discrepancy in $\beta_{M1}$. However it must be noted that this is not a pure prediction in DR, but relies on rather old fits to Compton data which are not as systematic as the recent chiral ones.

These discrepancies are very intriguing since 
all of these calculations provide a comparably good
description of the Compton-scattering observables
at energies up to 400 MeV. 
Our results for the polarised observables---beam asymmetry $\Sigma_3$ and beam-target asymmetries $\Sigma_{2x}$--- compare well with the experimental data from LEGS and with the recent results from A2@MAMI. 

The results
for $\Sigma_{2z}$ at zero scattering angle 
are in remarkable agreement with
the recent empirical evaluation based
on dispersive sum rules; see \Figref{S2z}.
This excellent agreement is 
due only to the pion-nucleon and pion-Delta
loop contributions, thus demonstrating the
importance of chiral dynamics in this observable.

Even greater dependence on the chiral dynamics
is seen in $\Sigma_{1z}$ (analogous to the $G$ asymmetry
in pion photoproduction); see \Figref{s3_s1z}.
In the range between 150 and 200 MeV the
dominant contribution to this observable
is seen to come from the chiral loops.
The  $\Sigma_{1z}$ asymmetry thus seems to be ideally
suited for testing chiral dynamics in Compton scattering,
and the various approaches are expected to differ here significantly. 
Any sufficiently precise measurement of this asymmetry  
is potentially very interesting.

In general, though as we are satisfied to see the
predictions of B$\chi$PT to agree with the available
data on proton Compton scattering, the range of predictions within 
the state-of-the-art approaches (including DRs and HB$\chi$PT)
is troublesome. We shall keep looking for
opportunities to put these approaches to further 
experimental tests. Given the ongoing experimental
programs at the HIGS and MAMI facilities, we are hopeful
that these issues will be resolved in the near future.

\section*{Acknowledgements}
We thank Barbara Pasquini and Aleksandrs Aleksejevs for providing us with the results of their calculations, and Philippe Martel for sending us the recent experimental data on $\Sigma_{2x}$.
V.~L.\ thanks Anatoly L'vov for stimulating discussions.

This work was supported by the Deutsche Forschungsgemeinschaft (DFG) through the
Collaborative Research Center ``The Low-Energy Frontier of the Standard Model'' (SFB 1044), by
the Russian Federation government grant No.\ NSh-3830.2014.2, by the MEPhI Academic Excellence Project (Contract No. 02.a03.21.0005), and the  UK Science and Technology Facilities
Council under grants  ST/J000159/1 and  ST/L005794/1. 
\appendix


\section{Transformation between the bases}
\label{app:matAtoA}
Here, we provide the matrices that transform between several different sets of amplitudes introduced in Sect.~\ref{sec:amp:tensor}.
In order to transform between any two of these sets, it is enough to know the elements of a chain of transformations that, combined, can
connect the two given sets. Here we show some of the minimal set of transformations that result in especially compact expressions, namely,
those that transform from $\MA_i$ to $T_i$ and to $A^\mathrm{L}_i$ (as defined in ref.~\cite{Babusci:1998ww}):
\def\ser{\\[1.2ex]} 
\beq
 C^{\MA\to T}=
\begingroup\small
\frac\MN{2\nu^2}\,\left(
\begin{array}{cccccc}
  -\dfrac{\eta-2 \nu ^2 }{\MN} & 0 & \dfrac{2 \nu ^2}{\MN} & 0 & 0 & 0 \ser
  -\dfrac{\eta  t}{2 \MN} & 0 & 0 & 0 & 0 & 0 \ser
  4 \nu  & \dfrac{t-4 \MN^2}{\MN^2} & 0 & 0 & \dfrac{\nu \left(4 
M^2-t\right)}{2 \MN^2} & \dfrac{2 \nu }{\MN}
    \ser
  2 \nu  \left(\eta - 2\nu ^2\right) & -2 \left(\eta -2 \nu ^2\right) & 
-4 \nu ^3 & 4 \nu ^2 & 0 &
    0 \ser
  2 \nu  (\eta +t) & -4 \left(\eta -2 \nu ^2\right) & 0 & 0 & \eta  \nu  
& \dfrac{\nu  t}{\MN} \ser
  2 \nu  (\eta -t) & -8 \nu ^2 & 0 & 0 & \eta  \nu & -\dfrac{\nu  t}{\MN} \ser
  -\eta  \nu  t & \eta  t & 0 & 0 & 0 & 0 \ser
  0 & 0 & 0 & 0 & -\dfrac{\nu ^2 t}{\MN} & 0 \ser
\end{array}
\right)^{\textstyle \top}\! ,
\endgroup
\eeq

\beq
C^{\MA\to A^\mathrm{L}}=
\begingroup\small
\frac\MN {2 \nu^3}\,\left(
\begin{array}{cccccc}
  -\dfrac{\nu  \left(4 \MN^2-t\right)}{4 \MN^3} & 0 & -\dfrac{\nu }{\MN} & 0 & 
0 & 0 \ser
  -\dfrac{\eta  \nu }{2 \MN} & 0 & -\dfrac{\nu  t}{2 \MN} & 0 & 0 & 0 \ser
  \dfrac{\nu ^2}{\MN^2} & 0 & 1 & 1 & \dfrac{t-4 \MN^2}{4 \MN^2} & \dfrac{t-4 
\MN^2}{4 \MN^2} \ser
  0 & -\dfrac{\nu ^2 \left(4 M^2-t\right)}{2 \MN^2} & \dfrac{1}{2} \left(4 
\nu ^2+t\right) &
    \dfrac{t}{2} & \dfrac{1}{2} \left(4 \nu ^2-\eta \right) & 
-\dfrac{\eta }{2} \ser
  \dfrac{\nu ^2 t}{2 \MN^2} & -\dfrac{\nu ^2 \left(4 \MN^2-t\right)}{2 \MN^2} 
& 2 \nu ^2+t & t & 2 \nu
    ^2-\eta  & 2 \nu ^2-\eta  \ser
  -\dfrac{\nu ^2 t}{2 \MN^2} & \dfrac{\nu ^2 \left(4 \MN^2-t\right)}{2\MN^2} 
& 2 \nu ^2 & 0 & -2 \nu ^2 &
    -2 \nu ^2 \ser
  0 & \eta  \nu ^2 & -\dfrac{1}{4} t \left(4 \nu ^2+t\right) & 
-\dfrac{t^2}{4} & \dfrac{\eta  t}{4} &
    \dfrac{\eta  t}{4} \ser
  0 & -\dfrac{2 \nu ^3}{\MN} & 0 & 0 & 0 & 0
\end{array}
\right)^{\textstyle \top}\!.
\endgroup
\eeq
In these equations the kinematical invariants $\nu$ and $\eta$ are
\beq
\nu = \frac{s-u}{4\MN},\qquad \eta=\frac{\MN^4-su}{\MN^2}\,.
\eeq
For completeness, we also give the transformation matrix between $T_i$ and $A^\mathrm{L}_i$~\cite{Babusci:1998ww}:
\beq
C^{ T\to A^\mathrm{L}}=
\begingroup\small\left(
\begin{array}{cccccc}
  \frac{1}{t} & \frac{\nu }{t} & \frac{1}{t} & \frac{\nu }{t} & 0 & 0\ser
  0 & \frac{\nu }{t} & 0 & \frac{\nu }{t} & \frac{2}{t} & 0\ser
  \frac{1}{\eta } & -\frac{t}{4 \eta  \nu } & -\frac{1}{\eta } & 
\frac{t}{4 \eta  \nu } & 0 & 0\ser
  0 & -\frac{t}{4 \eta  \nu } & 0 & \frac{t}{4 \eta  \nu } & 0 & \frac{2 
M}{\eta }\ser
  0 & \frac{1}{4 \nu } & 0 & \frac{1}{4 \nu } & 0 & 0\ser
  0 & \frac{1}{4 \nu } & 0 & -\frac{1}{4 \nu } & 0 & 0
\end{array}
\right)\endgroup\,.
\eeq
Due to its length we do not provide here the transformation from $A_i$ to $A^\mathrm{cm}_i$ (see \eqref{TviaAH}).  However
it can be found from 
$C^{A\to T}$ and the following~\cite{Babusci:1998ww}:
\beq
 C^{T\to A^{\mathrm{cm}}}=
\frac 1 {2\MN}\begin{pmatrix}
 0 & 0 & \frac{(W_+)^2-z(W_-)^2}{2W} & \omega (zW_- +W_+) & 0 & 0 \\
 \frac{(W_+)^2-z(W_-)^2}{2W(1-z^2)} & \frac{\omega (z W_- +W_+)}{1-z^2} & \frac{z (W_+)^2-z^2(W_-)^2}{2W(1-z^2)} &
   \frac{\omega z (z W_- +W_+)}{1-z^2} & 0 & 0 \\
 \omega-W_- & \omega W_- & z (\omega-W_-) & \omega z W_-  & 2 \omega & 2\omega W \\
 0 & 0 & \omega-W_- & \omega W_- & 0 & 0 \\
 \frac{z (W_--\omega)}{z^2-1} & \frac{\omega z W_- }{1-z^2} & \frac{z^2 (W_- -\omega)}{z^2-1} & \frac{\omega z^2 W_- }{1-z^2} & \frac{\omega}{1-z} &
   -\frac{\omega W}{ z+1} \\
 \frac{\omega-W_-}{z^2-1} & \frac{\omega W_-}{z^2-1} & \frac{z (\omega-W_-)}{z^2-1} & \frac{\omega z  W_- }{z^2-1} & \frac{\omega}{z-1} & -\frac{\omega
   W}{z+1}
\end{pmatrix}\,,
\eeq
where $W=\sqrt{s}$, $W_\pm=W\pm \MN$, and $\omega=(W^2-\MN^2)/(2W)$ and $z=\cos\theta_{\mathrm{cm}}$  refer to the cm-frame photon energy and scattering angle.

The transformation between $A_i$ and $A^\mathrm{Br}$ is given in the appendix of ref.~\cite{McGovern:2012ew}.

Finally, the transformation matrix between  $A^\mathrm{L}_i$ and the Breit-frame amplitudes $A^\mathrm{Br}_i$ is
\beq\label{eq:l'vov-Breit}
C^{A^\mathrm{L}\to A^{\mathrm{Br}}}=
\frac{\EN\w_B^2}{\MN}
\begin{pmatrix}
 (z-1) & 0 & -\frac{\EN^2  (z+1)}{\MN^2} & 0 & \frac{\w_B^2 (z-1)}{\MN^2} & - (z+1) \\
-1 & 0 & \frac{\EN^2}{\MN^2} & 0 & -\frac{\w_B^2}{\MN^2} & 1 \\
0 & \frac{\w_B (z-1)}{\EN} & 0 & \frac{\EN^2 \w_B (z+1)}{\MN^3} & \frac{\w_B (\MN(z+1)-2 \EN)}{\MN^2} & 0
  \\
0 & 0 & 0 & 0 & \frac{\w_B}{\MN} & -\frac{\w_B}{\MN} \\
0 & -\frac{\w_B}{2 \EN} & 0 & -\frac{\EN^2 \w_B}{2 \MN^3} & \frac{\w_B (\EN-\MN z)}{\MN^2 (z-1)} &
\frac{\w_B}{2 \MN} \\
0 & \frac{\w_B}{2 \EN} & 0 & -\frac{\EN^2 \w_B}{2 \MN^3} & \frac{\w_B (\MN-\EN)}{\MN^2 (z-1)} &
-\frac{\w_B}{2 \MN}
\end{pmatrix}\,,
 \eeq
where $\w_B$ and $z=\cos\th_B$ here refer to the Breit-frame photon energy and scattering angle, and $\EN$ is the nucleon energy (equal before and after the collision), with 
$\EN=\nu \MN/\w_B=\sqrt{\MN^2-t/4}$.

Note that the amplitudes $\MA_i$, $A^\mathrm{L}_i$ and $T_i$ are defined assuming a covariant spinor normalisation of $2\MN$, 
whereas the amplitudes in the  non-relativistic bases,   $A^\mathrm{Br}_i$ and  $A^\mathrm{cm}_i$, assume a Pauli spinor normalisation of unity.

\section{Matching of higher-order polarisabilities}
\label{app:hipols}

Here we provide the relations between the polarisabilities, static and dynamic, and the 
the L'vov amplitudes of ref.~\cite{Babusci:1998ww}. The latter can be expanded in the crossing-invariant variables
 $\nu=(s-u)/(2\MN)$ and $t$:
\begin{equation}
A^\mathrm{L}_i(\nu ,t)=a_i+a_{i,\nu}\nu^2+a_{i,t}t+\ldots.
\label{eq:lvov}
\end{equation}
The non-Born part of the Taylor coefficients
$a_i$, $a_{i,\nu}$, $a_{i,t}$ can be related to polarisabilities as follows.

We have already argued that crossing symmetry means that static polarisabilities 
should be defined in the Breit frame, so using 
$A_i^\mathrm{Br}=\sum_jC_{ij}^{A^\mathrm{L}\to A^{\mathrm{Br}}}A_i^\mathrm{L}$
with the transformation matrix defined in \eqref{l'vov-Breit} gives, for example,
\begin{align}
A_1^\mathrm{Br}=& \frac{\EN}{\MN}\,\left((a_1-a_3-a_6)
   z-a_1-a_3-a_6\right)\omega_\mathrm{B}^2\nonumber \\
&+\left(-a_{1,\nu }-a_{3,\nu }-a_{6,\nu }+
  2 \left(a_{1,t}+a_{3,t}+a_{6,t}\right)+
  z\left(a_{1,\nu }-a_{3,\nu }-a_{6,\nu}-4 a_{1,t}\right)\right.\nonumber \\ &\left.+
  2 z^2\left(a_{1,t}-a_{3,t}-a_{6,t}\right)
 +\textfrac1 {2 \MN^2}(a_3 z^2+2 a_5 z-a_3-2 a_5)
\right)\omega_\mathrm{B}^4+{\cal O}(\omega_\mathrm{B}^6).
\end{align}
where $\omega_\mathrm{B}$ and $z=\cos\theta_\mathrm{B}$ here refer to the Breit-frame photon energy and scattering angle.

By comparing these to the direct results of the effective Hamiltonians Eqs.~(\heffrefs) as given in \eqref{polamp}, we obtain
\begin{align} \label{eq:polsolbreit}
4\pi\alpha _{{E1}}&= -a_1-a_3-a_6,\qquad
4\pi\beta_{{M1}}= a_1-a_3-a_6,\nonumber \\
4\pi\gamma_{{E1E1}}&=
   \frac{a_2-a_4+2 a_5+a_6}{2 \MN},\qquad
4\pi\gamma_{{M1M1}}=
   \frac{-a_2-a_4-2 a_5+a_6}{2 \MN},\nonumber \\
4\pi\gamma_{{M1E2}}&=
   \frac{-a_2-a_4-a_6}{2 \MN},\qquad
4\pi\gamma_{{E1M2}}=
   \frac{a_2-a_4-a_6}{2 \MN},\nonumber \\
4\pi\alpha_{{E2}}&= -12 \left(a_{1,t}+a_{3,t}+a_{6,t}\right)+\frac{3a_3}{\MN^2},\nonumber \\
4\pi\beta_{{M2}}&= 12
   \left(a_{1,t}-a_{3,t}-a_{6,t}\right)+\frac{3 a_3}{\MN^2},\nonumber \\
4\pi\alpha_{{E1},\nu}&= 3 a_{1,t}-a_{1,\nu}
    +a_{3,t}-a_{3,\nu }+a_{6,t}-a_{6,\nu }+\frac{-{a_3}-4a_5}{4\MN^2},\nonumber \\
4\pi\beta_{{M1},\nu }&=
  -3a_{1,t}+a_{1,\nu }+a_{3,t}-a_{3,\nu }+a_{6,t}-a_{6,\nu}
   +\frac{4a_5-a_3}{4\MN^2},\nonumber \\
4\pi\gamma_{{E2E2}}&= \frac{a_{2,t}-a_{4,t}+3 a_{5,t}+2 a_{6,t}}{6 \MN}+
   \frac{a_2+2 a_4}{48\MN^3},\nonumber \\
4\pi\gamma_{{M2M2}}&=\frac{-a_{2,t}-a_{4,t}-3 a_{5,t}+2 a_{6,t}}{6 \MN}
   +\frac{2 a_4-a_2}{48\MN^3},\nonumber \\
4\pi\gamma_{{M2E3}}&= \frac{-a_{2,t}-a_{4,t}-a_{6,t}}{3 \MN}+
  \frac{2 a_4-a_2}{24\MN^3},\nonumber \\
4\pi\gamma_{{E2M3}}&= \frac{a_{2,t}-a_{4,t}-a_{6,t}}{3 \MN}+
   \frac{a_2+2 a_4}{24\MN^3},\nonumber \displaybreak[0]\\
4\pi\gamma_{{E1E1},\nu }&= \frac{-3 a_{2,t}+a_{2,\nu }+a_{4,t}-a_{4,\nu }-5a_{5,t}+2 a_{5,\nu }-
    2 a_{6,t} +a_{6,\nu }}{2 \MN}+
   \frac{-3 a_2-2 (a_4-2 a_5)}{16\MN^3},\nonumber \displaybreak[0] \\
4\pi\gamma_{{M1M1},\nu }&= \frac{3 a_{2,t}-a_{2,\nu
   }+a_{4,t}-a_{4,\nu }+5 a_{5,t}-2 a_{5,\nu }-2
   a_{6,t}+a_{6,\nu }}{2 \MN}+
  \frac{3 a_2-2 \left(a_4+2a_5\right)}{16 \MN^3},\nonumber \displaybreak[0]\\
4\pi\gamma_{{M1E2},\nu }&=\frac{18 a_{2,t}-5 a_{2,\nu }+2
   a_{4,t}-5 a_{4,\nu }+10 a_{5,t}+12 a_{6,t}-5 a_{6,\nu }}{10\MN}
   +\frac{9 a_2-2 \left(a_4+5a_5\right)}{40 \MN^3} ,\nonumber \\
4\pi\gamma_{{E1M2},\nu }&= \frac{-18 a_{2,t}+5 a_{2,\nu }+2 a_{4,t}-5
   a_{4,\nu }-10 a_{5,t}+12 a_{6,t}-5 a_{6,\nu }}{10 \MN}+
   \frac{-9 a_2-2( a_4-5 a_5)}{40 \MN^3}.
\end{align}
The first seven lines of this, namely the dipole polarisabilities and the spin-independent quadrupole polarisabilities, agree with Babusci \etal\ \cite{Babusci:1998ww}.  
The remaining terms, the spin-dependent quadrupole polarisabilities, were considered by Holstein \etal\ \cite{Holstein:1999uu}.  However those authors, while quoting the earlier paper for the spin-independent quadrupoles,
proceed in the centre of mass frame, and so the matching is not consistent.  It should also be noted that Holstein \etal\  projected their polarisabilities directly from the c.m. amplitudes $A^{\mathrm{cm}}_i$ 
as defined in \eqref{TviaAH}, and not, as is usual for the dynamical polarisabilities, from the Ragusa amplitudes which are related to the former by a factor of $\sqrt{s}/\MN$.

It should be noted that with the definitions above, the definition of $\bar\gamma_0$ in terms of quadrupole polarisabilities in \eqref{gamma-bar} is exact and not subject to recoil corrections, in contradistinction to that of ref.~\cite{Pasquini:2010zr}.

In order to match to the dynamical polarisabilities defined in the c.m.\ frame, we plug the same Taylor expansion of the L'vov amplitudes~\eqref{lvov} into the expressions given in appendix A of ref.~\cite{Hildebrandt:2003fm} for the helicity amplitudes and partial waves $f^{l\pm}_{EE}$ etc., then use the definitions of the dynamical polarisabilities of Eqs.~(\polrefs) above, to get for example
\begin{align}4\pi\alpha_{{E1}}(\omega)=&-(a_1 + a_3+ a_6) 
+ (a_1 - a_3 - a_6) \frac{\omega}{\MN}\nonumber\\ & - \bigl(
    7 a_1 + 5 a_3 + 8 a_5 + 3 a_6 + 8\MN^2 (a_{1,\nu} - 3 a_{1,t} +a_{3,\nu} -  a_{3,t}  +  a_{6,\nu}  - 
      a_{6,t} )\bigr)\frac{ \omega^2}{8 \MN^2} +\ldots
      \label{eq:lvov-dyn}
\end{align}
\eqref{polsolbreit} can be inverted to give the $a_i$ in terms of the static polarisabilities, and substituting into \eqref{lvov-dyn} gives:
\begin{align}
\alpha_{E1}(\omega)&=\alpha_{{E1}}
+\frac{\omega \beta_{{M1}}}{\MN}+\omega^2 \left(\alpha_{{E1},\nu }
+\frac{5 \alpha_{{E1}}-2 \beta_{{M1}}}{8 \MN^2}\right)+\ldots,\nonumber  \\ 
\beta_{M1}(\omega)&=\beta_{{M1}}
+\frac{\omega \alpha_{{E1}}}{\MN}+
\omega^2 \left(\beta_{{M1},\nu }-\frac{2 \alpha_{{E1}}-5 \beta_{{M1}}}{8 \MN^2}\right)+\ldots,\nonumber  \\ 
\gamma_{E1E1}(\omega)&= \gamma_{{E1E1}}
+\omega
  \left(\frac{4 \gamma_{{E1E1}}+7 \gamma_{{M1E2}}+5
   \gamma_{{M1M1}}}{8 \MN}+\frac{\beta_{{M1}}}{16
   \MN^2}\right)\nonumber  \\ & \quad
+\omega^2 \left(\gamma_{{E1E1},\nu }+
   \frac{12 \gamma_{{E1E1}}+9 \gamma_{{E1M2}}+4 \gamma_{{M1E2}}+3 \gamma_{{M1M1}}}{16 \MN^2}+\frac{\alpha_{{E1}}}{16 \MN^3}\right)+\ldots,\nonumber  \\ 
\gamma_{M1M1}(\omega)&=\gamma_{{M1M1}}
+\omega \left(\frac{5
   \gamma_{{E1E1}}+7 \gamma_{{E1M2}}+4 \gamma_{{M1M1}}}{8 \MN}+\frac{\alpha_{{E1}}}{16 \MN^2}\right)\nonumber  \\ &\quad
+\omega^2 \left(\gamma_{{M1M1},\nu }+\frac{3 \gamma_{{E1E1}}+4 \gamma_{{E1M2}}+9 \gamma_{{M1E2}}+12
   \gamma_{{M1M1}}}{16 \MN^2}+\frac{\beta_{{M1}}}{16
   \MN^3}\right)+\ldots,\nonumber  \\ 
\gamma_{{E1M2}}(\omega)&=\gamma_{{E1M2}}+
\omega \left(\frac{2 \gamma_{{E1M2}}+3 \left(\gamma_{{M1E2}}+\gamma_{{M1M1}}\right)}{4 \MN}-\frac{\beta_{{M1}}}{8\MN^2}\right)\nonumber  \\ & \quad
+\omega^2
   \left(\gamma_{{E1M2},\nu }-\frac{\alpha_{{E1}}}{8
   \MN^3}+\frac{23 \gamma_{{E1E1}}+34 \gamma_{{E1M2}}+9
   \gamma_{{M1E2}}+10 \gamma_{{M1M1}}}{40 \MN^2}\right)+\ldots,\nonumber  \\ 
\gamma_{{M1E2}}(\omega)&=\gamma_{{M1E2}}
+\omega \left(\frac{3 \gamma_{{E1E1}}+3 \gamma_{{E1M2}}+2 \gamma_{{M1E2}}}{4 \MN}-\frac{\alpha_{{E1}}}{8 \MN^2}\right)\nonumber  \\ & \quad
+\omega^2 \left(\gamma_{{M1E2},\nu }+\frac{10 \gamma_{{E1E1}}+9 \gamma_{{E1M2}}+34 \gamma_{{M1E2}}+23 \gamma_{{M1M1}}}{40 \MN^2}
-\frac{\beta_{{M1}}}{8 \MN^3}\right)+\ldots,
\label{eq:dpols:full}
\end{align}
and
\begin{align}
\alpha_{{E2}}(\omega)&=\alpha_{{E2}}
+\frac{3\beta_{{M1}}}{2 \MN^2}
+\omega \left(\frac{12 \alpha_{{E1},\nu }+2 \alpha_{{E2}}+\beta_{{M2}}}{2 \MN}
+\frac{3 \alpha_{{E1}}}{2 \MN^3}+\frac{3 \left(\gamma_{{E1E1}}-\gamma_{{E1M2}}\right)}{2 \MN^2}\right)+\ldots,\nonumber \displaybreak[0] \\ 
\beta_{{M2}}(\omega)&=\beta_{{M2}}
+\frac{3 \alpha_{{E1}}}{2 \MN^2}+
\omega \left(\frac{12 \beta_{{M1},\nu}+\alpha_{{E2}}+2 \beta_{{M2}}}{2 \MN}+\frac{3 \beta_{{M1}}}{2 \MN^3}
-\frac{3 \left(\gamma_{{M1E2}}-\gamma_{{M1M1}}\right)}{2 \MN^2}\right)+\ldots,\nonumber  \displaybreak[0]\\ 
\gamma_{{E2E2}}(\omega)&=
\frac{1}{\omega}\left(\frac{\alpha_{{E1}}}{16 \MN^2}+\frac{\gamma_{{E1E1}}-\gamma_{{E1M2}}}{8 \MN}\right)
+\gamma_{{E2E2}}\nonumber  \\ & \quad
+\frac{3 \gamma_{{E1E1}}-2 \gamma_{{E1M2}}+\gamma_{{M1E2}}+2 \gamma_{{M1M1}}}{48\MN^2}
+\frac{\beta_{{M1}}}{16\MN^3}+\ldots,\nonumber  \\ 
\gamma_{{M2M2}}(\omega)&=\frac{1}{\omega}\left(\frac{\beta_{{M1}}}{16\MN^2}-\frac{\gamma_{{M1E2}}-\gamma_{{M1M1}}}{8\MN}\right)
+\gamma_{{M2M2}}\nonumber  \\ & \quad
+\frac{2 \gamma_{{E1E1}}+\gamma_{{E1M2}}-2 \gamma_{{M1E2}}+3
   \gamma_{{M1M1}}}{48 \MN^2}+\frac{\alpha_{{E1}}}{16 \MN^3}+\ldots,\nonumber  \\ 
\gamma_{{E2M3}}(\omega)&=
\gamma_{{E2M3}}+
\frac{\gamma_{{E1M2}}+\gamma_{{M1E2}}+2 \gamma_{{M1M1}}}{24 \MN^2}+\ldots,\nonumber  \\ 
\gamma_{{M2E3}}(\omega)&=
\gamma_{{M2E3}}+
\frac{2 \gamma_{{E1E1}}+\gamma_{{E1M2}}+\gamma_{{M1E2}}}{24 \MN^2}+\ldots\,.
\label{eq:qpols:full}
\end{align}
Note the divergent $1/\omega$ pieces in the unmixed spin quadrupole polarisabilities, generated by the recoil terms.
As noted above, there are no recoil corrections to the $l=1$ polarisabilities at leading order in $\omega$.

To conclude this appendix, we provide here the numerical values of the non-Born coefficients $a_i$, $a_{i,\nu}$ and $a_{i,t}$ that result from our calculation,
see Table~\ref{tab:ai}. Our calculations have been performed with the use of computational packages FORM~\cite{Vermaseren:2000nd} and LoopTools~\cite{Hahn:1998yk}.

\begin{table}
\begin{tabular}{|r|r|r|r|r|r|r|} \hline
   &\multicolumn{2}{c|}{$\pi $N loop} &
   $\pi \Delta $ loop & $\Delta $ pole &
   \multicolumn{2}{c|}{total}\\ 
   & proton &  neutron & &  & proton & neutron \\\hline
 $a_1$ & -55.06 & -65.86 & -36.89 & 45.61 & -46.34 & -57.14 \\
 $a_2$ & -22.70 & -44.75 & 13.42 & -48.57 & -57.85 & -79.9 \\
 $a_3$ & 116.35 & 151.98 & -28.29 & -131.71 & -43.65 & -8.03 \\
 $a_4$ & 60.15 & 88.92 & -3.05 & -84.98 & -27.88 & 0.89 \\
 $a_5$ & -85.73 & -114.79 & 12.93 & -85.82 & -158.62 & -187.68 \\
 $a_6$ & -148.38 & -204.00 & 9.58 & 87.48 & -51.33 & -106.94 \\
 $a_{1,\nu}$ & -24.16 & -30.09 & -2.25 & 34.76 & 8.35 & 2.42 \\
 $a_{2,\nu}$  & -71.01 & -82.86 & 2.59 & -20.03 & -88.45 & -100.31 \\
 $a_{3,\nu}$ & 92.46 & 121.44 & -1.43 & -44.44 & 46.58 & 75.56 \\
 $a_{4,\nu}$ & 62.28 & 87.10 & -0.28 & -28.67 & 33.32& 58.15 \\
 $a_{5,\nu}$ & -79.57 & -105.54 & 0.59 & -28.96 &-107.94 & -133.91 \\
 $a_{6,\nu}$ & -111.01 & -148.18 & 0.65 & 26.08 & -84.28 & -121.46 \\
 $a_{1,t}$ & -11.49 & -11.07 & -3.04 & -2.65 & -17.18 &-16.76 \\
 $a_{2,t}$ & 0.94 & 2.02 & -0.14 & 2.31 & 3.10 & 4.18 \\
 $a_{3,t}$ & 5.15 & 1.68 & -1.13 & 7.01 & 11.03 & 7.56 \\
 $a_{4,t}$  & -1.83 & -4.81 & 0.08 & 5.58 & 3.84 & 0.85\\
 $a_{5,t}$ & -5.15 & -2.23 & 0.27 & 6.60 & 1.71 & 4.63 \\
 $a_{6,t}$ & -6.55 & -1.95 & 0.55 & -6.43 & -12.43 & -7.82 \\
\hline
\end{tabular} 
\caption{Predictions for the parameters of the Taylor expansion of the L'vov amplitudes $A^\mathrm{L}_i(\nu,t)$,
in units of $10^4 $~fm$^{-3}$ for $a_i$ and $10^4 $~fm$^{-5}$ for  $a_{i,t}$ and $a_{i,\nu}$}
\label{tab:ai}
\end{table}

\section{Expressions for static polarisabilities}
\label{app:pols}
Here we collect the analytic expressions for the various contributions computed in this work.
The corresponding numerical results are shown in 
Tables~\ref{tab:polsscalar}--\ref{tab:polsspin_n}. 
The spin polarisabilities $\gamma_1$--$\gamma_4$ provided here
are related to those defined in Sect.~\ref{sec:pols:static:def} by
\begin{equation}
\gamma_{E1E1}=-\gamma_1-\gamma_3,\quad \gamma_{M1M1}=\gamma_4,\quad \gamma_{E1M2}=\gamma_3,\quad \gamma_{M1E2}=\gamma_2+\gamma_4\,.
\nonumber
\end{equation}
\subsection{\texorpdfstring{$\piN$}{pi-N} loops}
\subsubsection{Proton}
The complete $O(p^3)$ chiral-loop contribution to the dipole, spin and quadrupole polarisabilities of the proton are 
(with $\mu=m_\pi/\MN$):
\beq
\begin{split}
\alpha_{E1}^{(\piN)}=&\frac{e^2 g_A^2}{192 \pi^3 f_\pi^2 \MN}\frac{1}{ \mu  \left(4-\mu^2\right)^{5/2}}\\
 \times &\left[
\vphantom{\left.\arccos\left(\frac{\mu ^2}{2}-1\right)\right)}
\mu  \left(-18 \mu ^6+157 \mu ^4-407 \mu ^2+304\right)
   \sqrt{4-\mu ^2} + \mu  \left(9 \mu ^4-20 \mu ^2+9\right) \left(4-\mu ^2\right)^{5/2}
   \log \mu ^2\right.\\
&\left.-\left(9 \mu ^{10}-110 \mu ^8+479 \mu ^6-870 \mu ^4+590 \mu ^2-80\right)
   \arccos\left(\frac{\mu ^2}{2}-1\right)\right]\,,
\end{split}
\eeq
\beq
\begin{split}
 \beta_{M1}^{(\piN)}=-&\frac{e^2 g_A^2}{192 \pi ^3 f_\pi^2 \MN}\frac{1}{\mu  \left(4-\mu^2\right)^{3/2}}\\
\times
& \left[
\vphantom{\left.\arccos\left(\frac{\mu ^2}{2}-1\right)\right)}
\mu  \left(-54 \mu ^4+235 \mu ^2-127\right) \sqrt{4-\mu ^2}-\mu 
   \left(27 \mu ^4-50 \mu ^2+9\right) \left(4-\mu ^2\right)^{3/2} \log \mu^2\right. \\
&\left. -\left(27 \mu ^8-212 \mu ^6+471 \mu ^4-246 \mu ^2+2\right) 
\arccos \left(\frac{\mu ^2}{2}-1\right)\right]\,,
\end{split}
\eeq
\beq
\begin{split}
\gamma_1^{(\piN)}=& \frac{e^2 g_A^2}{ 192 \pi ^3 f_\pi^2 \MN^2}\frac{1}{\mu ^2 \left(4-\mu ^2\right)^{5/2}}\\
\times & \left[
\vphantom{\left.\arccos\left(\frac{\mu ^2}{2}-1\right)\right)}
\sqrt{4-\mu ^2} \left(18 \mu ^8-173 \mu ^6+524 \mu ^4-452 \mu ^2-\left(4-\mu ^2\right)^2 \left(9 \mu ^4-28 \mu ^2+13\right) \mu ^2
 \log\mu ^2+32\right)\right.\\
&\left.+\mu  \left(9 \mu ^{10}-118 \mu ^8+563 \mu ^6-1150 \mu ^4+848 \mu ^2-104\right) \arccos\left(\frac{\mu^2}{2}-1\right)\right]\,,
\end{split}
\eeq
\beq
\begin{split}
\gamma_2^{(\piN)}=&\frac{e^2 g_A^2}{96 \pi ^3 f_\pi^2\MN^2} \frac{1}{ \mu ^2 \left(4-\mu ^2\right)^{5/2}}\\
\times & \left[
\vphantom{\left.\arccos\left(\frac{\mu ^2}{2}-1\right)\right)}
2 \sqrt{4-\mu ^2} \left(18 \mu ^8-149 \mu ^6+348 \mu ^4-179 \mu^2
-\left(4-\mu ^2\right)^2 \left(9 \mu ^4-16 \mu ^2+4\right) \mu ^2 \log \mu^2+4\right)\right.\\
&\left. 
+\mu  \left(18 \mu ^{10}-212 \mu ^8+868 \mu ^6-1400 \mu ^4+701 \mu^2-44\right) \arccos\left(\frac{\mu ^2}{2}-1\right)\right]\,,
\end{split}
\eeq
\beq
\begin{split}
\gamma_3^{(\piN)}=-&\frac{e^2 g_A^2}{384 \pi ^3 f_\pi^2 \MN^2}\frac{1}{\mu ^2 \left(4-\mu ^2\right)^{5/2}}\\
\times& \left[
\vphantom{\left.\arccos\left(\frac{\mu ^2}{2}-1\right)\right)}
\sqrt{4-\mu ^2} \left(18 \mu ^8-141 \mu ^6+284 \mu ^4-4 \mu ^2-3
   \left(\mu ^2-4\right)^2 \left(3 \mu ^4-4 \mu ^2-1\right) \mu ^2 \log \mu^2-16\right)\right.\\
&\left.+\mu  \left(9 \mu ^{10}-102 \mu ^8+387 \mu ^6-510 \mu ^4+46 \mu^2+56\right) \arccos \left(\frac{\mu ^2}{2}-1\right)\right]\,,
\end{split}
\eeq
\beq
\begin{split}
\gamma_4^{(\piN)}=&\frac{e^2 g_A^2 }{384 \pi ^3 f_\pi^2\MN^2}\frac{1}{ \mu ^2 \left(4-\mu ^2\right)^{5/2}}\\
\times &\left[
\vphantom{\left.\arccos\left(\frac{\mu ^2}{2}-1\right)\right)}
\sqrt{4-\mu ^2} \left(-126 \mu ^8+1047 \mu ^6-2462 \mu ^4+1308 \mu ^2+\left(\mu ^2-4\right)^2 \left(63 \mu ^4-114 \mu ^2+29\right) \mu ^2
\log\mu ^2-16\right)\right.\\
&\left.+\left(-63 \mu ^{11}+744 \mu ^9-3059 \mu ^7+4970 \mu ^5-2534 \mu ^3+152 \mu \right) \arccos\left(\frac{\mu^2}{2}-1\right)\right]\,,
\end{split}
\eeq
\beq
\begin{split}
\alpha_{E2}^{(\piN)}=& -\frac{3}{2\MN^2}\beta_{M1}^{(\piN)}+ \frac{e^2 g_A^2}{1920 \pi^3 f_\pi^2 \MN^3}\frac{1}{\mu^3\left(4-\mu ^2\right)^{7/2}}\\
&\times\left[
\vphantom{\left.\arccos\left(\frac{\mu ^2}{2}-1\right)\right)}
\mu \sqrt{4-\mu ^2} \Bigl(2610 \mu ^{10}-33069 \mu ^8+148813 \mu ^6-273436 \mu ^4+168848 \mu ^2-14144\right.\\
&\quad -3 \left(\mu^2-4\right)^3 \left(435 \mu ^4-944 \mu ^2+385\right) \mu ^2 \log \mu^2 \Bigr)\\
&\quad +3 \left(435 \mu ^{14}-7034 \mu ^{12}+44051 \mu^{10}-132370 \mu ^8+189490 \mu ^6-107960 \mu ^4+12576 \mu ^2+896\right)\\
&\qquad \left.\times   \arccos\left(\frac{\mu ^2}{2}-1\right)\right]\,,
\end{split}
\eeq
\beq
\begin{split}
\beta_{M2}^{(\piN)}=&-\frac{3}{2\MN^2} \alpha_{E1}^{(\piN)}+ \frac{e^2 g_A^2}{1920\pi^3 f_\pi^2 \MN^3}\frac{1}{\mu^3 \left(4-\mu ^2\right)^{5/2}}\\
&\times \left[
\vphantom{\left.\arccos\left(\frac{\mu ^2}{2}-1\right)\right)}
\mu\sqrt{4-\mu ^2} \Bigl(-4230 \mu ^8+35619 \mu ^6-86105 \mu ^4+49108 \mu ^2-368\right.\\
&\quad+3 \left(\mu ^2-4\right)^2 \left(705 \mu^4 -1354 \mu ^2+385\right) \mu ^2 \log \mu^2\Bigr)\\
&\quad\left.
-3 \left(705 \mu ^{12}-8404 \mu ^{10}+35075 \mu ^8-58570 \mu ^6+31630 \mu ^4
  -2200 \mu^2+96\right) \arccos\left(\frac{\mu ^2}{2}-1\right)\right]\,.
\end{split}
\eeq

\subsubsection{Neutron}
The complete $O(p^3)$ chiral-loop contribution to the dipole, spin and quadrupole polarisabilities of the neutron are 
(with $\mu=m_\pi/\MN$):
\beq
\begin{split}
\alpha_{E1}^{(\piN,n)}=&\frac{e^2 g_A^2}{192 \pi^3 f_\pi^2 M_N}\frac{1}{ \mu  \left(4-\mu ^2\right)^{5/2} }\\
\times &\left[\mu  \sqrt{4-\mu ^2} \left(-\mu ^2+3 \left(\mu ^2-4\right)^2 \log \mu ^2+16\right)
+\left(-3 \mu ^6+30 \mu ^4-98 \mu ^2+80\right) \arccos\left(\frac{\mu ^2}{2}-1\right)\right]\,,
\end{split}
\eeq
\beq
\begin{split}
 \beta_{M1}^{(\piN,n)}=-&\frac{e^2 g_A^2}{192 \pi ^3 f_\pi^2 M_N}\frac{1}{\mu \left(4-\mu ^2\right)^{3/2}}\\
 \times &\left[\mu  \sqrt{4-\mu ^2} \left(3 \left(\mu ^2-4\right) \log \mu ^2-11\right)-\left(3 \mu ^4-18 \mu ^2+2\right) \arccos\left(\frac{\mu ^2}{2}-1\right)\right]\,,
\end{split}
\eeq
\beq
\begin{split}
\gamma_1^{(\piN,n)}=& \frac{e^2 g_A^2 }{96 \pi ^3 f_\pi^2 M_N^2}\frac{1}{ \mu ^2 \left(4-\mu ^2\right)^{5/2}}\\
\times &
\left[
\vphantom{\left.\arccos\left(\frac{\mu ^2}{2}-1\right)\right)}
\sqrt{4-\mu ^2} \left(2 \left(2 \mu ^4-7 \mu ^2+8\right)-\mu ^2 \left(\mu ^2-4\right)^2 \log \mu ^2\right)\right.\\
&\left. 
+\mu  \left(\mu ^6-10 \mu ^4+24 \mu ^2-12\right) \arccos\left(\frac{\mu^2}{2}-1\right)\right]\,,
\end{split}
\eeq
\beq
\begin{split}
\gamma_2^{(\piN,n)}=&\frac{e^2 g_A^2}{192 \pi ^3 f_\pi^2 M_N^2}\frac{1}{ \mu ^2 \left(4-\mu ^2\right)^{5/2}}\\
\times &
\left[
\vphantom{\left.\arccos\left(\frac{\mu ^2}{2}-1\right)\right)}
\sqrt{4-\mu ^2} \left(6 \mu ^4-52 \mu ^2-3 \left(\mu ^2-4\right)^2 \mu ^2 \log \mu ^2+16\right)\right.\\
&\left.
+\mu  \left(3 \mu ^6-30 \mu ^4+94 \mu ^2-40\right) \arccos\left(\frac{\mu^2}{2}-1\right)\right]\,,
\end{split}
\eeq
\beq
\begin{split}
\gamma_3^{(\piN,n)}=&-\frac{e^2 g_A^2}{384 \pi ^3 f_\pi^2  M_N^2}\frac{1}{\mu ^2 \left(4-\mu ^2\right)^{5/2}}\\
&\times 
\left[
\vphantom{\left.\arccos\left(\frac{\mu ^2}{2}-1\right)\right)}
\sqrt{4-\mu ^2} \left(\mu ^2 \left(\mu ^2-4\right)^2 \log\mu ^2+2 \left(\mu ^4+10 \mu ^2-8\right)\right)\right.\\
&\left.
-\mu  \left(\mu ^6-10 \mu ^4+46 \mu ^2-40\right) \arccos\left(\frac{\mu^2}{2}-1\right)\right]\,,
\end{split}
\eeq
\beq
\begin{split}
\gamma_4^{(\piN,n)}=&\frac{e^2 g_A^2}{384 \pi ^3 f_\pi^2  M_N^2}\frac{1}{\mu ^2 \left(4-\mu ^2\right)^{5/2}}\\
\times &
\left[
\vphantom{\left.\arccos\left(\frac{\mu ^2}{2}-1\right)\right)}
\sqrt{4-\mu ^2} \left(-6 \mu ^4+76 \mu ^2+5 \left(\mu ^2-4\right)^2 \mu ^2 \log \mu ^2-16\right)\right.\\
&\left.
-\mu\left(5 \mu ^6-50 \mu ^4+166 \mu ^2-88 \right) \arccos\left(\frac{\mu^2}{2}-1\right)\right]\,,
\end{split}
\eeq
\beq
\begin{split}
\alpha_{E2}^{(\piN,n)}=& -\frac{3}{2\MN^2}\beta_{M1}^{(\piN,n)}
-\frac{e^2 g_A^2}{1920\pi ^3 f_\pi^2 M_N^3}\frac{1}{\mu ^3 \left(4-\mu ^2\right)^{7/2}}\\
&\times 
\left[
\vphantom{\left.\arccos\left(\frac{\mu ^2}{2}-1\right)\right)}
\mu  \sqrt{4-\mu ^2} \left(-617 \mu ^6+5796 \mu ^4-19728 \mu ^2+570 \left(\mu ^2-4\right)^3 \mu ^2 \log \mu+14144\right)\right.\\
&\left.-3 \left(95 \mu ^{10}-1330 \mu ^8+6570 \mu ^6-13320 \mu ^4+6816 \mu
   ^2+896\right) \arccos\left(\frac{\mu ^2}{2}-1\right)\right]\,,
\end{split}
\eeq
\beq
\begin{split}
\beta_{M2}^{(\piN,n)}=&-\frac{3}{2\MN^2} \alpha_{E1}^{(\piN,n)}+ 
\frac{e^2 g_A^2}{1920\pi^3 f_\pi^2 M_N^3}\frac{1}{\mu ^3 \left(4-\mu ^2\right)^{5/2}} \\
&\times 
\left[
\vphantom{\left.\arccos\left(\frac{\mu ^2}{2}-1\right)\right)}
\mu  \sqrt{4-\mu ^2} \left(-523 \mu ^4+3780 \mu ^2+570 
\left(\mu ^2-4\right)^2 \mu ^2 \log \mu-368\right)\right.\\
&\left. -3 \left(95 \mu ^8-950 \mu ^6+2890 \mu ^4-1400 \mu ^2+96\right) \arccos \left(\frac{\mu^2}{2}-1\right)\right]\,.
\end{split}
\eeq

\subsection{\texorpdfstring{$\pi \Delta$}{pi-Delta} loops}
Our expressions for the $\pi \Delta$-loop contributions are sufficiently compact when we leave one of the integrations
over the Feynman parameters undone. The integrands involve the following function of the Feynman parameter $x$:
\beq
D_\Delta(x)=x^2 + (1 + \delta)^2 - x (2 + 2 \delta + \delta^2 - \mu^2),
\eeq
where   $\mu=m_\pi/\MN$, $\delta=\varDelta/\MN$.  Furthermore, 
the divergent parts of the polarisabilities, absorbed by higher-order contact terms,
are renormalised according to the modified minimal subtraction ($\overline{\mathrm{MS}}$) scheme, by setting 
to 0 the factor arising in the dimensional regularisation: 
\beq
\Xi=\frac{2}{4-D}-\gamma_E+\log\frac{4\pi\Lambda^2}{\MN^2},
\eeq
with $D\simeq 4$ the number of dimensions, $\ga_E$ the Euler constant, and $\Lambda$ the renormalization scale.

With these definitions we obtain:\footnote{Here we correct the mistakes of ref.~\cite{Lensky:2009uv} in the
 expressions for $\alpha_{E1}^{(\pi \Delta)}$ and $\beta_{M1}^{(\pi \Delta)}$. The expressions for the other polarisabilities
 appear for the first time.}
\beq
\begin{split}
\alpha_{E1}^{(\pi \Delta)}=\frac{e^2 h_A^2 \MN}{1728 \pi ^3 f_\pi^2 M_{\Delta }^2}\left[8 \delta+\frac{35}{3}+
\int\limits^1_0\mathrm{d}x
 \left(\vphantom{\frac{18 (x-1)^2 x^4 (x+\delta+1)}{D_{\Delta }^2}}\right.\right.
  &\frac{18 (x-1)^2 x^4 (x+\delta+1)}{D_{\Delta }^2(x)}\\
  &-\frac{x^2 \left(120 x^3+4 (27 \delta-16) x^2 -  (166 \delta +115)x +57 (\delta +1)\right)}{D_{\Delta }(x)}\\
  &+6 x (x (22x+15 \delta-1)-12 (\delta +1)) \left[\Xi -\log D_{\Delta}(x)\right]
 \left.
   \left.
    \vphantom{\frac{18 (x-1)^2 x^4 (x+\delta+1)}{D_{\Delta }^2}}
   \right)
  \vphantom{\int\limits^1_0\mathrm{d}x\frac{18 (x-1)^2 x^4 (x+\delta+1)}{D_{\Delta }^2}}
 \right]\,,
\end{split}
\eeq
\beq
\begin{split}
 \beta_{M1}^{(\pi \Delta)}=\frac{e^2 h_A^2 \MN}{1728 \pi ^3 f_\pi^2 M_{\Delta }^2}
 \left[-8 \delta-\frac{65}{6}+\int\limits_0^1\mathrm{d}x
 \left(\vphantom{\frac{\left(24 x^2-32 x+9\right) x^2 (\delta +x+1)}{D_{\Delta }}}\right.\right.
 &\frac{\left(24 x^2-32 x+9\right) x^2 (x+\delta+1)}{D_{\Delta }(x)}\\
 &-6 (5 x-4) x (2x+3 \delta+3)\left[\Xi -\log D_{\Delta }(x)\right]
 \left.
  \left.
    \vphantom{\frac{\left(24 x^2-32 x+9\right) x^2 (\delta +x+1)}{D_{\Delta }}}
  \right)
 \right]\,,
\end{split}
\eeq
\beq
\begin{split}
\gamma_1^{(\pi \Delta)}=\frac{e^2 h_A^2}{1728 \pi ^3 f_\pi^2 M_{\Delta }^2}
 \left[2 \delta+\frac{7}{4} +\int\limits_0^1\mathrm{d}x\left(
\vphantom{\frac{8 (x-1)^3 x^5 (\delta +x+1)}{D_{\Delta}^3(x)}} \right.\right.
&\frac{8 (x-1)^3 x^5 (x+\delta+1)}{D_{\Delta}^3(x)}\\
&-\frac{2 (x-1)^2 x^3 \left(22 x^2+8 \delta  x+3 \delta +3\right)}{D_{\Delta}^2(x)}\\
&+\frac{(x-1) x^2 \left(105 x^2-2 (5 \delta +38) x+33 (\delta +1)\right)}{D_{\Delta}(x)}\\
&+2 x (x (-16 x+9 \delta +21)-12 (\delta +1)) \left[\Xi -\log D_{\Delta }(x)\right]
\left.
 \left.
  \vphantom{\frac{8 (x-1)^3 x^5 (\delta +x+1)}{D_{\Delta}^3(x)}}
 \right)
 \vphantom{\int\limits_0^1\frac{8 (x-1)^3 x^5 (\delta +x+1)}{D_{\Delta}^3(x)}}
\right]\,,
\end{split}
\eeq
\beq
\begin{split}
\gamma_2^{(\pi \Delta)}=\frac{e^2 h_A^2}{1728 \pi ^3 f_\pi^2 M_{\Delta }^2}
\left[\frac{5}{12}+\int\limits_0^1\mathrm{d}x\left(
 \vphantom{\frac{2 \left(4 x^2-7 x+3\right) x^3 (x+\delta+1)}{D_{\Delta}^2(x)}}\right.\right.
 &\frac{2 \left(4 x^2-7 x+3\right) x^3 (x+\delta+1)}{D_{\Delta}^2(x)}\\
 &-\frac{x^2 \left(21 x^3-4 (3 \delta +5) x^2+(37 \delta +26) x-24 (\delta +1)\right)}{D_{\Delta }(x)}\\
 &+12 (3 x-2) x^2 \left[\Xi -\log D_{\Delta} (x)\right]
 \left.
    \left.
    \vphantom{\frac{2 \left(4 x^2-7 x+3\right) x^3 (x+\delta+1)}{D_{\Delta}^2(x)}}
    \right)
    \vphantom{\int\limits_0^1\frac{2 \left(4 x^2-7 x+3\right) x^3 (x+\delta+1)}{D_{\Delta}^2(x)}}
 \right]\,,
\end{split}
\eeq
\beq
\begin{split}
\gamma_3^{(\pi \Delta)}=\frac{e^2 h_A^2}{1728 \pi ^3 f_\pi^2 M_{\Delta }^2}
\left[-\delta-\frac{2}{3}+\int\limits^1_0\mathrm{d}x \left(
 \vphantom{\frac{(x-1) x^5 (x+\delta+1)}{D_{\Delta }^2(x)}}\right.\right.
 &\frac{(x-1) x^5 (x+\delta+1)}{D_{\Delta }^2(x)}\\
 &-\frac{x^2 \left(24 x^3+(2 \delta -11) x^2+(8 \delta -1) x-9 (\delta +1)\right)}{2D_{\Delta}(x)}\\
 &+ x (x (34x-9 \delta-33)+12 (\delta +1)) \left[\Xi -\log D_{\Delta }(x)\right]
 \left.
    \left.
    \vphantom{\frac{(x-1) x^5 (x+\delta+1)}{D_{\Delta }^2(x)}}
    \right)
    \vphantom{\int\limits_0^1\frac{(x-1) x^5 (x+\delta+1)}{D_{\Delta }^2(x)}}
 \right]\,,
 \end{split}
\eeq
\beq
\begin{split}
\gamma_4^{(\pi \Delta)}=\frac{e^2 h_A^2}{1728 \pi ^3 f_\pi^2 M_{\Delta }^2}
\left[\delta +\frac{2}{3}+ \int\limits_0^1\mathrm{d}x\left(
 \vphantom{-\frac{\left(x^3+5 x^2-12 x+6\right) x^3 (x+\delta+1)}{D_{\Delta }^2(x)}}\right.\right.
 &-\frac{\left(x^3+5 x^2-12 x+6\right) x^3 (x+\delta+1)}{D_{\Delta }^2(x)}\\
 &+\frac{x^2 \left(48 x^3-(22 \delta +59) x^2+(80 \delta +71)x-57 (\delta +1)\right)}{2D_{\Delta}(x)}\\
 &+x (x ( -34 x+9 \delta+33)-12 (\delta+1)) \left[\Xi -\log D_{\Delta }(x)\right]
 \left.
    \left.
    \vphantom{-\frac{\left(x^3+5 x^2-12 x+6\right) x^3 (x+\delta+1)}{D_{\Delta }^2(x)}}
    \right)
    \vphantom{\int\limits_0^1-\frac{\left(x^3+5 x^2-12 x+6\right) x^3 (x+\delta+1)}{D_{\Delta }^2(x)}}
\right]\,,
\end{split}
\eeq
\beq
\begin{split}
\alpha_{E2}^{(\pi\Delta)}=& \frac{e^2 h_A^2}{17280\pi^3 f_\pi^2 M_\Delta^2 \MN}\\
\times\left[
\vphantom{\int\limits^1_0\frac{80 (x-1)^2 x^5 (13 x-9) (x+\delta+1)}{D_{\Delta}^3(x)}}
\right.
&-\frac{325}{2}-120\delta\\
&+\int\limits_0^1\mathrm{d}x\left(\frac{80 (x-1)^2 x^5 (13 x-9) (x+\delta+1)}{D_{\Delta}^3(x)}\right.\\
&\quad-\frac{2x^3\left(2235 x^4+(2095 \delta-2569) x^3-3 (1518 \delta +563) x^2+15 (199 \delta +167) x-570 (\delta +1)\right)}{D_{\Delta}^2(x)}\\
&\quad+\frac{15 x^2 \left(201 (\delta +1)+494 x^3+4 (101 \delta -72) x^2-(608 \delta +407) x\right)}{D_{\Delta}(x)}\\
&\quad-90 x (5 x-4) (2x+3 \delta+3) \left[\Xi -\log D_{\Delta }(x)\right]
\left.
\vphantom{\frac{80 (x-1)^2 x^5 (13 x-9) (x+\delta+1)}{D_{\Delta}^3(x)}}
\right)
\left.
\vphantom{\int\limits^1_0\frac{80 (x-1)^2 x^5 (13 x-9) (x+\delta+1)}{D_{\Delta}^3(x)}}
\right]-\frac{3}{2}\frac{\beta_{M1}^{(\pi\Delta)}}{\MN^2}\,,
\end{split}
\eeq
\beq
\begin{split}
\beta_{M2}^{(\pi \Delta)}=&\frac{e^2 h_A^2}{5760\pi^3 f_\pi^2 M_\Delta^2 \MN}\\
\times
\left[
\vphantom{\int\limits^1_0\frac{80 (x-1)^2 x^5 (13 x-9) (x+\delta+1)}{D_{\Delta}^3(x)}}
\right.
&\frac{175}{3} + 40\delta\\
&+\int\limits_0^1\mathrm{d}x\left(\frac{4 x^3 \left(75 x^3-146 x^2+80 x-15\right) (x+\delta+1)}{D_{\Delta}^2(x)}\right.\\
&\quad-\frac{5 x^2 \left(286 x^3+(320 \delta-84 ) x^2-5 (94 \delta +71) x+153 (\delta +1)\right)}{D_{\Delta}(x)}\\
&\quad+30 x \left(22 x^2+(15 \delta -1) x-12 (\delta +1)\right)\left[\Xi -\log D_{\Delta }(x)\right]
\left.
\vphantom{\frac{80 (x-1)^2 x^5 (13 x-9) (x+\delta+1)}{D_{\Delta}^3(x)}}
\right)
\left.
\vphantom{\int\limits^1_0\frac{80 (x-1)^2 x^5 (13 x-9) (x+\delta+1)}{D_{\Delta}^3(x)}}
\right]-\frac{3}{2}\frac{\alpha_{E1}^{(\pi\Delta)}}{\MN^2}\,.
\end{split}
\eeq

\subsection{ \texorpdfstring{$\Delta$}{Delta} excitation}
Introducing $M_+ = \MN + M_\Delta$ and recalling that $\varDelta = M_\De-\MN$, the contribution of the Delta(1232) electromagnetic excitation  to the dipole,
spin and quadrupole polarisabilities of the proton are, respectively:
\bea
\alpha_{E1}^{(\Delta)}&=&-\frac{e^2}{4 \pi}\frac{ 2 g_E^2}{   M_+^3}\,,\\
\beta_{M1}^{(\Delta)}&=&\hphantom{-}\frac{e^2}{4 \pi}\frac{ 2 g_M^2}{ M_+^2 \varDelta   }\,,
\eea
\bea
\gamma_1^{(\Delta)} &=&\hphantom{-}\frac{e^2}{4 \pi}\frac{1}{  M_+^2 \MN}\left(\frac{g_M^2}{\varDelta }-\frac{2 g_E g_M  }{  M_+}
+\frac{g_E^2  M_\Delta}{  M_+^2}\right)\,,\\
\gamma_2^{(\Delta)} &=&-\frac{e^2}{4 \pi }\frac{1}{ M_+^2 \MN}\left(\frac{g_M^2 M_\Delta}{\varDelta^2}+\frac{g_E g_M}{\varDelta}\right)\,,\\
\gamma_3^{(\Delta)} &=&-\frac{e^2}{4 \pi} \frac{1}{ M_+^2 \MN}\left(\frac{g_M^2}{2\varDelta}-\frac{3g_E g_M}{2 M_+}\right)\,,\\
\gamma_4^{(\Delta)} &=&\hphantom{-}\frac{e^2}{4 \pi}\frac{1}{ M_+^2 \MN}
\left(\frac{ g_M^2 M_\Delta}{\varDelta^2}-\frac{g_E g_M}{2   \varDelta}+\frac{g_E^2}{2  M_+}\right)\,,
\eea
\bea
\alpha_{E2}^{(\Delta)}&=&-\frac{3}{2\MN^2}\beta_{M1}^{(\Delta)}+\frac{e^2}{4\pi} \frac{3}{ M_+^2 \MN^2}
\left(\frac{2g_M^2}{\varDelta}
-\frac{g_E g_M}{ M_+}
+\frac{g_E^2\left(3M_\Delta M_+ +\varDelta^2\right)}{M_+^2 \varDelta }\right)\,,\quad\\
\beta_{M2}^{(\Delta)}&=&-\frac{3}{2\MN^2}\alpha_{E1}^{(\Delta)}-\frac{e^2}{4\pi}\frac{3}{ M_+^2 \MN^2}
\left(\frac{g_M^2( 3M_\Delta \varDelta + M_+^2)}{\varDelta^2 M_+}
-\frac{g_E g_M}{\varDelta}+\frac{2g_E^2}{ M_+}\right)\,.
\eea

\section{Details of higher-order Delta self-energy}
\label{sec:remarks-details}
For the benefit of future workers, we here provide the details of the corrections to the  Delta self-energy which enter beyond the order to which we work in this paper, as discussed in Section \ref{sec:remarks}.

We write the Delta self-energy
in the form:
\beq
\Sigma(\slashed{p})=\sigma(s)+\tau(s)(\slashed{p}-M_\Delta)\,,
\eeq
where the one-loop contributions with the intermediate nucleon or Delta are given by~\cite{Pascalutsa:2005vq}
\begin{equation}
\begin{split}
\sigma(s)_{N,\Delta}&=-\frac{C_{N,\Delta}}{2(8\pi f_\pi)^2}\int\limits_0^1 dx(x M_\Delta + M_{N,\Delta})\mathcal{M}^2_{N,\Delta}\left[-\Xi-1+\log\mathcal{M}^2_{N,\Delta}\right]\,,\\
\tau(s)_{N,\Delta}&=-\frac{C_{N,\Delta}}{2(8\pi f_\pi)^2}\int\limits_0^1 dx\, x\mathcal{M}^2_{N,\Delta}\left[-\Xi-1+\log\mathcal{M}^2_{N,\Delta}\right]\,,
\end{split}
\end{equation}
with $\Xi$ the constant representing the ultraviolet divergence, $C_N=h_A^2$, $C_\Delta=(5/3 H_A)^2$, and
\begin{equation}
\mathcal{M}^2_{N,\Delta}=m_\pi^2 x+M_{N,\Delta}^2(1-x)-s\, x(1-x).
\end{equation}
For the $\pi \Delta$ axial coupling we adopt the large-$N_C$ value $H_A=9/5 g_A$.
The renormalised functions $\sigma_R(s)$ and $\tau_R(s)$ are obtained after two and one subtractions, respectively:
\beq
\sigma_R(s)=\sigma(s)-\re\sigma(M_\Delta^2)-(s-M_\Delta^2)\re\sigma'(M_\Delta^2)\,,\quad \tau_R(s)=\tau(s)-\re\tau(M_\Delta^2)\,.
\eeq
The running Delta mass is then given by $M_\Delta(s)=M_\Delta +\re\sigma_R(s)$,
while the Delta width is $\Gamma(s)=-2\im\sigma(s)$.

While $\sigma_R(s)$ is finite, $\tau_R(s)$ still contains a divergent piece which is to be 
renormalized by an appropriate low-energy
constant. In the absence of further information on such a constant, we neglect it and 
merely set $\Xi=0$.
Note that the expressions for $\sigma(s)$ and $\tau(s)$ above do not contain the extra factor of $s/M_\Delta^2$ seen in ref.~\cite{Pascalutsa:2005vq}.
This is due to the fact that each of the leading $\gamma$N$\Delta$ vertices in the first graph in Fig.~\ref{fig:piDloops} brings in an extra
Delta momentum giving a total factor of $s$ (see ref.~\cite{Pascalutsa:2002pi}). When the self energy graphs are resummed in the
intermediate state, the factor $s/M_\Delta^2$ is taken out from the self energy in order to be restored in front of the total result, hence
this factor is absent in the expressions we use here. 

We also give here the dispersion integral that simplifies the calculation
of the running couplings $g_{M,E}(s)$. Namely,
the running due to the leading loop corrections can be calculated
from the already employed imaginary parts of these corrections by the once-subtracted integral
\beq
\re g_{M,E}(s) =\re g_{M,E}(M_\Delta^2) + (s-M_\Delta^2)\int\limits_{(M_N+m_\pi)^2}^\infty \frac{ds}{\pi}\frac{\im g_{M,E}(s')}{(s'-M_\Delta^2)(s'-s)}\,,
\eeq
where the integral is understood as the Cauchy principal value; note the explicit $(s-M_\Delta^2)$ factor.

\end{document}